\documentclass[11pt,a4paper]{article}
\usepackage{amsmath, amsfonts,amssymb,mathtools,nicefrac,nccmath,cases}
\usepackage{microtype,booktabs,multirow}
\usepackage[usenames,dvipsnames,table]{xcolor}
\usepackage{colortbl}
\usepackage{tikz}
\usetikzlibrary{shapes,arrows}
\usetikzlibrary{calc,shapes.callouts,shapes.arrows,bending,intersections}
\usetikzlibrary{shapes.geometric}
\usetikzlibrary{arrows.meta,arrows}
\usetikzlibrary{decorations.pathmorphing}
\usetikzlibrary{decorations.markings}
\usetikzlibrary{shapes.multipart}
\usetikzlibrary{tikzmark}
\usepackage{arydshln}

\usepackage[nosort]{cite}
\usepackage{colortbl}

\usepackage{xcolor}
\usepackage{lipsum}
\newcommand*{\mybox}[2]{\colorbox{#1!30}{\parbox{.98\linewidth}{#2}}}

\usepackage[linktocpage=true]{hyperref}
\hypersetup{colorlinks=true,linkcolor=nicecolor,citecolor=nicecolor,urlcolor=nicecolor}
\definecolor{nicecolor}{rgb}{0.1, 0.3, 0.4}

\definecolor{blue}{rgb}{0.06, 0.3, 0.57}
\definecolor{Gray}{gray}{0.4}

\definecolor{nicecolor}{rgb}{0.1, 0.3, 0.4}
\definecolor{blue}{rgb}{0.06, 0.3, 0.57}
\definecolor{Gray}{gray}{0.4}
\colorlet{tableheadcolor}{gray!15} 
\colorlet{tablerowcolor}{gray!7} 

\def\hybrid{\topmargin -20pt    \oddsidemargin 0pt
	\headheight 0pt \headsep 0pt
	\textwidth 6.5in        
	\textheight 9in         
	\textwidth 6.25in       
	\textheight 9 in       
	\marginparwidth .875in
	\parskip 5pt plus 1pt 
	\jot = 1.5ex
}
\hybrid
\numberwithin{equation}{section}
\numberwithin{table}{section}
\usepackage{float}
\usepackage{tabularx}
\newcolumntype{D}{>{\centering\arraybackslash}X}
\newcolumntype{L}{>{$}l<{$}}
\newcolumntype{R}{>{$}r<{$}}
\newcolumntype{C}{>{$}c<{$}}
\usepackage{IEEEtrantools}
\usepackage{makecell}

\newcommand{\beq}{\begin{equation}}  \newcommand{\eeq}{\end{equation}}
\newcommand{\bal}{\begin{aligned}}   \newcommand{\eal}{\end{aligned}}
\newcommand{\bea}{\begin{eqnarray}}  \newcommand{\eea}{\end{eqnarray}}
\def\beqa{\begin{eqnarray}}
\def\eeqa{\end{eqnarray}}

\newcommand{\bmat}{\left(\begin{array}}
\newcommand{\emat}{\end{array}\right)}

\newcommand{\bbC}{\mathbb{C}}
\newcommand{\bbR}{\mathbb{R}}

\newcommand{\cO}{\mathcal{O}}

\newcommand{\cE}{\mathcal{E}}

\newcommand{\cK}{\mathcal{K}}
\newcommand{\cN}{\mathcal{N}}

\newcommand{\cR}{\mathcal{R}}

\newcommand{\cV}{\mathcal{V}}

\newcommand{\cM}{\mathcal M}

\newcommand{\I}{\text{Im}\,}
\newcommand{\R}{\text{Re}\,}

\usepackage{cancel}

\newcommand{\be}{\begin{equation}}
\newcommand{\ee}{\end{equation}}

\newcommand{\slt}{\mathfrak{sl}(2, \bbC)}

\newcommand{\rI}{\mathrm{I}}
\newcommand{\rII}{\mathrm{II}}
\newcommand{\rIII}{\mathrm{III}}
\newcommand{\rIV}{\mathrm{IV}}
\newcommand{\rV}{\mathrm{V}}

\newcommand{\bbZ}{\mathbb{Z}}
\newcommand{\subs}{\subset}

\newcommand{\spanC}[1]{\mathrm{span}_\bbC\{#1\}}

\newcommand{\conj}[1]{\overline{#1}}

\newcommand{\pair}[1]{\ensuremath \langle {#1} \rangle}

\newcommand*\Bell{{\ensuremath{\boldsymbol\ell}}}

\newcommand{\nmod}{{\hat n}}

\definecolor{Gray}{gray}{0.95}

\begin{document}

\baselineskip=14pt
\parskip 5pt plus 1pt

\vspace*{-1.5cm}
\begin{flushright}    
  {\small 
  
  }
\end{flushright}

\vspace{2cm}
\begin{center}        

  {\huge Asymptotic Flux Compactifications\\[.2cm]
   and the Swampland}
  
\end{center}

\vspace{0.5cm}
\begin{center}        
{\large  Thomas W.~Grimm$^{1}$, Chongchuo Li$^1$, Irene Valenzuela$^2$}
\end{center}

\vspace{0.15cm}
\begin{center}        
\emph{$^1$ Institute for Theoretical Physics \\
Utrecht University, Princetonplein 5, 3584 CE Utrecht, The Netherlands}
 \\[.3cm]
  \emph{$^2$Jefferson Physical Laboratory, Harvard University, Cambridge, MA 02138, USA}
             \\[0.15cm]
 
\end{center}

\vspace{2cm}


\begin{abstract}
\noindent
We initiate the systematic study of flux scalar potentials and their 
vacua by using asymptotic Hodge theory. To begin with, we consider 
F-theory compactifications on Calabi-Yau fourfolds with four-form flux. We 
argue that a classification of all scalar potentials  
can be performed when focusing on regions in the field space in which one 
or several fields are large and close to a boundary. To exemplify the constraints on such 
asymptotic flux compactifications, we explicitly determine this classification for situations in which 
two complex structure moduli are taken to be large. Our classification 
captures, for example, the weak string coupling limit and the large complex structure limit.   
We then show that none of these scalar potentials admits de Sitter critical points at parametric control, formulating a new no-go theorem valid beyond weak string coupling.  We also check that the recently proposed asymptotic de Sitter conjecture is satisfied near any infinite distance boundary.
Extending this strategy further, we generally identify the type of fluxes that 
induce an infinite  series of Anti-de Sitter critical points, thereby generalizing the well-known Type IIA settings.  
Finally, we argue that also the large field dynamics of any axion in complex structure moduli 
space is universally constrained. Displacing such an axion by large field values 
will generally lead to severe backreaction effects destabilizing other directions. 

\end{abstract}

\thispagestyle{empty}
\clearpage

\setcounter{page}{1}


\newpage

\begingroup
  \flushbottom
  \setlength{\parskip}{0pt plus 1fil} 
  \tableofcontents
  \newpage
\endgroup


\section{Introduction}
\label{sec:intro}

The search for a landscape of de Sitter vacua is one of the most fundamental
tasks in string theory. The Dine-Seiberg problem \cite{Dine:1985he} together with well known no-go theorems \cite{Hertzberg:2007wc,Flauger:2008ad,Wrase:2010ew} for weakly coupled classical vacua in Type II compactifications (see also \cite{Banlaki:2018ayh,Roupec:2018mbn,Junghans:2018gdb,Andriot:2019wrs} for recent progress in this direction) suggest that a de Sitter vacuum 
will require to consider quantum and/or non-perturbative corrections that will move us away from the regimes in which we have asymptotic perturbative control of the effective theory. Note that these no-go results are formulated as observations  
on certain string theory configurations and based on a study of examples. Embedding them into universal constraints arising from consistency with quantum gravity is at the heart of the swampland program (see \cite{Palti:2019pca} for a recent review). In particular, the recent asymptotic de Sitter conjecture \cite{Ooguri:2018wrx} (see \cite{Obied:2018sgi,Garg:2018reu} for previous formulations) claims a universal bound on the potential that forbids de Sitter vacua when approaching any infinite distance limit in field space and hence implies that there is a sort of Dine-Seiberg problem for any scalar field near any infinite distance limit. 
The universality claim is motivated in \cite{Ooguri:2018wrx} by a connection to another conjecture, the so-called Swampland Distance Conjecture \cite{Vafa2005,Ooguri:2006in}, which asserts the universal existence of an infinite tower of massless states at any infinite field distance limit. Crucial to this present paper will be the fact that in the search of evidence 
for the Swampland Distance Conjecture the works \cite{GPV,Grimm:2018cpv,Corvilain:2018lgw} uncovered a universal structure in any large field limit in geometric moduli spaces. 
It turns out that this structure also constraints the form of the flux-induced scalar potentials and provides us a tool to systematically classify such potentials at any large field limit and promote the above conjectures into precise statements linked to this universal structure.  We will not only provide significant evidence for the asymptotic de Sitter conjecture \cite{Ooguri:2018wrx}, but also bring a new angle to the
origin of the set of seemingly infinite number of Anti-de Sitter vacua of \cite{DeWolfe:2005uu} and get general constraints on axion scalar potentials relevant for backreaction issues  in axion monodromy \cite{Baume:2016psm,Valenzuela:2016yny,Blumenhagen:2017cxt} that are related to the refined Distance Conjecture \cite{Ooguri:2006in,Klaewer:2016kiy}. 

To answer systematically questions about the scalar potentials arising in string theory, we initiate the general study of flux compactifications in any region of field space that involves a large field limit. We call such settings  \textit{asymptotic flux compactifications} in the following. These compactifications will share the common feature 
that they capture limits that occur when approaching the boundary of the field space which, however, is not constrained to be of infinite distance in the field space metric. 
Asymptotic flux compactifications often describe an effective theory in which, at least in a dual description, a small coupling constant ensures that the leading perturbative expansion suffices to study the properties of the system. 
Two famous examples are 
Type IIB orientifold flux compactifications carried out at small string coupling, and Type IIA flux compactifications studied in the large volume regime \cite{Grana:2005jc,Douglas:2006es,Denef:2008wq}. We argue in this work that also the flux scalar potential in more general asymptotic limits can be systematically studied by using F-theory compactified on a Calabi-Yau fourfold with $G_4$-form flux. The complex structure moduli space of such fourfolds has a very rich structure, which allows us, among others, to recover flux potentials encountered at weak string coupling or large volume.  Clearly, interpreting the various limits might require to move to a dual frame, as we will show by relating the flux scalar potentials in F-theory, Type IIB orientifolds and Type IIA orientifolds via mirror symmetry. Although, in general, such a dual description does not necessarily correspond to a perturbative string theory. It turns out that considering all possible asymptotic flux compactifications of F-theory goes beyond these well-known settings and yields a set of new characteristic scalar potentials. 
These insights then allow us to generalize the no-go theorems for flux-induced de Sitter vacua to more general asymptotic regimes beyond string weak coupling. Let us remark that our results also go beyond the Maldacena-Nu\~nez no-go theorem \cite{Maldacena:2000mw} as the F-theory potential also includes the contribution from higher derivative terms and more exotic seven-branes, such as the ones combining into orientifold planes.

The mathematical machinery that we will employ is part of asymptotic Hodge theory, which in particular implies that there exists a so-called limiting mixed Hodge structure at any asymptotic limit to the boundary of the moduli space. These mixed Hodge structures encode crucial information about 
the behavior of the $(p,q)$-decomposition of forms on the compactification manifold in the asymptotic limits in complex structure moduli space. In particular, asymptotic Hodge theory provides an asymptotic expression of the Hodge norm \cite{CKS} that we will use heavily in this work. It also allow us to discuss the conditions on self-dual fluxes in the 
asymptotic regime. 
Furthermore, it is crucial that all allowed limiting mixed Hodge structures can be classified by using the underlying $\slt$-representation theory \cite{Kerr2017}, as has been done for Calabi-Yau threefolds in \cite{Grimm:2018cpv}.
Our analysis aims to give the first steps towards a classification of asymptotic regimes in Calabi-Yau fourfolds and subsequently all asymptotic flux-induced scalar potentials induced by $G_4$-flux.  
Let us note that this machinery has been proven useful to test the Swampland Distance Conjecture and the Weak Gravity Conjecture \cite{ArkaniHamed:2006dz}
in Calabi-Yau string compactifications \cite{GPV,Grimm:2018cpv,Corvilain:2018lgw,Font:2019cxq,Grimm:2019wtx}\footnote{See \cite{Palti:2015xra,Baume:2016psm,Valenzuela:2016yny,Bielleman:2016olv,Blumenhagen:2017cxt,Palti:2017elp,Hebecker:2017lxm,Cicoli:2018tcq,Blumenhagen:2018nts,Gonzalo:2018guu,Buratti:2018xjt,Lee:2018spm,Lee:2018urn,Font:2019cxq,Marchesano:2019ifh,Lee:2019wij,Lee:2019xtm} for other works testing the Swampland Distance Conjecture in the context of asymptotic string compactifications.}. 

In this paper we will study asymptotic flux compactifications with $G_4$ with a focus on asymptotic limits given by only two fields becoming large. In other words, we will consider regions near codimension-two boundary loci in the complex structure moduli space and leave the generalization to higher codimensions for future work. We classify all possible asymptotic two-variable large field 
limits in general Calabi-Yau fourfolds, both at finite and infinite field distance. We then focus on the \textit{strict asymptotic regime} in which two fields and their ratio are large. Physically this implies a suppression of certain perturbative corrections, while mathematically it corresponds to using the so-called $\mathrm{sl(2)}$-orbit approximation. It is then possible to explicitly derive the  asymptotic scalar potential for all such strict asymptotic regimes (see table \ref{potentials_list1}). This allows us to study the structure of flux vacua and obtain a no-go theorem that forbids the presence of de Sitter vacua at parametric control near any large field limit of two fields parametrizing complex structure deformations. The details of the no-go and assumptions can be found in section \ref{sec:deSitter}. The list of scalar potentials also allows us to explicitly test the asymptotic Sitter conjecture of \cite{Ooguri:2018wrx} and to show that it is satisfied if we are dealing with infinite distance limits. Note that we do not discuss the stabilization of K\"ahler structure moduli. Remarkably, our findings can be interpreted as stating that 
a subsector of moduli, which one aims to stabilize near the boundary, already imposes strong constraints on the vacuum structure. This becomes more apparent 
when considering the flux scalar potentials in a more general context in which it yields a generalization of the Type IIA no-go of \cite{Hertzberg:2007wc}.  

Crucially, we therefore also consider a more general class of flux scalar potentials that capture, in particular, the 
 potentials found in Type IIA flux compactifications \cite{Grimm:2004ua}. These potentials are, in contrast to the standard F-theory flux potentials 
not positive definite and hence can admit Anti-de Sitter vacua. In particular, it was argued in \cite{DeWolfe:2005uu} that a seemingly 
infinite series of flux vacua exists in Type IIA at weak string coupling and large volume. We identify the special fluxes that are necessary to generate such sequences and check if they can exist at the various limits in moduli space. More precisely, such fluxes are necessarily having vanishing Hodge norm in the asymptotic limit and drop out from the tadpole constraint. This implies that they cannot correspond to self-dual fluxes and hence would induce a backreaction on the geometry in the F-theory context. Remarkably, their construction and existence seems deeply related to the infinite charge orbits presented in \cite{GPV,Grimm:2018cpv} in the study of the Swampland Distance Conjecture.

Our approach also allows us to generally analyze the backreaction effects of axion monodromy inflationary models in Calabi-Yau manifolds, in which the role of the inflaton is played by an axion with a flux-induced potential. It was shown in \cite{Baume:2016psm,Valenzuela:2016yny,Blumenhagen:2017cxt} for particular examples that displacing an axion for large field values implies in turn a displacement of the saxionic fields which backreacts on the kinetic term of the axion such that the proper field distance grows only logarithmically with the inflaton vev. This further implies that the cut-off scale set by the infinite tower of states of the Distance Conjecture also decreases exponentially in terms of the axionic field distance and invalidates the effective theory. It was argued \cite{Baume:2016psm,Valenzuela:2016yny,Blumenhagen:2017cxt}  that for closed string axions with a flux-induced potential generated at weak coupling and large volume, these backreaction effects cannot be delayed but become important at transplanckian field values, disfavoring certain models of large field inflation. However, it remained as an open question if the backreaction can be delayed in other setups by generating a mass hierarchy between the axion and the saxions \cite{Valenzuela:2016yny,Landete:2017amp} (see also \cite{Hebecker:2014kva,Blumenhagen:2014nba}). Here, we will show with complete generality that the backreaction cannot be delayed for any axion belonging to the complex structure moduli space of F-theory Calabi-Yau compactifications in the asymptotic regimes analyzed in this paper, as long as we move along a gradient flow trajectory. The reason is that the parameter that controls the backreaction becomes independent of the fluxes at large field for any two-moduli asymptotic limit of the moduli space of a Calabi-Yau fourfold. Interpreted in the Type IIB context this result implies that neither closed 
string complex structure deformations, nor open-string seven-brane deformations can provide axions where backreaction effects can be made small. This provides new evidence for the refined Distance Conjecture \cite{Ooguri:2006in,Klaewer:2016kiy}.

The outline of the paper goes as follows. We start in section \ref{sec:fluxcompactifications} by reviewing the scalar potential of $\cN=1$ compactifications of M/F-theory on a Calabi-Yau fourfold with $G_4$ flux, and the chain of dualities that reduce the setting to four dimensional Type IIB and IIA flux compactifications. In section \ref{sec:asymptotic}, we will introduce the machinery to study these flux compactifications in the asymptotic regimes of the moduli space. Key results are the asymptotic decomposition of the fluxes adapted to the different limits and the asymptotic behavior of the Hodge norm, which allow us to determine the universal leading behavior of the flux-induced scalar potential at the asymptotic limits. In section \ref{sec:sugra3} we explain this structure in the context of an $\cN=1$ supergravity embedding and its relation to the dual description of the scalar potential in terms of three-form gauge fields. A complete classification of all possible two-moduli asymptotic limits in the Calabi-Yau fourfold is performed in section \ref{general-two-moduli} together with the flux-induced scalar potential arising in each case. In section \ref{Fluxvacua} we analyze the vacua structure of this potential and get a new no-go theorem for de Sitter as well as new insights regarding infinite sets of families of AdS vacua. The analysis of the axion dependence of the scalar potential and the implications for axion monodromy models are discussed in section \ref{sec:axionDependence}, while section \ref{sec:con} contains our conclusions.

\section{Flux compactifications on Calabi-Yau fourfolds\label{sec:fluxcompactifications}}

In this section we introduce the setup that we investigate in detail in this work. Concretely, we will be interested in 
flux compactifications of F-theory and Type IIB orientifolds that can be studied via the duality to M-theory. We will thus 
first recall in subsection \ref{sc+tad} the scalar potential $V_{\rm M}$ of M-theory compactified on a Calabi-Yau fourfold with $G_4$ flux
and introduce the tadpole cancellation condition \cite{Duff:1995wd,Sethi:1996es}. We briefly comment on how $V_{\rm M}$ lifts to a four-dimensional 
scalar potential of an $\cN=1$ compactification of F-theory on an elliptically fibered $Y_4$. 
In subsection~\ref{orientifold_vacua} we then recall how the F-theory setting reduces to a four-dimensional flux compactified  
Type IIB on an orientifold background. Restricting the allowed background fluxes we also show how a specific scalar potential \eqref{IIAscalar_simp} in Type IIA flux compactification can be described within this setting and we will later on analyze generalizations of such potential by loosening the correlation between the coefficients in the remaining sections.

\subsection{Four-form flux and the scalar potential} \label{sc+tad}

Compactifications for M-theory, or rather eleven-dimensional supergravity, on a Calabi-Yau fourfold leads to a
three-dimensional effective supergravity theory with $\cN=2$ supersymmetry. This theory is characterized
by a K\"ahler potential, determining the metric of the dynamical scalars, and a superpotential, inducing a non-trivial 
scalar potential for these fields. In case one is considering a smooth Calabi-Yau fourfold the superpotential 
is only induced by four-form fluxes $G_4$, which parametrize non-vanishing vacuum expectation values of 
the field-strength $\hat G_4$ of the M-theory three-form $\hat C_3$ through four-cycles of the internal space $Y_4$. 
Such fluxes $\hat G_4$ can also induce a gauging of the theory \cite{Berg:2002es,Grimm:2010ks}, but we will not discuss this part of the 
effective action in any detail in the following. We will also be not concerned with the quantization of fluxes, since this discreteness property
will not be of significance in the later analysis.

Performing the dimensional reduction the  three-dimensional scalar potential in the Einstein frame takes the form
\beq \label{VM}
V_{\rm M}=\frac{1}{\cV_4^3}\Big(\int_{Y_4} G_4\wedge \star G_4-\int_{Y_4} G_4\wedge G_4 \Big)
\eeq
where $\cV_4$ is the volume of $Y_4$ and $\star$ is the Hodge-star on $Y_4$. Note that the derivation of $V_{\rm M}$ requires to perform a 
dimensional reduction with a non-trivial warp-factor and higher-derivative terms \cite{Becker:1996gj,Dasgupta:1999ss,Grimm:2014xva,Grimm:2014efa,Grimm:2015mua,Grimm:2017pid}. 
The warp-factor equation integrated over $Y_4$ furthermore induces a non-trivial 
consistency condition linking flux and curvature. This tadpole cancellation condition takes the form
\beq \label{tadpole}
   \frac{1}{2}\int_{Y_4} G_4 \wedge G_4 = \frac{\chi(Y_4)}{24}\ ,
\eeq 
where $\chi(Y_4) = \int_{Y_4} c_4 (Y_4)$ is the Euler characteristic of $Y_4$. The condition \eqref{tadpole} 
has to be used crucially in the derivation of \eqref{VM} and leads to the second term.

The scalar potential \eqref{VM} depends via the Hodge-star and $\cV_{4}$ both on the complex structure 
moduli and K\"ahler structure moduli of $Y_4$. 
Since our main target will be to investigate the vacua in complex structure moduli space, it is convenient 
to split the scalar potential with respect to these two sets of moduli. We will do that by demanding that the 
flux under consideration satisfies the primitivity condition 
\beq J\wedge G_4 = 0\ ,  \label{J-primitive}
\eeq 
which should hold in cohomology and defines the primitive cohomology $H^4_{\rm p}(Y_4,\bbR)$.
This condition forces the scalar potential 
induced by this flux to only depend on the complex structure moduli and the overall volume factor. 
In fact, 
one shows \cite{Haack:2001jz} that it then can be written as 
\beq \label{VM-KW}
    V_{\rm M}=  e^{K} G^{I\bar J} D_I W   \overline{D_J W}\ ,
\eeq
where $K$ is a K\"ahler potential, determining the metric $G_{I \bar J}$ and its inverse $G^{I \bar J}$, and $W$ a holomorphic 
superpotential. The derivative appearing in \eqref{VM-KW} are given by $D_I W = \partial_I W + (\partial_I K) W$, with 
$\partial_I$ are derivatives with respect to the complex structure moduli fields of $Y_4$. Note that a term proportional to 
$|W|^2$ does not arise due to the no-scale condition for the K\"ahler moduli.  

Let us introduce the various quantities appearing in expression \eqref{VM-KW} in more detail. 
Firstly, we have introduced the K\"ahler potential $K = - 3 \log  \cV_4  + K^{\rm cs}$, which absorbs 
the overall volume factor and depends on the K\"ahler potential $K^{\rm cs}$. The latter determines 
the metric $G_{I\bar J} = \partial_{z^I} \partial_{\bar z^J} K^{\rm cs}$ on the  complex structure moduli 
space $\cM^{\rm cs}$ of $Y_4$. In general, $K^{\rm cs}$
is a very non-trivial function of the complex structure moduli $z^I$, $I=1,\ldots,h^{3,1}(Y_4)$. 
Explicitly it can be written as
\beq \label{Kcs}
    K^{\rm cs}(z,\bar z)  =  - \log \int_{Y_4}  \Omega(z) \wedge \bar \Omega(\bar z)  \ ,
\eeq
where $\Omega$ is the, up to rescalings, unique $(4,0)$-form on $Y_4$. Note that $\Omega$ varies 
holomorphically in the fields $z^J$.
Secondly, we have used that the superpotential depending on the complex structure moduli takes the form \cite{Gukov:1999ya}
\beq \label{Wcs}
     W(z)  = \int_{Y_4}  G_4 \wedge \Omega(z)\ . 
\eeq

In order to simplify the notation, let us introduce a bilinear form $\langle \,\cdot\, ,\,\cdot\, \rangle$ 
and the Hodge norm $\|\cdot \|$ by defining 
\beq
    \langle v, v' \rangle \equiv \int_{Y_4} v \wedge v'\ , \qquad     \|  v \|^2  \equiv   \int_{Y_4}  v \wedge \star\bar  v\ ,
\eeq
Note that $\langle \,\cdot\,, \,\cdot\, \rangle$ is symmetric for Calabi-Yau fourfolds. Using this notation one finds that
\eqref{Kcs} and \eqref{Wcs} reduce to 
\beq  \label{Kcs-Wcs-compact}
  K^{\rm cs}=   - \log \langle \Omega , \bar \Omega \rangle = - \log \| \Omega \|^2 \ , \qquad W = \langle G_4 , \Omega \rangle\ , 
\eeq
where we have used that $\star\Omega = \Omega$. 
Furthermore, we can write the scalar potential \eqref{VM} elegantly as
\beq \label{VMsimp}
   V_{\rm M}=\frac{1}{\cV_4^3}\Big(\| G_4\|^2 - \langle G_4 , G_4 \rangle\Big)= \frac{1}{2 \cV_4^3} \| G_4 - \star G_4\|^2 \ .
\eeq
 
It will be crucial for our later discussion to recall some well-known features 
of the vacua of \eqref{VM}, \eqref{VM-KW}. If we look for supersymmetric vacua one has 
to demand $D_I W =0 $ and $ W = 0$, where the later condition arises when 
considering a $W$ independent of the K\"ahler structure moduli. Hence, in 
the $(p,q)$-Hodge decomposition of the primitive cohomology $H^4_{\rm p}(Y_4, \mathbb{C})$, defined by the vanishing of 
the wedge product of these forms with $J$ as in \eqref{J-primitive}, 
supersymmetric fluxes are of type $(2,2)$. Clearly, the potential \eqref{VM-KW} is vanishing for these 
vacua. In fact, it is important to stress that if one demands that the equations 
of motion for a background solution are strictly satisfied, one has 
\beq \label{self}
   G_4 = \star G_4 \ , 
\eeq
and the scalar potential \eqref{VM} vanishes identically. Therefore, in order to obtain non-trivial Anti-de Sitter or de Sitter solutions 
we have to violate \eqref{self} in the vacuum. In order that this does not destabilize the solution, this has to be done in a controlled way, as 
we discuss in more detail below. 

Let us close the recap of the fourfold compactifications by noting that the scalar potential \eqref{VMsimp}
admits a lift to four-dimensional F-theory compactifications if $Y_4$ is elliptically fibered with a threefold base $B_3$ \cite{Vafa:1996xn}. In order 
to discuss this up-lift in some more detail, we note that the restriction to primitive fluxes $G_4$ is important in 
the following discussion. In fact, in contrast to some of the K\"ahler moduli, 
the complex structure moduli of $Y_4$ will equally be complex scalar fields in a four-dimensional F-theory compactification. 
Therefore, for primitive flux the combination $\| G_4\|^2- \langle G_4 , G_4 \rangle$ in \eqref{VMsimp} 
will lift directly to four dimensions. The overall volume, however, has to be split into a volume of the base $B_3$ denoted by $\cV_{\rm b}$ and the volume of 
the fiber as discussed in \cite{Grimm:2010ks}. Identifying the fiber volume with the radius of the circle connecting M-theory and F-theory, we then 
obtain the F-theory scalar potential
\beq \label{VFsimp}
   V_{\rm F}=\frac{1}{\cV_{\rm b}^2}\Big(\| G_4\|^2 - \langle G_4 , G_4 \rangle\Big)\ .
\eeq
Crucially, this result contains the volume $\cV_{\rm b}$ of a Type IIB compactification performed in ten-dimensional Einstein frame
and no further dilation factors appear in the overall prefactor. In the next subsection we will discuss how \eqref{VFsimp}
reduces to the flux potential of a Type IIB orientifold compactification. The latter then relates to a Type IIA flux potential via mirror symmetry.

\subsection{Relation to flux vacua in Type IIB and Type IIA orientifolds} \label{orientifold_vacua}

In this section we briefly discuss how the $G_4$ flux compactifications introduced in section \ref{sc+tad} 
are linked with flux compactifications of Type IIB and Type IIA orientifolds. In particular, we will recall the well-known 
results about Type IIA flux vacua following \cite{Grimm:2004ua,DeWolfe:2005uu,Hertzberg:2007wc}.
This will make it easier to compare later on our results to previous no-go theorems found in the literature.

Let us first discuss how the first term in the F-theory scalar potential \eqref{VFsimp} given by the Hodge norm of $G_4$ reduces to the well known flux induced scalar potential of Type IIB Calabi-Yau orientifold compactifications. This requires to perform Sen's weak coupling limit \cite{Sen:1996vd}, 
which is a well-know limit in complex structure moduli space and will arise as a special case of the more general discussion introduced in the 
next section. Concretely, it requires to send the imaginary part of one of the complex structure moduli, namely the one corresponding to the complex structure modulus of the generic elliptic fiber of $Y_4$, to be very large. Denoting this modulus by $S$ one then identifies $S=C_0 + i e^{-\phi_{\rm B}}$, where $\phi_{\rm B}$ is the ten-dimensional dilaton.
This implies that $\I S \gg 1$ is indeed the weak string coupling limit. The flux $G_4$ splits as $G_4= H_3 \wedge dy + F_3 \wedge dx$, 
where $dx$ and $dy$ are the two one-forms on the generic elliptic fiber and $H_3$ and $F_3$ are NS-NS and R-R fluxes in Type IIB, respectively. 
Inserting this form of $G_4$ into the F-theory potential \eqref{VFsimp} and using the standard torus metric, one finds that 
Type IIB orientifold flux potential takes the form 
\beq
V_{\rm IIB}=\frac{e^{3\phi_{\rm B}} }{4 (\cV_s^{\rm B} )^2}  \Big[ e^{-\phi_{\rm B}} \int_{Y_3} H_3\wedge \star H_3+ e^{\phi_{\rm B}}   \int_{Y_3} F_3\wedge \star F_3  -  \int_{Y_3} F_3 \wedge H_3 \Big]\ .
\label{IIB}
\eeq
Note that $\cV^{\rm B}_{s}$ is the volume of the Calabi-Yau threefold emerging in the orientifold limit in the ten-dimensional string frame. 
The volume is related to $\cV_{\rm b}$ via $\cV_{s}^{\rm B}=\cV_{\rm b} e^{3\phi_{\rm B}/2}$
and one has  $B_3= Y_3/\mathbb{Z}_2$.
This implies also that the Hodge norm in \eqref{IIB} now only includes the dependence on the complex structure moduli of the threefold $Y_3$, which were part of the  complex structure moduli of the fourfold $Y_4$. It is straightforward to express \eqref{IIB} in terms of the complex flux $F_3-S H_3$ and then determine the well-known orientifold flux superpotential. 

Let us now turn to discussing Type IIA orientifold compactifications with fluxes. 
Their effective action can also be determined by 
direct dimensional reduction from massive IIA supergravity \cite{Grimm:2004ua}. However, we can alternatively 
use mirror symmetry to derive the effective theory of Type IIA on the mirror Calabi-Yau orientifold. 
By mirror symmetry, the complex structure moduli are mapped to K\"ahler moduli in Type IIA, while the four dimensional Type 
IIB dilaton $e^{D_{\rm B}}=e^{\phi_{\rm B}}/\sqrt{\cV^{\rm B}_s}$ gets mapped to the Type IIA dilaton 
$e^{D_{\rm A}} = e^{\phi_{\rm A}}/\sqrt{\cV^{\rm A}_s} $. It will be convenient for us to define \footnote{Note that $s = \R C Z^0 $ in the notation of \cite{Grimm:2004ua}. The 
factor $i\int \Omega \wedge \bar \Omega$ was not included in \cite{Hertzberg:2007wc}.}
\beq \label{def-taurho}
   s = e^{-\phi_{\rm A}}  \frac{(\cV^{\rm A}_s)^{1/2}}{|\Omega^A|} \ , \qquad u = (\cV_s^{\rm A})^{1/3}\ ,
\eeq
where we defined $|\Omega^A|^2 \equiv i \int_{\tilde Y_3}\bar \Omega^{\rm A} \wedge \Omega^{\rm A}$.
The mirror identification of the fields implies 
\beq
   e^{-\phi_{\rm B}}  \quad \leftrightarrow \quad s\ , \qquad \quad \cV_{s}^{\rm B}  \quad \leftrightarrow \quad  |\Omega^A|^2\ , \qquad \quad |\Omega^B|^2 \quad \leftrightarrow \quad \cV_s^A\ , 
\label{dilaton}
\eeq
with the definition $|\Omega^B|^2 \equiv i \int_{ Y_3}\bar \Omega^{\rm B} \wedge \Omega^{\rm B}$.

The different components of the R-R three-form fluxes map to Type IIA R-R $p$-form fluxes with $p=0,2,4,6$. The NS-NS flux, though, can yield different components mapping to NS-NS flux, metric fluxes or non-geometric fluxes in IIA. For simplicity in this section, let us illustrate the result only for the R-R fluxes and the NS-NS component which maps to a NS-NS flux in IIA. Using \eqref{def-taurho} and \eqref{dilaton} the Type IIA scalar potential dual to \eqref{IIB} reads 
\beq \label{IIAscalar}
V_{\rm IIA}= \frac{1}{4 s^3 |\Omega^A|^4} \Big(\frac{s}{u^3} |\Omega^A|^2  \int_{\tilde Y_3} H_3\wedge \star H_3 + \frac{1}{su^3} \sum_{p}\int_{\tilde Y_3} F_p\wedge \star F_p  -  \int_{\rm O6/D6} F_0 H_3 \Big)\ .
\eeq
In performing this duality one has to realize that also the Hodge star maps non-trivially under mirror symmetry (see e.g.~\cite{Grimm:2004ua,Grimm:2005fa} for a more detailed discussion).
Interestingly, not only all these fluxes have the same M-theory origin in terms of $G_4$, but also the contribution from O6-planes can be derived from the second 
term in \eqref{VM}. Since the orientifold planes are geometrised in M-theory, they will contribute to the Euler characteristic of the fourfold which appears in the tadpole cancellation condition \eqref{tadpole}. This term is topological so the only moduli dependence arises from the overall volume factor. 
Hence, there is an additional factor $1/s^3$ when comparing the Type IIB/F-theory scalar potential \eqref{VFsimp} and Type IIA scalar potential \eqref{IIAscalar}
 arising from the change to 
the string frame and the use of the mirror map. For later reference it will be useful to write \eqref{IIAscalar} in a more compact form 
in the case one has only one volume modulus, namely $u$. In this case one show that $\int_{\tilde Y_3} F_p\wedge \star F_p \propto u^{6-2p}$
and \eqref{IIAscalar} becomes
\beq \label{IIAscalar_simp}
   V_{\rm IIA}= \frac{1}{4 s^3} \Big(\frac{s}{u^3}\hat A_{H_3} + \frac{u^3}{s}  \hat A_{F_0} +\frac{u}{s}  \hat A_{F_2}
   +\frac{1}{s u}  \hat A_{F_4}  +\frac{1}{su^3}  \hat A_{F_6}  -  \hat A_{\rm loc} \Big)\ ,
\eeq
where we have absorbed $ |\Omega^A|^4$ in the definitions in the coefficients $\hat A_{H_3}, \hat A_{F_0},\ldots,\hat A_{F_6}\geq 0$ and $\hat A_{\rm loc}$.

The typical advantage of working using the M-theory language is that, as we have seen, Type II objects with different nature are described in a unified way in M-theory. However, this is not the only advantage.
Notice that the volume and dilaton fields in Type IIA map to complex structure and dilaton in Type IIB respectively, and both lift to complex structure of the fourfold in M-theory. By studying different points in the complex structure moduli space of the fourfold we are, therefore, considering different limits for the volume and dilaton in Type IIA. Only a very special point in this complex structure moduli space corresponds to the large volume and small coupling limit in Type IIA, and only near this special point we can follow the chain of dualities by staying within the regime in which the Type IIA supergravity description is under control. Therefore, another clear advantage of studying these effective theories in the M-theory language, is that we can in fact move to other points in the complex structure moduli space of the fourfold in a controlled way, which allows us to study the effective theory beyond the large volume and weak coupling limit of Type IIA.

The question that drives our work is whether the conclusions and no-go's obtained from studying the structure of flux vacua at large volume and weak coupling limits are also valid when exploring other infinite distance limits of the moduli space. For this purpose, we will introduce a mathematical machinery that will allow us to compute the asymptotic splitting of $G_4$ into different components adapted to each type of infinite distance singularity. In the well known case of the large complex structure point, this asymptotic splitting of $G_4$ corresponds to the different components that map to the RR and NS fluxes in Type IIA. However, this may vary at other special points of the moduli space. Together with this asymptotic splitting we will provide the moduli dependence of each component, which will allow us to study the asymptotic structure of flux vacua in general grounds in section \ref{Fluxvacua}.

\section{Asymptotic flux potential\label{sec:asymptotic}}

In this section we discuss flux compactifications restricted to the asymptotic regime in the complex structure moduli space 
of a Calabi-Yau fourfold $Y_4$. The moduli space regions of interest are near limits in moduli space in which 
$Y_4$ becomes singular. To begin with, we first explain in section \ref{asympt-limits} how the moduli dependence of the 
the $(4,0)$-form $\Omega$ can be approximated in each asymptotic regime when knowing the monodromy matrices and a limiting 
four-form $a_0$ associated to the singular locus. We also briefly discuss how this data can be used to 
classify the limits. Furthermore, we then sketch in section~\ref{flux-split} that the same data defines, very non-trivially, 
an orthogonal split of the fourth cohomology group, and hence the flux space, into smaller vector spaces $V_\Bell$ with
certain remarkable properties. In fact, in section \ref{aym_HodgeNorm} we show that it can be used to 
give an asymptotic approximation to the Hodge norm in \eqref{VM} and hence the flux scalar potential itself.
Using these insights, we are then able to show in section \ref{self-dual_asy} that self-dual fluxes take a particularly simple form in the strict asymptotic regime.
In addition we define a certain new class of fluxes in section \ref{unboundedflux}, which 
are relevant in determining the scaling limits of the scalar potential. 

\subsection{Asymptotic limits in Calabi-Yau fourfolds} \label{asympt-limits}

In the following we will discuss the considered limits in the complex structure moduli space $\mathcal{M}^{\rm cs}(Y_4)$. 
The limits of interest are taken to reach the boundary of $\mathcal{M}^{\rm cs}(Y_4)$ 
at which $Y_4$ becomes singular. Of particular interest will be the ones which lead us to 
points that are of infinite geodesic distance in the metric $G_{I \bar J}$ derived from  \eqref{Kcs}. 
A well-known example of such a degeneration point is the large complex structure point, but the following statements 
apply to all infinite distance points that can also lie on higher-dimensional degeneration loci. 
One describe the degeneration loci of $Y_4$ locally as the vanishing locus of $n$ 
coordinates $z^1=\cdots=z^\nmod=0$.\footnote{This 
equation describes the intersection of $\hat{n}$ divisors in a blown-up version of the complex structure moduli space.} 
We can also introduce new coordinates $t^j = \frac{1}{2 \pi i} \log z^j $, such that the  limits of interest are given by
\beq \label{limit}
    t^j \ \rightarrow \ i\infty\ ,\qquad j=1,\ldots,\hat n\ , 
\eeq 
with all other coordinates $\zeta^\kappa$ finite. 
In the following we will set 
\beq
    t^j = \phi^j + i s^j\ , 
\eeq
such that \eqref{limit} corresponds to sending $s^j \rightarrow \infty$, while the $\phi^j$ approach any finite values. 

Since we will be interested in the region close to the degeneration locus of $Y_4$, we will consider large values 
of $s^1,\ldots,s^\nmod$. In this case we can use a result of \cite{Schmid} that the limiting behavior of 
$\Omega$ is approximated by the so-called nilpotent orbit $\Omega_{\rm nil}$ which takes 
a much simpler form than the general  $\Omega$ and will be introduced next. 
Firstly, $\Omega_{\rm nil}$ depends on the monodromy matrix $T_j$ associated to the $t^j = i \infty$ point. To define the monodromy matrix, one needs to choose a flat 
basis for the four-form cohomology $H_{\rm p}^4(Y_4, \bbR)$ and identify the $(4, 0)$-form $\Omega$ with its period vector $\mathbf{\Pi}$ under such an integral basis. This period vector $\mathbf{\Pi}$ solves the Picard-Fuchs equations associated to the complex structure deformation. Then the monodromy matrix appears if one asks how the period vector $\mathbf{\Pi}$ transforms under $t^j \to t^j + 1$, i.e.~it is defined via 
\beq \label{monodromy}
    \mathbf{\Pi}(\ldots, t^j + 1,\ldots)=T_j^{-1} \mathbf{\Pi}(\ldots, t^j,\ldots)\ ,
\eeq
where the appearance of the inverse of $T_j$ is purely conventional. 
In the following we will use a shorthand notation writing a matrix action on a form. This is always understood 
as having the matrix acting on the integral basis of four-forms. For example, equation \eqref{monodromy} is then expressed as
\beq
   \Omega(\ldots, t^j + 1,\ldots)=T_j \Omega(\ldots, t^j,\ldots)\ ,
\eeq
where the inverse arises due to the action on the basis rather than on the coefficient vector. 

If $T_j$ possesses a non-trivial unipotent part, it 
defines a nilpotent matrix \footnote{In the following we will assume that we have transformed the variables $z^j$ and $t^j$, such that 
only the unipotent part of $T_j$ is relevant in the transformation \eqref{monodromy}. This procedure causes us to lose some of the information about the monodromies of orbifold singularities, but the aspects crucial to the infinite distances are retained.}
\beq \label{N=logT}
    N_j = \log T_j \ . 
\eeq
The $N_j$ form a commuting set of matrices and one has $\langle N_j \,\cdot\,,\,\cdot\, \rangle = - \langle  \,\cdot\,,N_j \,\cdot\, \rangle $. 
The nilpotent orbit theorem of \cite{Schmid} states that  
$\Omega$ is approximated by the nilpotent orbit \footnote{Note that this 
statement is true up to an overall holomorphic rescaling of $\Omega$. Such rescalings yield to a K\"ahler transformation of 
$K$ given in \eqref{Kcs}. Unless otherwise indicated the following discussion is invariant under such rescalings.}
\beq \label{Pinil}
     \Omega ( t,\zeta)   =  \underbrace{e^{t^i N_i} a_0(\zeta)}_{ \Omega_{\rm nil} ( t,\zeta)} + \cO(e^{2\pi i t^j})\ ,
\eeq
where we sum in the exponential over $i=1,\ldots,\nmod$. Here 
$a_0$ is a holomorphic function in the coordinates that are not send to a limit \eqref{limit}. 
Note here that the exponential yields a polynomial in $t^i$, since the $N_i$ are nilpotent matrices. 
The 
important statement of \eqref{Pinil} is that the vector $\Omega_{\rm nil}$ 
approximates $\Omega$ up to corrections that are suppressed  by $e^{2\pi i t^j}$ 
in the limit of large $s^1,\ldots,s^\nmod$.  
The nilpotent orbit is the starting point for our analysis of the asymptotic regions in $\cM^{\rm cs}$. 

Let us note that all possible nilpotent matrices $N$, defined via \eqref{N=logT}, arising from the degeneration limits \eqref{limit} of
Calabi-Yau fourfolds can be classified systematically \cite{Kerr2017}. This classification proceeds analogously to the 
one of singularity types occurring for Calabi-Yau threefolds discussed in \cite{Kerr2017,Grimm:2018cpv}. In the fourfold case 
one distinguishes five general types denoted by $\rI, \rII, \rIII, \rIV,$ and $\rV$. 
Following a similar strategy as for Calabi-Yau threefolds we enumerate all singularity types of the primitive middle Hodge 
numbers $(1, h^{3,1}, \hat m, h^{3,1}, 1)$, where $\hat m$ denotes the dimension of the primitive part $H^{2,2}_{\rm p}(Y_4)$ of $H^{2,2}(Y_4)$.

One way of distinguishing these cases is by 
asking what the highest power of $N$ is that does not annihilate $a_0$, i.e.~one determines the integer $d$ satisfying
\beq \label{def-dpower}
    N^d a_0 \neq 0\ , \quad N^{d+1} a_0 = 0 \ .  
\eeq
Since $d\leq 4$, one finds exactly five cases, $d=0,\ldots,4$ corresponding to the singularity types $\rI,\ldots,\rV$. 
As for Calabi-Yau threefolds each of these types has further sub-types. For fourfolds one can label 
them by two indices and write:
\beq \label{listtypes}
   \begin{array}{lll}\toprule \\[-.4cm]
      \text{I}_{a,a'}   & \quad 0 \leq a \leq a' \leq h^{3, 1}         & \quad 2a' - a \leq \hat{m}    \\[.1cm]
      \text{II}_{b,b'}  & \quad 0 \leq b \leq b' \leq h^{3, 1} - 1     & \quad 2b' - b \leq \hat{m}    \\[.1cm]
      \text{III}_{c,c'} & \quad 0 \leq c \leq c' \leq h^{3, 1} - 1     & \quad 2c' - c \leq \hat{m} - 2\\[.1cm]
      \text{IV}_{d,d'}  & \quad 1 \leq d + 1 \leq d' \leq h^{3, 1} - 1 & \quad 2d' - d \leq \hat{m}    \\[.1cm]
      \text{V}_{e,e'}   & \quad 1 \leq e \leq e' \leq h^{3, 1}         & \quad 2e' - e \leq \hat{m}    \\[.1cm]\bottomrule
   \end{array}
\eeq
The precise connection of $N$ to the singularity type is summarized in table \ref{TypeTable}.

\begin{table}[!ht]
	\centering
  \begin{tabularx}{\textwidth}{|D|c|DDDD|}\hline
	  \rule[-.20cm]{0cm}{.65cm}\multirow{2}{*}{Type} & Action on $a_0$               & \multicolumn{4}{c|}{Rank of}\\
    \rule[-.20cm]{0cm}{.6cm}                       & highest $d:\, N^d a_0 \neq 0$ & $N$           & $N^2$   & $N^3$ & $N^4$\\\hline 
	  \rule[-.20cm]{0cm}{.7cm}  $\rI_{a,a'}$         & $d = 0$                       & $2a' - a$     & $a$     & $0$   & $0$\\
	  \rule[-.20cm]{0cm}{.7cm}  $\rII_{b,b'}$        & $d = 1$                       & $2b' - b + 2$ & $b$     & $0$   & $0$\\
	  \rule[-.20cm]{0cm}{.7cm}  $\rIII_{c,c'}$       & $d = 2$                       & $2c' + 4$     & $c + 2$ & $0$   & $0$\\
	  \rule[-.20cm]{0cm}{.7cm}  $\rIV_{d,d'}$        & $d = 3$                       & $2d' + 4$     & $d + 4$ & $2$   & $0$\\
	  \rule[-.20cm]{0cm}{.7cm}  $\rV_{e,e'}$         & $d = 4$                       & $2e' + 2$     & $e + 2$ & $2$   & $1$\\\hline
	\end{tabularx}
	\caption{Classification of the arising limits and singularities occurring in the complex 
		moduli space of Calabi-Yau fourfolds.}
	\label{TypeTable}
\end{table}

\subsection{Asymptotic split of the flux space} \label{flux-split}

In the following we want to introduce a basis of $G_4$ fluxes, 
which is adapted to the limits \eqref{limit} discussed in the previous subsection. 
It turns out that in order to use the mathematical machinery that we will introduce next, one has to first divide the space into  
growth sectors. One such growth sector is given by 
\begin{equation}\label{growthsector}
    \mathcal{R}_{12 \cdots \nmod} = \Big\{t^j = \phi^j + i s^j \Big| \frac{s^1}{s^2} > \gamma , \ldots, \frac{s^{\nmod - 1}}{s^\nmod} > \gamma , s^\nmod > \gamma, \phi^j < \delta \Big\}\ ,
\end{equation}
where we can chose arbitrary positive $\gamma,\delta$. Other growth sectors can be obtained by the same expression but with permuted 
$s^j$.

Let us now introduce a basis for the $G_4$. It will depend on the following set of 
data: (1) the monodromy matrices $N_i$ and the vector $a_0$ appearing in \eqref{Pinil}, (2) 
the growth sector \eqref{growthsector} which one considers. Given this data 
it was shown in \cite{CKS} that one can always find an associated set of 
\beq  \label{triples}
   \text{commuting}\ \mathfrak{sl}(2,\mathbb{C})\text{-triples}: \qquad (N_i^-,N_i^+,Y_i)\ , \quad i=1,\ldots,\nmod\ ,
\eeq 
which captures the asymptotic behavior of the \eqref{Pinil}  and its derivatives. 
These triples satisfy the standard commutation relations 
\beq
[Y_i,N_i^{\pm} ]= \pm 2N_i^{\pm}\ , \qquad [N_i^+,N_i^-]=Y_i\ . 
\eeq
In practice it it non-trivial to construct these $\mathfrak{sl}(2,\mathbb{C})$-triples starting with the 
data defining the nilpotent orbit  \eqref{Pinil}. For Calabi-Yau threefolds an explicit example 
was worked out in \cite{Grimm:2018cpv}. In the following we will assume that 
the steps summarized in \cite{Grimm:2018cpv} have been performed and the commuting triples to 
the considered limit are known.  

The  $\mathfrak{sl}(2,\mathbb{C})$-triples can now be used to split the primitive cohomology group  
$H^4_{\rm p}(Y_4,\mathbb{R})$ into eigenspaces of $Y_i$. Let us introduce
\beq \label{split-Vell}
   H^4_{\rm p}(Y_4,\mathbb{R}) = \bigoplus_{\Bell\in \cE} V_{\Bell}\ ,\qquad  \Bell=(\ell_1, \ldots, \ell_\nmod)\ ,
\eeq
where $\ell_i \in \{0,\ldots,8 \}$ are integers representing the eigenvalues of $Y_{(i)}=Y_1 + \cdots + Y_i$, i.e.
\beq\label{eigenvalue}
  v_{\Bell} \in V_{\Bell} \quad \iff \quad Y_{(i)} v_{\Bell} = (\ell_i - 4) v_{\Bell}\ .
\eeq
In writing \eqref{split-Vell} we have introduced the set $\cE$ of all possible vectors $\Bell$ labelling non-trivial $V_{\Bell}$ and 
collecting all eigenvalue combinations of $(Y_{(1)},\ldots,Y_{(\nmod)})$. The allowed vectors in $\cE$ are determined 
by investigating the properties of the singularity occurring in the limit \eqref{limit} and we will see in more detail below.  
The $\mathfrak{sl}(2,\mathbb{C})$-algebra allows to derive several interesting properties 
of the vector spaces $V_{\Bell}$. For example, one finds that  
\beq \label{dim-eq}
   \text{dim}\, V_{\Bell} =   \text{dim}\, V_{\mathbf{8} - \Bell} \ ,
\eeq
where we abbreviated $\mathbf{8} = (8,\ldots,8)$, which implies that $V_{\Bell} \cong V_{\mathbf{8} - \Bell} $.
Furthermore, the spaces $V_{\Bell}$ satisfy the 
orthogonality property 
\beq \label{V-orth}
    \langle V_{\Bell}, V_{\Bell'} \rangle =  0 \quad \text{unless} \quad \Bell+\Bell' = \mathbf{8}\ ,
\eeq
as can be inferred by using the fact that $ \langle \,\cdot\, , Y_{(i)} \,\cdot\, \rangle = - \langle Y_{(i)} \,\cdot\, ,  \,\cdot\, \rangle $. In other words using 
 \eqref{split-Vell} one finds a decomposition of an element $H^4_{\rm p} (Y_4,\bbR)$ into sets of pairwise orthogonal components.

Applied to the fluxes  $G_4 \in H^{4}_{\rm p} (Y_4,\bbR)$, this decomposition implies an asymptotic split of the flux space into orthogonal components
\beq\label{flux_split}
  \qquad G_4 = \sum_{\Bell \in \cE} G_4^{\Bell},\quad \textrm{ where } G_4^{\Bell} \in V_{\Bell} \textrm{ for every } \Bell \in \cE.
\eeq
  This flux splitting will be the key of our starting program to classify possible flux scalar potentials in string compactifications. 
    
\subsection{The asymptotic behavior of the Hodge norm} \label{aym_HodgeNorm}

In the following we will introduce one of the most non-trivial consequences of the splitting \eqref{split-Vell}, by arguing that 
it determines the asymptotic behavior of the Hodge norm. To begin with let us recall some facts about the Hodge star operator $\star$. 
To define its action on the primitive middle cohomology $H_{\rm p}^{4}(Y_4,\mathbb{C})$ we can introduce the Hodge decomposition 
\beq \label{H4-decomp}
     H^{4}_{\rm p}(Y_4,\bbC) = H^{4,0} \oplus H^{3,1} \oplus H^{2,2}_{\rm p} \oplus H^{1,3} \oplus H^{0,4}\ . 
\eeq
As long as the manifold $Y_4$ is non-singular the action of $\star$ is simply given by $\star v^{p,q} = i^{p-q} v^{p,q}$, 
for any element $v^{p,q} \in H^{p,q}$. Clearly, since the $(p,q)$-split in \eqref{H4-decomp} depends on the 
choice of complex structure, it will vary when changing the complex structure moduli. This is the origin of the complex structure 
moduli dependence in \eqref{VM}. Close to a degeneration point of $Y_4$ we expect that also $\star$ takes 
a simplified form, just as the $(4,0)$-form $\Omega$ simplifies as discussed in section~\ref{asympt-limits}. In fact, we stated around 
\eqref{Pinil} that $\Omega$ simplifies, when dropping exponentially suppressed corrections, to the nilpotent orbit $\Omega_{\rm nil}$.
This approximation can also be applied to the Hodge star operator $\star$ as we will discuss in the following. 

Let us start with a general element 
of $G_4 \in H^{4}_{\rm p} (Y_4,\bbR)$, which we can consider to be our $G_4$-flux. 
We want to evaluate the Hodge metric by using $\Omega_{\rm nil}$ rather than the complete $(4,0)$-form $\Omega$. This can 
be done systematically, when extending the nilpotent orbit construction to the whole cohomology as we discuss in appendix \ref{app:B}. 
In this way one finds 
\beq \label{v-exp_nil}
    \| G_4 \|^2 = \int_{Y_4} G_4 \wedge \star  G_4 =   \langle C_{\rm nil} G_4 , G_4\rangle+ \cO(e^{2\pi i t^j})\ ,  
\eeq
where $C_{\rm nil}(t,\zeta)$, the \textit{Weil operator} associated to the nilpotent orbit, captures the moduli dependence on the fields $t^j$ through terms involving $e^{t^i N_i}$ as
appearing in the nilpotent orbit \eqref{Pinil}. We will introduce $C_{\rm nil}$ properly in appendix \ref{app:B}. Crucially, 
due to the fact that the $t^j$ dependence of $C_{\rm nil}$ is simplified due to the nilpotent orbit approximation, 
we find that its dependence on the axions $\phi^i=\R t^i$ can be made explicit by writing  
\beq \label{C-nil_exp}
        C_{\rm nil} (t,\zeta) =  e^{\phi^i N_i}  \hat C_{\rm nil}  e^{-\phi^i N_i}\  , \qquad \hat C_{\rm nil}\equiv  C_{\rm nil}(\phi^i = 0)\ .
\eeq
We can use this identity by defining $\rho(G_4, \phi) = e^{-\phi^i N_i} G_4 $ and deduce that \eqref{v-exp_nil} becomes
\beq \label{norm_hatCnil}
    \| G_4 \|^2 = \int_{Y_4} G_4 \wedge \star  G_4  =   \langle \hat C_{\rm nil} \rho , \rho \rangle + \cO(e^{2\pi i t^j})\ ,
\eeq
where crucially all $\phi^i$ dependence is now captured by $\rho(G_4,\phi)$ when neglecting the exponentially suppressed 
corrections. Let us note that we can expand the $\phi^i$-dependent vectors $\rho$ in any basis $v^A, A=1,\ldots,\text{dim} H^4_{\rm p}(Y_4,\bbR)$ as   $\rho = \varrho_A v^A$. If we also give the basis expression $Z^{AB} =  \langle \hat C_{\rm nil} v^A ,  v^B \rangle$  for the inner product, we can write \eqref{norm_hatCnil}
as 
\beq
\label{G4Zrho}
    \| G_4 \|^2 = Z^{AB}\varrho_A \varrho_B+ \cO(e^{2\pi i t^j})\ .
\eeq
It turns out that there is a clever choice of basis $v^A$, which allows us to also make the field dependence on the scalars $s^i = \I t^i$ explicit. This basis is adapted to the splitting \eqref{split-Vell} as we will discuss in the following.

Let us consider the real four-form $G_4$ and determine its split into vector spaces $V_{\Bell}$ by 
expanding $G_4 = \sum_{\Bell \in \cE} G_4^{\Bell}$ as in \eqref{flux_split}. These vector spaces further satisfy to be orthogonal with respect to the inner product 
\beq\label{orth}
   \langle C_\infty V_\Bell , V_{\Bell'} \rangle = 0, \quad \text{for} \ \ \Bell' \neq \Bell\ . 
\eeq
where $C_\infty$ is the Weil operator inducing a natural limiting Hodge norm
\beq
   \| v \|_\infty = \langle C_\infty v , v\rangle\ ,
\eeq
which is defined using only the structure at the limiting locus \eqref{limit}. It is therefore independent of the coordinates $t^1,\ldots,t^\nmod$, while non-trivially varying with the remaining coordinates $\zeta^\kappa$. The operator $C_\infty$ will be introduced in more detail in appendix \ref{app:B}. We also point out that equation \eqref{V-orth} and \eqref{orth} imply the following behavior of this Weil operator
\begin{equation} \label{eqn:CinftySwapV}
  C_\infty: V_{\Bell} \to V_{\mathbf{8} - \Bell},
\end{equation}
for all $\Bell$, i.e., $C_\infty$ exchanges $V_{\Bell}$ and $V_{\mathbf{8} - \Bell}$.  For the purpose of this section, it is enough to remark that the flux norm satisfies the following direct sum decomposition on the split \eqref{split-Vell},
\beq \label{norm_infty_split}
  \|G_4 \|^2_\infty   =\sum_{{\Bell} \in \cE} \|G_4^{ \Bell} \|^2_\infty \ , 
\eeq
thanks to the orthogonality property \eqref{orth} which forces all non-diagonal terms to vanish.
  
 The next step is to move a bit away from the singular loci in order to recover the dependence on the scalar fields of the Hodge norm.
First, in order to explicitly keep the axion dependence, we use \eqref{C-nil_exp} to also include the exponential $e^{-\phi^i N_i}$ 
and expand 
\beq \label{exp-xN}
  \rho(G_4,\phi) \equiv e^{-\phi^i N_i} G_4= \sum_{\Bell \in \cE} \rho_{\Bell}(G_4,\phi)\ ,
  \qquad \rho_{\Bell}(G_4,\phi^i=0) =G_4^{\Bell} \ ,
\eeq
where $\rho_{\Bell}$ is the restriction of $e^{-\phi^i N_i} G_4$ to the vector space $V_\Bell$. Notice that the components $\rho_\Bell$ satisfy the same asymptotic orthogonality properties as $G_4^\Bell$, regardless of the value of the axions.
We can then use this expansion to get an 
asymptotic expression for the Hodge norm  \cite{CKS,MR817170}  with all dependence on $\phi^i$ and $s^i$ being explicit. 
More precisely, one has
\begin{equation} \label{general-norm-growth}
\| G_4 \|^2\ \sim\   \|G_4\|^2_{\rm sl(2)} = \sum_{\Bell \in \cE} \Big( \frac{s^1}{s^2}\Big)^{\ell_{1}-4} \cdots \Big(\frac{s^{\nmod-1}}{s^\nmod} \Big)^{\ell_{\nmod-1}-4}
 (s^\nmod)^{\ell_{\nmod}-4}\, \| \rho_{\Bell}(G_4,\phi) \|^2_\infty \ .
\end{equation}
where we have introduced the Weil operator $C_{\rm sl(2)}$ by setting 
\beq \label{def-Csl2}
     \| G_4 \|_{\rm sl(2)}^2 \equiv \langle C_{\rm sl(2)} G_4 , G_4 \rangle\ .
\eeq 
More detailed discussion on the operator $C_{\rm sl(2)}$ can be found in appendix \ref{app:B}. This operator captures the leading dependence on the saxionic coordinates $s^i$ but neglects all sub-leading polynomial corrections of the form $s^i/s^{i+1}$ for the corresponding growth sector \eqref{growthsector}. Hence, it is only a good approximation once a growth sector is selected and provides the asymptotic form of the Hodge norm along when considering the $s^i$ in the growth sector with $\gamma \gg 1$. From now on, we will denote this regime of validity the \emph{strict asymptotic regime}, in opposition to the \emph{asymptotic regime} which captured all polynomial corrections and neglected only the exponentially suppressed terms of order  $\cO(e^{2 \pi it^j})$. In the mathematical terminology, the latter corresponds to the nilpotent orbit result while the strict asymptotic regime is given by the sl(2)-orbit approximation. We have summarized the different approximations of the Hodge operator and their regime of validity in table \ref{tab:norms}. 
 
 This strict asymptotic behavior of the Hodge norm is a very powerful result that will allow us to classify all possible flux scalar potentials and their vacua arising in the asymptotic regions of string compactifications. All we need is to provide a list of all possible values of the integer vector $\Bell=(\ell_1,\ell_2,\ldots,\ell_{\hat n})\in \cE$ associated to the different singular limits. This classification will be performed in section \ref{sec:singularitiesAndEnhancements} for the case of two moduli becoming large in a Calabi-Yau fourfold. Notice also that the operator $C_{\rm sl(2)}$ still satisfies the same orthogonality properties as $C_\infty$ with respect to the vector spaces $V_\Bell$, implying that the flux scalar potential will be simply given by a sum of squares, simplifying enormously the analysis of flux vacua.
 
 We close this subsection by stressing that the symbol $\sim$ in \eqref{general-norm-growth} indicates that this expression 
displays the strict asymptotic behavior of the Hodge norm. In fact, this statement is actually well-defined. The expression \eqref{general-norm-growth} implies that for $s^1/s^2,\ldots,s^{\nmod-1}/s^\nmod>\gamma$, i.e.~in a growth sector \eqref{growthsector}, there exist two positive constants $\alpha, \beta$
such that 
\beq \label{norm-bound}
  \alpha \|G_4\|^2_{\rm sl(2)}\  \leq\ \| G_4 \|^2\ \leq \ \beta \|G_4\|^2_{\rm sl(2)}\ .
\eeq 
The constants $\alpha,\beta$ do, in general, depend on $\gamma$, but are independent of $G_4$.
Note that this inequality has the immediate consequence that we have to be careful when approximating 
the Hodge norm $ \|G_4\|^2$ with $ \|G_4\|^2_{\rm sl(2)}$, since it limits our ability to 
infer detailed information about $ \|G_4\|^2$ from the much simpler norm $ \|G_4\|^2_{\rm sl(2)}$.
In general, only in the limit $\gamma \rightarrow \infty$ the constants 
$\alpha,\ \beta$ will approach each other and the norm $ \|G_4\|^2$ converges to $\|G_4\|^2_{\rm sl(2)}$.
However, there can be particular situations in which $\|G_4\|^2_{\rm sl(2)}$ provides the full result for the Hodge norm up to exponentially suppressed corrections, as we will explain more carefully when discussing the supergravity embedding in section \ref{sec:sugra}.

\begin{table}[H]
\centering
\begin{tabular}{|c||c|c|c|}\hline
  \rule[-.25cm]{0cm}{.8cm} Regime of                  & Asymptotic                  & Strict asymptotic                            & At boundary  \\ 
    \rule[-.3cm]{0cm}{.6cm}   validity: & $s^i$ large                & $\cR_{1\cdots\hat{n}}$ with $\gamma \gg 1$   & $s^i = \infty$ \\
 \hline
  \rule[-.3cm]{0cm}{.8cm} Approx.~Hodge-operator:            & $C_{\rm nil}$              & $C_{\rm sl(2)}$                              & $C_\infty$ \\\hline
  \rule[-.3cm]{0cm}{.8cm} Corrections dropped: & drop $\cO(e^{2 \pi it^j})$ & drop sub-leading $\frac{s^i}{s^{i+1}}$-polys & $t^i$-independent \\\hline
\end{tabular}
\caption{Weil operators and their regime of validity.} \label{tab:norms}
\end{table}

\subsection{Self-dual fluxes in the strict asymptotic regime} \label{self-dual_asy}

In this subsection we discuss a first way of finding vacua of the potential \eqref{VM} by restricting to asymptotically self-dual fluxes. Note that this potential 
is positive definite when written in the form \eqref{VM-KW} and vanishes when considering vacua 
in which the flux $G_4$ satisfies the self-duality condition \eqref{self}.  
Recall that the self-duality condition is a necessity if we want the vacuum to solve the equations of motion of the eleven-dimensional supergravity. This condition fixes the moduli, since 
it involves the moduli-dependent Hodge star $\star$. As in the previous subsection, we can thus ask the question 
if, at least in the asymptotic regime, one can give an explicit moduli dependence of the self-duality condition and eventually fix the moduli explicitly. 

In order to study moduli stabilization we thus replace $\star$ with its asymptotic counterparts $C_{\rm nil}$, defined in \eqref{v-exp_nil}, and
$C_{\rm sl(2)}$, defined in \eqref{def-Csl2}. In the former case one neglects exponentially suppressed corrections in the 
variables $t^i$ that are taken to the limit. Using \eqref{C-nil_exp}, we find that the self-duality condition \eqref{self} is approximated by 
\beq
     \hat C_{\rm nil} \, \rho(G_4, \phi) = \rho(G_4, \phi)\ .
\eeq
Expanded into a basis this equation gives still a very complicated set of equations even in the $t^i$. 
To further decouple these equations we will move deeper into the asymptotic regime as in section \ref{aym_HodgeNorm}. 
Let us thus consider the asymptotic expression of \eqref{self} using $C_{\rm sl(2)}$. In this case we 
can exploit the fact that everything splits into the $V_{\Bell}$ and we can extract the explicit $t^i$ moduli 
dependence. We thus consider the asymptotic self-duality condition 
\beq \label{self-dual_sl2}
      C_{\rm sl(2)} G_4= G_4\ . 
\eeq

In order to separate this condition into multiple equations we introduce a basis for the $V_{\Bell}$ as
\beq \label{def-vbasis}
     v^{\Bell}_{i_\Bell}\ : \quad \text{span}_\bbR \big\{ v^{\Bell}_{1} , \ldots,  v^{\Bell}_{\text{dim} V_\Bell} \big\} = V_\Bell\ , 
\eeq 
where $\Bell \in \cE$ is a vector as before. We normalize these basis vectors with respect to the inner product, such 
that 
\beq \label{v-orth<>}
   \langle v^{\Bell}_{i_\Bell} ,  v^{\mathbf{8}-\Bell}_{j_{\mathbf{8}-\Bell}}  \rangle= \delta_{i_\Bell j_{\mathbf{8} -\Bell}}\ , \qquad  
    \langle v^{\Bell}_{i_\Bell} ,  v^{\Bell'}_{j_{\Bell'}}  \rangle= 0 \ \ \text{for} \ \ \Bell \neq \mathbf{8}-\Bell'\ ,
\eeq
where we recall that the orthogonality \eqref{V-orth} of the $V_\Bell$ enforces all other products to vanish. We 
also abbreviate the inner product between the basis vectors as
\beq \label{v-orth<C>}
     \cK_{i_\Bell j_\Bell}^\Bell  =  \langle C_{\infty} v^{\Bell}_{i_\Bell} , v^{\Bell}_{j_{\Bell}}  \rangle\ ,\qquad  \langle C_{\infty} v^{\Bell}_{i_\Bell} , v^{\Bell'}_{j_{\Bell'}} \rangle=0 \ \ \text{for} \ \ \Bell \neq \Bell'\ ,
\eeq
where we note that $\langle C_\infty \,\cdot\, , \,\cdot\, \rangle$ is block-diagonal on the $V_{\Bell}$ as noted in \eqref{norm_infty_split}.
Now we can expand 
\beq \label{G4basis_exp}
 G_4 =\sum_{\Bell \in \cE}  G_4^\Bell  = \sum_{\Bell \in \cE}  \sum_{i_\Bell} g_\Bell^{i_{\Bell}}  v^{\Bell}_{i_\Bell} \ ,
 \eeq
 with $g_\Bell^{i_{\Bell}} $ being the `flux quanta' of the $G_4$.

With these preliminaries we can now split \eqref{self-dual_sl2} into scalar equations. 
We first evaluate the product of \eqref{self-dual_sl2} with the basis $\{ v^{\mathbf{m}}_{i_{\mathbf{m}}} \} $ 
introduced in \eqref{def-vbasis}. 
Using the orthogonality conditions \eqref{v-orth<>} and \eqref{v-orth<C>} we find 
\beq
         \Big( \frac{s^1}{s^2}\Big)^{m_{1}-4} \cdots \Big(\frac{s^{\nmod-1}}{s^\nmod} \Big)^{m_{\nmod-1}-4}
 (s^\nmod)^{m_{\nmod}-4}\, \langle  C_\infty \rho_{\mathbf{m}}, v_{\mathbf{m}}^{i_{\mathbf{m}} }\rangle = \langle \rho_{\mathbf{8} - \mathbf{m}} ,  v_{\mathbf{m}}^{i_{\mathbf{m}}} \rangle\ .
\eeq
In order to interpret this expression, we set for the moment $\phi^i=0$, which implies that this expression reduces to 
\beq \label{eqn:SDcondition}
   \Big( \frac{s^1}{s^2}\Big)^{4- m_{1}} \cdots \Big(\frac{s^{\nmod-1}}{s^\nmod} \Big)^{4-m_{\nmod-1}}
 (s^\nmod)^{4-m_{\nmod}}= \frac{ g_{\mathbf{m}}^{i_{\mathbf{m}}} \cK_{i_{\mathbf{m}} j_{\mathbf{m}}}^{\mathbf{m}} }{g_{\mathbf{8} - \mathbf{m}}^{j_{\mathbf{m}}}}, \quad \mathbf{m} \textrm{ not summed},
\eeq
with $ \cK_{i_{\mathbf{m}} j_{\mathbf{m}}}^{\mathbf{m}}$ and $g_{\mathbf{m}}^{i_{\mathbf{m}}}$ defined in \eqref{v-orth<C>}, \eqref{G4basis_exp}, respectively. Note that the
right-hand-side only depends on the fluxes $g_{\mathbf{m}}^{i_{\mathbf{m}}}$, $g_{\mathbf{8} -\mathbf{m}}^{i_{\mathbf{m}}}$ and, via $  \cK_{i_{\mathbf{m}} j_{\mathbf{m}}}^{\mathbf{m}}(\zeta)$, the coordinates $\zeta^\kappa$ not taken to a limit. 
This implies that the 
combination of the $s^i$ appearing on the left-hand side are fixed when imposing the asymptotic 
self-duality condition \eqref{self-dual_sl2}. Whether or not this fixes a particular $s^i$, or even all of them, depends on the vectors
$\Bell \in \cE$, and we will determine all possible sets for two $s^1,s^2$ in section \ref{mainexample}.

\subsection{Unbounded asymptotically massless fluxes} \label{unboundedflux}

In this subsection we want to define a specific type of four-form flux 
that will be relevant in finding vacua in an asymptotic flux compactification. 
The basic idea is to identify a flux $\hat G_4$ that does not 
contribute to the tadpole cancellation condition \eqref{tadpole} and thus, at lease taking into
account only this constraint, can be made arbitrary large. However, it is clear that such a flux cannot satisfy the self-duality condition \eqref{self}  and hence violates the equations of motion. We
therefore also require the flux to have an asymptotically 
vanishing norm  $||\hat{G}_4 ||^2$. As we will discuss in detail below, precisely such fluxes  
enable us to construct vacua that are under parametric control.
 
Let us stress that the complete 
flux under consideration will be of the form
\beq
   G_ 4 = \hat G_4 + G_4^0.
\eeq
Here the flux $\hat G_4$ is defined to have the following properties 
\bea
     (1\,\text{a}) && \langle \hat{G}_4 , \hat{G}_4 \rangle = 0 \ , \qquad (1\,\text{b})\quad  \langle \hat{G}_4 , G^0_4 \rangle = 0\ , \label{no-tadpole} \\
    (2) && \| \hat{G}_4 \| \rightarrow 0 \quad \text{on every path with $t^1,\ldots,t^\nmod \rightarrow i \infty$ in \eqref{growthsector}}\ .  \label{massless}
\eea
while the \emph{rest} of the fluxes will be part of $G_4^0$.
In the following we will call the fluxes satisfying $(1\,\text{a})$ and $(1\,\text{b})$ to be \textit{unbounded}, since they are not restricted by the 
tadpole condition \eqref{tadpole}. The fluxes satisfying $(2)$ will be called  \textit{asymptotically massless} in the following.
As explained above, this latter condition  has been introduced to ensure that the self-duality condition \eqref{self} is only violated 
mildly and restored in the limit. In fact, $\langle \hat{G}_4 , \hat{G}_4 \rangle = 0$ implies that $\hat G_4$
cannot be self-dual at any finite value of the moduli, since otherwise $\langle \hat{G}_4 ,  \hat{G}_4 \rangle = \|  \hat{G}_4 \|^2>0 $. 
In the following, we will explain how to identify these unbounded asymptotically massless fluxes $\hat G_4$ in complete generality. 

The split of the fourth cohomology into $V_{\rm \Bell}$ as in \eqref{split-Vell} and the
general growth property of the Hodge norm \eqref{general-norm-growth} gives us a powerful tool to 
specify the fluxes that satisfy the condition \eqref{massless}. 
Recall that the asymptotic form of the Hodge norm was given in \eqref{general-norm-growth} and takes the form  
\beq \label{G4-growth_gen}
   \| G_4 \|^2\ \sim\ \sum_{\Bell \in \cE} \Big( \frac{s^1}{s^2}\Big)^{\ell_{1}-4} \cdots \Big(\frac{s^{\nmod-1}}{s^\nmod} \Big)^{\ell_{\nmod-1}-4}
 (s^\nmod)^{\ell_{\nmod}-4}\,  A_{\Bell}\ , 
\eeq 
where we have set 
\beq
\label{Al}
    A_{\Bell} \equiv  \| \rho_\Bell (G_4,\phi)  \|^2_\infty  > 0 \ .
\eeq
Let us use this to identify the asymptotically massless part $\hat G_4$. Since by definition 
$ \rho_\Bell (G_4,\phi)  \in V_{\Bell}$ we directly infer from \eqref{G4-growth_gen} that a sufficient condition that $ \| \hat{G}_4 \|^2 \rightarrow 0$ on all paths with $t^1,\ldots,t^\nmod \rightarrow i \infty$ in \eqref{growthsector} is that $\hat G_4$ has 
only components in the $V_{\Bell}$ with  
$\ell_1 ,\ldots,\ell_{\nmod-1} \leq 4$ and $\ell_\nmod <4$. To see this one can use that in \eqref{growthsector} all fractions $(s^1/s^2)^{-1}$,\ldots,$(s^{\nmod-1}/s^\nmod)^{-1}$ are bounded and the power $(s^\nmod)^{\ell_\nmod-4}$ ensures that  $ \| \hat{G}_4 \|^2$ vanishes asymptotically. 

Note that this analysis suggests that it is natural to split the vector space $H^{4}_{\rm p}(Y_4,\bbR)$ into three vector spaces as
\beq
   H^{4}_{\rm p}(Y_4,\bbR) = V_{\rm light} \oplus V_{\rm heavy} \oplus V_{\rm rest}\ , 
\eeq
where we define 
\bea
    V_{\rm light} &=&  \bigoplus_{\Bell \in \cE_{\rm light}} V_{ \Bell}\ , \qquad \quad \cE_{\rm light} =  \{ \ell_1,\ldots,\ell_{\nmod-1} \leq 4 , \ell_\nmod <4 \}\ , \\
     V_{\rm heavy} &=&  \bigoplus_{\Bell \in \cE_{\rm heavy}} V_{ \Bell}\ , \qquad \quad \cE_{\rm heavy} =  \{ \ell_1,\ldots,\ell_{\nmod-1} \geq 4 , \ell_\nmod >4 \}\ . 
\eea
Using the growth result \eqref{general-norm-growth} one infers that $ G_4 \in V_{\rm light}$ is equivalent to the 
statement that $\| G_4 \| \rightarrow 0$ on every path with $t^1,\ldots,t^\nmod \rightarrow i \infty$ in \eqref{growthsector}.
Similarly, one sees that $ G_4 \in V_{\rm heavy}$ is equivalent to demanding $\|  G_4 \| \rightarrow \infty$ on every path to the limit in the considered growth sector.
It is not difficult to see from \eqref{V-orth} and \eqref{dim-eq} that 
\beq
   \langle V_{\rm light} , V_{\rm light} \rangle  =  0\ , \qquad  \langle V_{\rm heavy} , V_{\rm heavy} \rangle = 0\ , 
\eeq
and  that $V_{\rm light}$ and $ V_{\rm heavy}$ can be identified as vector spaces. With these observations 
at hand the asymptotically massless fluxes satisfy 
\beq \label{hatG_inVlight}
     \hat  G_4  \in V_{\rm light}\ . 
\eeq
Note that this identification immediately implies also condition $ (1\,\text{a}) $. In contrast, condition 
$(1\,\text{b})$ should be read as a constraint on both $\hat G_4$ and $G_4^0$.
In fact, we will see that for a given choice of fluxes in $G_4^0$ we will have to switch off components in $\hat G_4$ to select only those asymptotically massless fluxes $\hat G_4$  which have a vanishing inner product with  $G_4^0$  to find a solution to both $(1),(2)$. 

Finally, let us notice that the condition \eqref{hatG_inVlight} is equivalent to the condition imposed over the charge lattice of BPS states in \cite{GPV,Grimm:2018cpv,Corvilain:2018lgw} to find a tower of states that become massless at the singular loci, as predicted by the Swampland Distance Conjecture. Analogously, the condition to be unbounded resembles to the condition of stability \cite{GPV}. A BPS state in a monodromy orbit cannot fragment into two BPS states if they are mutually local, i.e. if the inner product $(1\,\text{b})$ vanishes. Therefore, the same element in $H^{4}_{\rm p}(Y_4,\bbR)$ that generated a tower of massless stable BPS states at the singular loci gives rise here to an unbounded asymptotically massless flux  which is necessary to construct vacua at parametric control. This puts manifest an intriguing relation between the Swampland Distance Conjecture and the presence of vacua at parametric control which deserves further investigation in the future.

\section{Supergravity embedding and three-forms\label{sec:sugra3}}

In this section we will study the $\cN=1$ supergravity embedding of the scalar potential at the asymptotic limits of the moduli space. We will provide the asymptotic form of the K\"ahler potential and superpotential arising in these limits in section \ref{sec:sugra} and explain what the strict asymptotic approximation taken in the previous section means for these supergravity quantities. In section \ref{3-forms}, we will relate our results to the dual field theory description in terms of three-form gauge fields commonly used for axion monodromy models. This will allow us to provide a geometric meaning to the underlying structure  revealed by the the three-form gauge fields in string flux compactifications. The reader only interested in the results of our analysis of flux vacua can safely skip this section.

\subsection{Asymptotic limits and the $\cN=1$ supergravity data\label{sec:sugra}}

Equivalently to studying the asymptotic limits of the scalar potential we can also determine the asymptotic 
behavior of the K\"ahler potential \eqref{Kcs} and flux superpotential \eqref{Wcs}. This analysis 
will highlight various properties of the asymptotic limits and clarify our approximation taken in the strict asymptotic regime. 

Let us begin by investigating the K\"ahler potential \eqref{Kcs}, which can be written in a more compact form 
as indicated in \eqref{Kcs-Wcs-compact}. The moduli dependence in $K^{\rm cs}$ on $t^i,\zeta^\kappa$ arises through the appearance of $\Omega$.
 As a first approximation when taking the limit $t^i \rightarrow i \infty$, we will replace $ \Omega $ with the nilpotent orbit $ \Omega_{\rm nil}$
 as discussed around \eqref{Pinil}. Inserting the expression for $\Omega_{\rm nil}$ we can use the 
 properties of $N_i$ in $\langle \,\cdot\, , \,\cdot\, \rangle$ to write 
\beq  \label{Knil}
  K^{\rm cs}_{\rm nil} =   - \log \langle \Omega_{\rm nil} , \bar{\Omega}_{\rm nil} \rangle = - \log \langle e^{2i s^j N_j}a_0(\zeta) ,\bar{a}_0 (\bar \zeta)\rangle \ .
\eeq
Since $N_j$ are nilpotent operators, the exponential in \eqref{Knil} can always be expanded to get a polynomial with a finite number of terms.
This implies that $K^{\rm cs}$, in the nilpotent orbit approximation with all exponential corrections $e^{2\pi it^j}$ dropped, is 
the logarithm of a polynomial in the $s^i$ and is independent of the axions $\phi^i$.  
$K^{\rm cs}_{\rm nil}$ still depends on a considered variable $s^i$ if $N_i a_0 \neq 0$. This latter condition is a necessary condition for the limit 
to be at infinite distance in the metric derived from $K^{\rm cs}$. The appearance of the continuous shift symmetries $\phi^i \rightarrow \phi^i + c^i$ at infinite distance singularities
was recently discussed in \cite{GPV} in the context of the Swampland Distance Conjecture. 
It is important to stress that $K^{\rm cs}_{\rm nil}$ given in \eqref{Knil} is not yet the strict asymptotic expression obtained by using the growth result \eqref{general-norm-growth}.
In fact, to apply this growth estimate one first has to fix a growth sector \eqref{growthsector} 
and expand \eqref{Knil} in powers of the ratios $s^i/s^{i+1}$ to obtain
\beq \label{Ksl2}
   K^{\rm cs}_{\rm sl(2)} \sim - \log\Big[ \Big( \frac{s^1}{s^2}\Big)^{d_{1}} \cdots \Big(\frac{s^{\nmod-1}}{s^\nmod} \Big)^{d_{\nmod-1}}
 (s^\nmod)^{d_{\nmod}} f(\zeta,\bar \zeta) \Big]\ ,
\eeq
where $d_i$ is the highest power of $N_1+\cdots+N_i$ acting on $a_0$ that is non-zero as in \eqref{def-dpower}. 
In other words, the estimate \eqref{Ksl2} extracts the leading power of the coordinates $s^i$ from the general expression \eqref{Knil}
in a sector \eqref{growthsector}. This implies that not only exponential corrections are omitted, but also sub-leading polynomial corrections 
in the coordinates $s^i$. 

In a next step we look at the flux superpotential $W$ introduced in \eqref{Wcs}. The approximation of neglecting exponential corrections 
is again implemented by replacing $\Omega$ with $\Omega_{\rm nil}$ in the asymptotic regime. Using the shorthand notation \eqref{Kcs-Wcs-compact}
we thus find 
\beq
\label{Wnil}
    W_{\rm nil} = \langle G_4, \Omega_{\rm nil} \rangle =  \langle \rho(G_4,\phi), e^{i s^j N_j} a_0 \rangle\ ,
\eeq
where $\rho(G_4,\phi)$ was defined in \eqref{exp-xN}. Despite the fact that we have dropped exponential corrections, this 
expression captures the field dependence in a non-trivial way. Let us expand $\rho$ into some basis. To be concrete we 
use the basis associated to the splitting of $H^4_{\rm p} (Y_4,\bbR)$ given by the $V_{\Bell}$, and denote it by
\beq
\label{basis}
     v^{\Bell}_{i_\Bell}\ : \quad \text{span}_\bbR\big\{ v^{\Bell}_{1} , \ldots,  v^{\Bell}_{\text{dim} V_\Bell} \big\} = V_\Bell\ , 
\eeq 
where $\Bell \in \cE$ is a vector as before. 
We thus write $\rho = \varrho^{i_\Bell}_\Bell  v^{\Bell}_{i_\Bell}$, such that \eqref{Wnil} takes the form
\beq 
W_{\rm nil} = \sum_{\Bell \in \cE} \sum_{i_\Bell}  \varrho^{i_\Bell}_\Bell (G_4,\phi)\,  \Gamma^{\Bell}_{i_\Bell}(s,\zeta) \ , \qquad           \Gamma^{\Bell}_{i_\Bell} =  \langle v^{\Bell}_{i_\Bell} , e^{i s^j N_j} a_0 \rangle \ .
\eeq
The remarkable fact about this expansion is, on the one hand, that we succeeded to separate the $\phi^i$ and $s^i$ dependence. On the other hand, we
have done this cleverly, such that the $ \Gamma^{\Bell}_{i_\Bell} $ are polynomials of a highest $s^i$-power determined by $\Bell$, and the singularity type. Concretely they admit the expansion
\beq
 \Gamma^{\Bell}_{i_\Bell}  =  (is^1)^{d_1-4+\ell_1} (is^2)^{d_2-d_1+\ell_2-\ell_1}\cdots (i s^{\hat n})^{d_{\hat n}-d_{\hat n-1}+\ell_{\hat n}-\ell_{\hat n-1}} \  \widehat \Gamma^{\Bell}_{i_\Bell} \Big(\frac{s^1}{s^2},\frac{s^2}{s^3},\ldots,s^{\hat n} \Big)
  \label{gamma}
\eeq
where $\widehat{\Gamma}^{\Bell}_{i_\Bell} \Big(\frac{s^1}{s^2},\frac{s^2}{s^3},\ldots,s^{\hat n} \Big)$ involves subleading polynomial corrections in the coordinates $s^i$. To determine the 
sl(2)-approximation, denoted for us as the strict asymptotic result,  we have to further  drop out  the subleading polynomial corrections in the coordinates ratios $s^i/s^{i+1}$ in \eqref{gamma}, so that  $\widehat{\Gamma}^{\Bell}_{i_\Bell}$ becomes just a constant  $\widehat{\Gamma}^{\Bell}_{i_\Bell}\sim c^{\Bell}_{i_\Bell}$ and the superpotential reads\footnote{
It is possible to get the same result for the superpotential if first extracting the leading dependence on the coordinates $s^i$ and denoting
\begin{equation*}
\varrho_{\Bell}\equiv \langle \rho(G_4,\phi),N_1^{ (d_1-4+l_1)/2 }N_2^{(d_2-d_1+l_2-l_1)/2}\cdots  N_{\hat n}^{(d_{\hat n}-d_{\hat n-1}+l_n-l_{\hat n-1})/2} a_0 \rangle\ .
\end{equation*}}

\beq \scalebox{0.96}{$
W_{\rm sl(2)} = \sum\limits_{\Bell \in \cE} \sum\limits_{i_\Bell}  \varrho^{i_\Bell}_\Bell (G_4,\phi)\, c^{\Bell}_{i_\Bell}(s,\zeta) (is^1)^{d_1-4+\ell_1} (is^2)^{d_2-d_1+\ell_2-\ell_1}\cdots (i s^{\hat n})^{d_{\hat n}-d_{\hat n-1}+\ell_{\hat n}-\ell_{\hat n-1}} \ .
$} \eeq

This, together with \eqref{Ksl2}, will give rise to the leading growth of the scalar potential given in \eqref{general-norm-growth}.
We will see in section \ref{3-forms} that this expansion also allows us to extract the crucial 
information when formulating the theory using Minkowski three-form gauge fields.  

To sum up, the strict asymptotic approximation consists of neglecting, not only the exponentially suppressed corrections, but also subleading polynomial terms in the coordinates $s^i$. This can be done in a consistent way near any singular limit of the moduli space and provides the leading behavior of the scalar potential for each growth sector \eqref{growthsector}. In terms of the supergravity embedding, it corresponds to considering a factorizable K\"ahler potential that keeps only the leading term, i.e. the logarithm of a monomial of degree $d_{\hat n}$, and a superpotential where each axionic function  $\varrho^{i_\Bell}_\Bell $ is multiplied by a single saxionic monomial of degree $d_{\hat n}-4+l_{\hat n}$. This yields a scalar potential that can be expressed as a sum of squares as in \eqref{general-norm-growth}.

Let us close this section by noting that the expressions arising in the strict asymptotic approximation can have a clear physical interpretation 
as neglecting some perturbative and non-perturbative corrections. This is for example the case in the famous Sen's weak coupling limit in Type IIB and the mirror Type IIA duals at large volume, in which the dependence on the dilaton can be factorized in the K\"ahler potential to leading order in $\alpha'$. Hence, the sl(2)-norm provides the correct dilaton-dependence of the scalar potential at tree level and neglects $\alpha'$-corrections that will mix the dilaton and the K\"ahler moduli. However, such an interpretation fails in other types of limits, where the subleading polynomial corrections have nothing to do with $\alpha'$-corrections. It remains as an open question for the future to study how sensitive to this approximation our results are for the flux vacua presented in the next sections.

\subsection{Relation to Minkowski three-form gauge fields} \label{3-forms}

The asymptotic flux splitting and the nilpotent orbit result for the scalar potential at the large field limits derived in section \ref{sec:asymptotic} have a very intuitive physics interpretation in terms of the dual formulation of Minkowski three-form gauge fields, as we will explain in the following. 

First, let us notice that each infinite distance limit of the form \eqref{limit} is characterized by the appearance of some axions $\phi^i = \text{Re}\,t^i$ whose discrete axionic shift symmetry is inherited from the monodromy transformation $T_i$ around the singular locus located at $s^i = \text{Im}\,t^i\rightarrow \infty$. 
In the context of the complex structure moduli space of Calabi-Yau compactifications, the axions  can receive a flux-induced scalar potential which is multi-branched, i.e.  only the combined discrete transformation of the axion and the fluxes leave invariant the effective theory.

The scalar potential of an axion can always be described in a dual picture by means of a coupling to the field strength of a space-time three-form gauge field $F_4=dC_3$ \cite{Dvali:2005an,Kaloper:2008fb,Kaloper:2011jz}. Allowing for the presence of multiple axions and three-forms gauge fields, the scalar potential reads
\beq
V=-Z_{AB}(s^i)F_4^AF_4^B+ F_4^A \varrho_A(\phi^i)\ 
\label{V4form}
\eeq
where $A,B$ run over the number of three-form gauge fields.
Here $Z_{AB}(s)$ is the kinetic matrix of the three-form gauge fields and is parametrized by the saxions, while all the dependence on the axion appears only through the shift symmetric functions $\varrho_A(\phi)$. In particular, it has been shown in \cite{4forms,Carta:2016ynn} that the flux induced scalar potential of Type II compactifications can always be brought to the above form, where $Z_{AB}$ and $\varrho_A$ were derived by dimensional reduction from ten-dimensional Type~II supergravity to four dimensions.\footnote{Note a three-form with action \eqref{V4form} naturally arises 
when computing the Type IIA scalar potential \cite{Grimm:2004ua}. Furthermore, three-forms are essential when studying the couplings to D-branes \cite{Grimm:2008dq,Grimm:2011dx,Kerstan:2011dy}.} The functions $\varrho_A$ are a shift symmetric combination of the internal fluxes and the axions that can be generically expressed \cite{Herraez:2018vae} as
\beq
\label{rhoq}
\varrho_A=(e^{-\phi^i N_i})^B_A\ q_B
\eeq 
where $N_i$ are nilpotent matrices associated to the discrete axionic symmetries and $q_B$ a vector of internal fluxes.

Upon integrating out the three-form gauge fields via their equations of motion,
\beq
\label{on-shell}
\star F_4^A=Z^{AB}\varrho_B
\eeq
the scalar potential becomes
\beq
\label{4forms2}
V=Z^{AB}(s)\varrho_A(\phi)\varrho_B(\phi)
\eeq
which corresponds to a quadratic form on $\varrho_A$. It was also shown in \cite{4forms} (see also \cite{Carta:2016ynn,Herraez:2018vae,Farakos:2017ocw,Bandos:2018gjp,Escobar:2018rna,Lanza:2019xxg,Marchesano:2019hfb}) that the above scalar potential reproduces the usual form of the scalar potential derived from the  $\cN=1$ supergravity formulae in four dimensional flux compactifications when combined with the contribution of localized sources.

Interestingly, the form \eqref{4forms2} is the same expression for the scalar potential found in \eqref{G4Zrho} upon applying the nilpotent orbit theorem in the asymptotic regime. 
Each flux component in \eqref{exp-xN} corresponds to the on-shell result of a four-form \eqref{on-shell} and the nilpotent matrices in \eqref{rhoq} are the same nilpotent operators $N_i=\log T_i$ of \eqref{N=logT} in which the entire formalism is based on. 
This is expected from the fact that both formalisms rely on the presence of axionic shift symmetries inherited from the monodromy transformations and, therefore, become manifest in these asymptotic regimes. Let us remark that, even if the discrete shift symmetries are valid everywhere in the moduli space, the notion of an axion as a scalar field enjoying an approximate continuous shift symmetry is only valid in these asymptotic regimes.
Let us also notice that this dual description in terms of four-form fields is independent of supersymmetry and can in principle even describe non-perturbative potentials \cite{Garcia-Valdecasas:2016voz}. It would be thus very interesting to further explore how the asymptotic Hodge theory approach can be interlinked with the use of four-forms and how much of the structure derived with the four-forms has in fact a geometric counterpart. 
To give another example, the flux sublattice of dynamical fluxes found in \cite{Lanza:2019xxg} has a deep relation with the massless components in the asymptotic flux splitting of section \ref{flux-split} which would be interesting to further investigate in the future.

Finally, we would like to remark that the strict asymptotic approximation taken in \eqref{general-norm-growth} allows us to further express the potential as the sum of asymptotically orthogonal flux components at the large field limit.
In other words, it is always possible to find a basis such that the kinetic matrix $Z^{AB}$ of the four-forms is nearly block-diagonal in the sense that the non-diagonal terms are subleading with respect to the diagonal ones. The strict asymptotic approximation consists of neglecting these non-diagonal terms so that the potential becomes a sum of squares, 
\beq
V=\sum_{\Bell\in \cE}Z^\Bell(s)\|\rho_\Bell(G_4,\phi)\|^2\ =\sum_{\Bell\in \cE} \sum_{i_\Bell, j_\Bell}Z_{i_\Bell j_\Bell}^\Bell(s)\ \varrho^{i_\Bell}_\Bell(\phi) \varrho^{j_\Bell}_\Bell(\phi)\ 
\label{V4forms}
\eeq
with the exception of a possible remnant coming from tadpole cancellation. Here, we have used again the expansion \eqref{basis} into a basis of vectors associated to the flux splitting into orthogonal $V_\Bell$ vector spaces, such that $\rho = \varrho^{i_\Bell}_\Bell  v^{\Bell}_{i_\Bell}$.
Using the growth theorem \eqref{general-norm-growth} we can infer the leading behavior of each block diagonal piece of the inverse metric $Z^{AB}(s)$,
\beq
\label{Zsl2}
Z_{i_\Bell j_\Bell}^\Bell\sim \Big( \frac{s^1}{s^2}\Big)^{\ell_{1}-4} \cdots \Big(\frac{s^{\nmod-1}}{s^\nmod} \Big)^{\ell_{\nmod-1}-4}
 (s^\nmod)^{\ell_{\nmod}-4} \cK_{i_\Bell j_\Bell}^\Bell\ 
 \eeq
where
    $ \cK_{i_\Bell j_\Bell}^\Bell$ was defined in \eqref{v-orth<C>}.  This is something that could not be determined only in terms of the four-forms. Hence, our classification of the asymptotic flux splittings at the large field limits of Calabi-Yau manifolds can allow us to derive the three-form gauge field metrics and with them, the axionic monodromic potential, at other types of singularities beyond the typical case of the large complex structure limit. Furthermore, this  monodromic potential written \`a la Dvali-Kaloper-Sorbo in terms of four-forms is useful to construct axion inflationary models and study the viability of large field ranges. In section \ref{sec:axionDependence} we will exploit our formalism to derive general conclusions about backreaction issues and large field ranges in axion monodromy models.
    
Let us finally mention that this bilinear form of the potential has been proven to be very useful to minimize the scalar potential of weakly coupled Type IIA flux compactifications and study the vacua structure in a systematic way \cite{Valenzuela:2016yny,Marchesano:2019hfb}. In fact, the ansatz assumed in \cite{Marchesano:2019hfb} is precisely guaranteed by the strict asymptotic approximation yielding \eqref{Zsl2}. In this paper, we will exploit the algebraic structure arising in the strict asymptotic regime to study the vacua structure at any asymptotic limit of the complex structure moduli space of a Calabi-Yau manifold.

\section{General two-moduli limits} \label{general-two-moduli}

In this section we apply the machinery introduced in the previous sections for two-moduli families of Calabi-Yau fourfolds.
More precisely, we investigate the limits \eqref{limit} with $\nmod=2$ in the complex structure moduli space of any Calabi-Yau fourfold $Y_4$ with
$h^{3, 1} = 2$. We first set up notations in order to get familiar with the asymptotic splitting of $H^4_{\rm p}(Y_4, \bbR)$
in the two-moduli setting in subsection \ref{flux-splitting2}. Then, in subsection \ref{sec:singularitiesAndEnhancements}, we list all possible
limits and corresponding singularity types that can occur in this moduli space. 
As a consequence, we are able to infer the asymptotic splitting of the flux space for each limit. To exemplify the use of these results,
we focus in subsection \ref{mainexample} on a particular limit and discuss all possible decompositions $G_4 = \hat G_4 + G_4^0$,
with $\hat G_4$ being an unbounded asymptotically massless flux.
This data will be used in the next section to establish universal no-go results on flux vacua.

\subsection{Asymptotic flux splitting and scalar potential} \label{flux-splitting2}

Let us consider the complex structure moduli space of a Calabi-Yau fourfold $Y_4$ with $h^{3, 1} = 2$.
We are interested in the case $\nmod=2$ in \eqref{limit} sending both coordinates to a limit.
Around such a limit we introduce local coordinates $t^1,t^2$ denoted by
\beq
\label{2fieldlimit}
  t^1 = \phi^1 + i s   \ , \qquad t^2 = \phi^2 + i u \ ,
\eeq
such that $Y_4$ becomes singular at $s, u \to \infty$.
For any chosen positive $\gamma, \delta$ we can consider two growth sectors \eqref{growthsector}
given by
\begin{equation} \label{eqn:2moduliGrowthSector}
  \cR_{12} = \left\{(t^1, t^2) \middle\vert \frac{s}{u} > \gamma, u > \gamma, \phi^i < \delta\right\}\ , \quad \cR_{21} =\left\{(t^1, t^2) \middle\vert \frac{u}{s} > \gamma, s > \gamma, \phi^i < \delta\right\}\ .
\end{equation}
The first sector $\cR_{12}$ can be interpreted as capturing paths in which $s$ grows faster than $u$ when approaching the limit $s,u\to \infty$, while $\cR_{21}$ exchanges the roles of $s$ and $u$. Let us consider $\cR_{12}$, after possible renaming, and divide the limit into two
steps. We first go to the singular locus $s \to \infty$ and call the arising singularity type $\mathsf{Type\ A}$, where we necessarily find one of
the types listed in \eqref{listtypes}. In a second step we send $u \to \infty$ arriving at singularity type $\mathsf{Type\ B}$ from the list \eqref{listtypes}. In this situation, we say that the sector $\cR_{12}$ is associated to the singularity enhancement
\beq \label{AtoB}
     \mathsf{Type\ A}\ \rightarrow \mathsf{Type\ B}\ .
\eeq
Importantly, we are able to classify all possible singularity types, as already discussed in section~\ref{asympt-limits}, and determine all allowed
enhancements, as discussed in section \ref{sec:singularitiesAndEnhancements}.

Associated to an enhancement $\mathsf{Type\ A} \to \mathsf{Type\ B}$, there is an asymptotic splitting of $H^4_{\rm p}(Y_4, \bbR)$ introduced in \eqref{split-Vell}. In the two-moduli case it takes the form
\begin{equation} \label{eqn:2moduliAsymptoticSplitting}
  H^4_{\rm p}(Y_4, \bbR) = \bigoplus_{\Bell=(m,n) \in \cE} V_{mn}\ ,
\end{equation}
where we explicitly spelled out the indices on the subspaces $V_{mn}$. The set $\cE$ depends on the enhancement $\mathsf{Type\ A} \to \mathsf{Type\ B}$ and will be given explicitly in subsection \ref{sec:singularitiesAndEnhancements} for each possible enhancement.
Using \eqref{eqn:2moduliAsymptoticSplitting} a general flux $G_4$ can be decomposed as
\beq
    G_4 = \sum_{(m,n) \in \cE} G_4^{mn}\ ,\qquad G_4^{mn} \in V_{mn}\ ,
\eeq
and we also introduce the expansion
\beq \label{xNG_2}
   e^{-\phi^i N_i} G_4 =  \sum_{(m,n) \in \cE} \rho_{mn}\ ,\qquad \rho_{mn} \in V_{mn}\ ,
\eeq
Then the growth of the norm $\Vert G_4 \Vert^2$ can be inferred from \eqref{G4-growth_gen} and reads
\begin{equation} \label{G4-growth_two-mod}
  \Vert G_4 \Vert^2\ \sim\ \sum_{(m, n) \in \cE} \Big( \frac{s}{u}\Big)^{m-4} u^{n-4}\ A_{mn}(G_4,\phi)\ ,
\end{equation}
where we defined $A_{mn} = \Vert \rho_{mn} \Vert^2_\infty > 0$. Inserting this asymptotic growth into the M-theory scalar potential, we have
\begin{equation} \label{VM-growth}
  V_{\rm M} \sim \frac{1}{\cV_4^3}\Big(\sum_{(m,n) \in \cE} s^{m-4} u^{n-m}\,  A_{mn} - A_{\rm loc}\Big),
\end{equation}
where we have set $A_{\rm loc} \equiv \langle G_4, G_4\rangle$, which is independent of the moduli. The
scalar potential \eqref{VM-growth} will be the starting point of our study of flux vacua in section \ref{Fluxvacua}.

In the next section we aim to establish no-go results for vacua of \eqref{VM-growth} that are under parametric control. This control
is encoded in the fluxes and hence the coefficients $A_{mn}$. Whether or not a flux can be made very large is determined by
the tadpole constraint \eqref{tadpole}. In section \ref{unboundedflux} we have introduced a type of flux, denoted by $\hat G_4$, that does not
contribute to the tadpole constraint and has an asymptotically vanishing contribution to the scalar potential.
Clearly, the determination of the allowed splits $\hat G_4$, $G_4^0$ depends crucially on the set of possible indices $\cE$ appearing in the asymptotic splitting \eqref{eqn:2moduliAsymptoticSplitting}. In particular, we recall from \eqref{hatG_inVlight} that
\beq
     \hat G_4 \ \in\  V_{\rm light} = \bigoplus_{(m,n) \in \cE_{\rm light}} V_{mn}\ ,
\eeq
and hence the vectors in $\cE_{\rm light}$ crucially determine the allowed $\hat G_4$.
It is the power of the used formalism that we can classify systematically all possible singularities and hence all possible splits \eqref{eqn:2moduliAsymptoticSplitting}. In the next subsection, we will show a full classification of singularity types of Calabi-Yau fourfolds with $h^{3, 1} = 2$. We determine all possible singularity enhancements, the associated asymptotic splittings, and the form of the sets $\cE$ and $\cE_{\rm light}$. For each
of these cases one can then determine all possible $\hat G_4$ as we exemplify for an example in section \ref{mainexample}.

\subsection{Classification of two-moduli limits and enhancements therein} \label{sec:singularitiesAndEnhancements}

In this section we summarize the classification of all possible singularity types that can arise in a Calabi-Yau fourfold with
$h^{3,1}=2$, when both complex structure variables tend to a limit \eqref{limit}.  Following a similar strategy as for Calabi-Yau threefolds, as discussed in detail in \cite{Kerr2017,Grimm:2018cpv}, we enumerate all singularity types of the primitive middle Hodge numbers $(1, 2, \hat m, 2, 1)$. Here we denoted by $\hat m$ the dimension of the primitive part $H^{2,2}_{\rm p}(Y_4)$ of $H^{2,2}(Y_4)$. As explained in section~\ref{asympt-limits}, there are five major types $\rI, \rII, \rIII, \rIV$, and $\rV$. Each type is supplemented by two indices as shown in \eqref{listtypes}. The classification is summarized in table \ref{tab:singularityTypes}. In fact, the appearance of each type depends on the primitive Hodge number $\hat m$. When $0 \le \hat m \le 3$, not all types can occur. To avoid singling out special cases, we will assume $\hat m \ge 4$. The cases dropped with this assumption admit the same features as some of the cases we consider here and thus will not alter our conclusions.

\begin{table}[H]
\centering
\begin{tabular}{|c|l|}\hline
  \rule[-.25cm]{0cm}{.7cm} $\rI$      & $\rI_{0, 0},\  \rI_{0, 1},\ \rI_{0, 2},\ \rI_{1, 1},\ \rI_{1, 2}, \
  \rI_{2, 2}$\\\hline
  \rule[-.25cm]{0cm}{.7cm} $\rII$     & $\rII_{0, 0},\ \rII_{0, 1},\ \rII_{1, 1}$\\\hline
  \rule[-.25cm]{0cm}{.7cm} $\rIII$    & $\rIII_{0, 0},\ \rIII_{0, 1},\ \rIII_{1, 1}$\\\hline
  \rule[-.25cm]{0cm}{.7cm} $\rIV$     & $\rIV_{0, 1}$\\\hline
  \rule[-.25cm]{0cm}{.7cm} $\rV$      & $\rV_{1, 1},\ \rV_{1, 2},\ \rV_{2, 2}$\\\hline
\end{tabular}
\caption{Table showing all $16$ singularity types that can occur in a two-moduli family of Calabi-Yau fourfolds with primitive Hodge number $\hat m \ge 4$.} \label{tab:singularityTypes}
\end{table}

\begin{figure}[H]
\centering
\includegraphics[width=\textwidth]{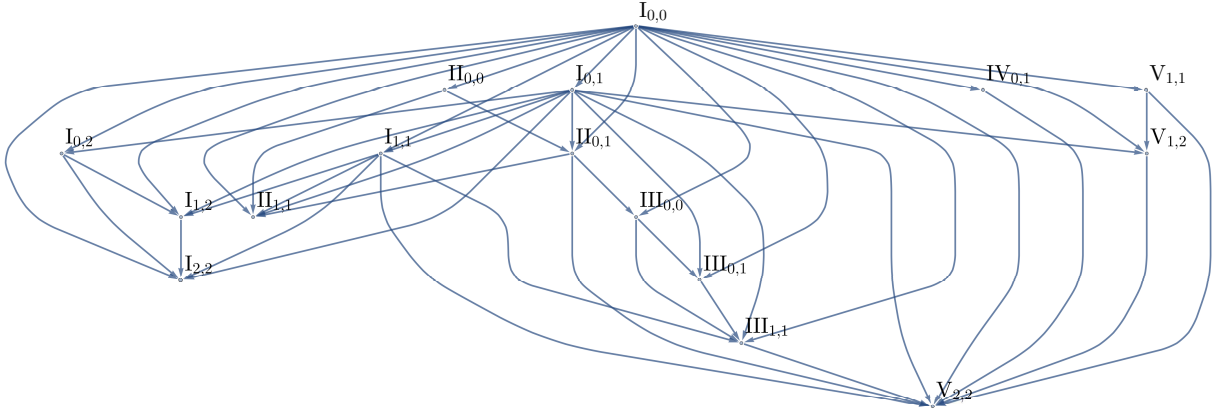}
\caption{The enhancement network of Calabi-Yau fourfolds with primitive middle Hodge numbers $(1, 2, \hat m, 2, 1)$ with $\hat m \ge 4$. In this graph, an edge $\mathsf{Type\ A} \to \mathsf{Type\ B}$ indicates an enhancement of singularity type from $\mathsf{Type\ A}$ to $\mathsf{Type\ B}$. Note that the enhancement relation is not transitive, as can be easily checked in, e.g., the enhancement chain $\rII_{0, 0} \to \rII_{0, 1} \to \rIII_{0, 0}$.} \label{fig:enhancement2moduli}
\end{figure}

Given the list of allowed singularity types in table \ref{tab:singularityTypes}, we can now check which singularities
can occur in an enhancement where we send $s,u$ to infinity successively. As in \eqref{AtoB} we can send
$s \to \infty$ to get a singularity type $\mathsf{Type\ A}$ and then $u \to \infty$ to get a singularity type $\mathsf{Type\ B}$. We say $\mathsf{Type\ A}$ gets enhanced to $\mathsf{Type\ B}$. There are intricate rules guarding the possible enhancements among different singularity types. And these rules determine the asymptotic splitting directly. These rules are described in \cite{Kerr2017} following the classic work \cite{CKS}, and its application in Calabi-Yau threefold degenerations can be found in \cite{Kerr2017,Grimm:2018cpv}. Following the same procedure as in \cite{Grimm:2018cpv}, we determine the enhancement network among the types given in table \ref{tab:singularityTypes}. The result is shown in figure \ref{fig:enhancement2moduli}.\footnote{
It was recently pointed out in \cite{Grimm:2019bey} that this strategy, applied to the K\"ahler moduli side, can be employed to classify Calabi-Yau manifolds.}

It is worth pointing out that the type II enhancements occur, for example, at the Sen's weak coupling limit when the Calabi-Yau fourfold
is used as an F-theory background. This has been discussed in detail in \cite{Donagi:2012ts}. In a two-moduli limit as discussed
here, one can combine the weak coupling limit with another limit in complex structure moduli space. In fact, as we will discuss below
an example enhancement that occurs when combining Sen's weak coupling limit with another limit to reach the large complex structure
point of $Y_4$. Concretely one finds in this case
\beq \label{special_cases}
    \rII_{0, 1} \ \rightarrow \ \rV_{2, 2} \ ,
\eeq
where we have displayed the enhancement for which we first send $s\to \infty$
and then $u\to \infty$ as required for the growth sector $\cR_{12}$ in \eqref{eqn:2moduliGrowthSector}. The limit $s \rightarrow \infty$ corresponds to the weak coupling limit. 

Having determined all possible enhancements we can also compute for each case the associated asymptotic splitting \eqref{eqn:2moduliAsymptoticSplitting}. The results are shown in table  \ref{tab:asymptoticSplittings}.
We will demonstrate the usage of this table in the following subsection in which we discuss one case in detail and determine the allowed unbounded asymptotically massless fluxes $\hat G_4$.

\begin{table}[p]
\centering
\begin{tabular}{|Rp{1cm}L|C|C|C|}\hline
  \multicolumn{3}{|c|}{\rm Enhancements}                                                                                        & \cE_{\rm light}                   & \cE_{\rm rest}                   & \cE_{\rm heavy}\\\hline\hline
  \multirow{11}{*}{$\rI_{0, 1}$}  & \multirow{11}{*}{\tikzmark{j}} & \tikzmark{j1}\rI_{0, 2}                & (3, 3), (4, 3)                    & \mathbf{(4, 4)}                  & \mathbf{(4, 5)}, \mathbf{(5, 5)}\\\cline{4-6}
                                            &                                & \tikzmark{j2}\rI_{1, 1}                    & (3, 2)                            & (3, 4), \mathbf{(4, 4)}, (5, 4)  & \mathbf{(5, 6)}\\\cline{4-6}
                                            &                                & \tikzmark{j3}\rI_{1, 2}                & (3, 2), (4, 3)                    & (3, 4), \mathbf{(4, 4)}, (5, 4)  & \mathbf{(4, 5)}, \mathbf{(5, 6)}\\\cdashline{4-6}
                                            &                                & \tikzmark{j11}\rI_{1, 2}               & (3, 3), (4, 2)                    & \mathbf{(4, 4)}                  & \mathbf{(4, 6)}, \mathbf{(5, 5)}\\\cline{4-6}
                                            &                                & \tikzmark{j10}\rI_{2, 2}                   & (3, 2), (4, 2)                    & (3, 4), \mathbf{(4, 4)}, (5, 4)  & \mathbf{(4, 6)}, \mathbf{(5, 6)}\\\cline{4-6}
                                            &                                & \tikzmark{j4}\rII_{0, 1}               & (3, 3), (4, 3)                    & \mathbf{(4, 4)}                  & \mathbf{(4, 5)}, \mathbf{(5, 5)}\\\cline{4-6}
                                            &                                & \tikzmark{j5}\rII_{1, 1}                   & (3, 2), (4, 3)                    & (3, 4), \mathbf{(4, 4)}, (5, 4)  & \mathbf{(4, 5)}, \mathbf{(5, 6)}\\\cline{4-6}
                                            &                                & \tikzmark{j6}\rIII_{0, 1}              & (3, 3), (4, 2)                    & \mathbf{(4, 4)}                  & \mathbf{(4, 6)}, \mathbf{(5, 5)}\\\cline{4-6}
                                            &                                & \tikzmark{j7}\rIII_{1, 1}              & (3, 2), (4, 2)                    & (3, 4), \mathbf{(4, 4)}, (5, 4)  & \mathbf{(4, 6)}, \mathbf{(5, 6)}\\\cline{4-6}
                                            &                                & \tikzmark{j8}\rV_{1, 2}                & (3, 3), (4, 0), (4, 2)            & \mathbf{(4, 4)}                  & (4, 6), \mathbf{(4, 8)}, \mathbf{(5, 5)}\\\cline{4-6}
                                            &                                & \tikzmark{j9}\rV_{2, 2}                    & (3, 2), (4, 0), (4, 2)            & (3, 4), \mathbf{(4, 4)}, (5, 4)  & (4, 6), \mathbf{(4, 8)}, \mathbf{(5, 6)}\\\hline\hline
  \multirow{2}{*}{$\rI_{0, 2}$}   & \multirow{2}{*}{\tikzmark{k}}  & \tikzmark{k1}\rI_{1, 2}                & (3, 2), (3, 3)                    & (3, 4), \mathbf{(4, 4)}, (5, 4)  & \mathbf{(5, 5)}, \mathbf{(5, 6)}\\\cline{4-6}
                                            &                                & \tikzmark{k2}\rI_{2, 2}                    & (3, 2)                            & (3, 4), \mathbf{(4, 4)}, (5, 4)  & \mathbf{(5, 6)}\\\hline\hline
  \multirow{5}{*}{$\rI_{1, 1}$}       & \multirow{5}{*}{\tikzmark{l}}  & \tikzmark{l1}\rI_{1, 2}                & (2, 2), (4, 3)                    & \mathbf{(4, 4)}                  & \mathbf{(4, 5)}, \mathbf{(6, 6)}\\\cline{4-6}
                                            &                                & \tikzmark{l2}\rI_{2, 2}                    & (2, 2), (4, 2)                    & \mathbf{(4, 4)}                  & \mathbf{(4, 6)}, \mathbf{(6, 6)}\\\cline{4-6}
                                            &                                & \tikzmark{l3}\rII_{1, 1}                   & (2, 2), (4, 3)                    & \mathbf{(4, 4)}                  & \mathbf{(4, 5)}, \mathbf{(6, 6)}\\\cline{4-6}
                                            &                                & \tikzmark{l4}\rIII_{1, 1}              & (2, 2), (4, 2)                    & \mathbf{(4, 4)}                  & \mathbf{(4, 6)}, \mathbf{(6, 6)}\\\cline{4-6}
                                            &                                & \tikzmark{l5}\rV_{2, 2}                    & (2, 2), (4, 0), (4, 2)            & \mathbf{(4, 4)}                  & (4, 6), \mathbf{(4, 8)}, \mathbf{(6, 6)}\\\hline\hline
  \rI_{1, 2}                      & \tikzmark{m}                   & \tikzmark{m1}\rI_{2, 2}                    & (2, 2), (3, 2)                    & (3, 4), \mathbf{(4, 4)}, (5, 4)  & \mathbf{(5, 6)}, \mathbf{(6, 6)}\\\hline\hline
  \multirow{2}{*}{$\rII_{0, 0}$}      & \multirow{2}{*}{\tikzmark{a}}  & \tikzmark{a1}\rII_{0, 1}               & (3, 3), (4, 3)                    & \mathbf{(4, 4)}                  & \mathbf{(4, 5)}, \mathbf{(5, 5)}\\\cline{4-6}
                                            &                                & \tikzmark{a2}\rII_{1, 1}                   & (3, 3), (4, 2)                    & \mathbf{(4, 4)}                  & \mathbf{(4, 6)}, \mathbf{(5, 5)}\\\hline\hline
  \multirow{4}{*}{$\rII_{0, 1}$}  & \multirow{4}{*}{\tikzmark{b}}  & \tikzmark{b1}\rII_{1, 1}                   & (3, 2), (3, 3)                    & (3, 4), \mathbf{(4, 4)}, (5, 4)  & \mathbf{(5, 5)}, \mathbf{(5, 6)}\\\cline{4-6}
                                            &                                & \tikzmark{b2}\rIII_{0, 0}              & (3, 2)                            & (3, 4), \mathbf{(4, 4)}, (5, 4)  & \mathbf{(5, 6)}\\\cline{4-6}
                                            &                                & \multirow{2}{*}{\tikzmark{b3}$\rV_{2, 2}$} & \multirow{2}{*}{$(3, 0), (3, 2)$} & (3, 4), (3, 6), \mathbf{(4, 4)}, & \multirow{2}{*}{$(5, 6), \mathbf{(5, 8)}$}\\
                                            &                                &                                                  &                                   & (5, 2), (5, 4)                   &\\\hline\hline
  \multirow{2}{*}{$\rIII_{0, 0}$} & \multirow{2}{*}{\tikzmark{d}}  & \tikzmark{d1}\rIII_{0, 1}              & (2, 2), (4, 3)                    & \mathbf{(4, 4)}                  & \mathbf{(4, 5)}, \mathbf{(6, 6)}\\\cline{4-6}
                                            &                                & \tikzmark{d2}\rIII_{1, 1}              & (2, 2), (4, 2)                    & \mathbf{(4, 4)}                  & \mathbf{(4, 6)}, \mathbf{(6, 6)}\\\hline\hline
  \rIII_{0, 1}                    & \tikzmark{e}                   & \tikzmark{e1}\rIII_{1, 1}              & (2, 2), (3, 2)                    & (3, 4), \mathbf{(4, 4)}, (5, 4)  & \mathbf{(5, 6)}, \mathbf{(6, 6)}\\\hline\hline
  \rIII_{1, 1}                    & \tikzmark{f}                   & \tikzmark{f1}\rV_{2, 2}                    & (2, 0), (2, 2), (4, 2)            & (2, 4), \mathbf{(4, 4)}, (6, 4)  & (4, 6), (6, 6), \mathbf{(6, 8)}\\\hline\hline
  \rIV_{0, 1}                     & \tikzmark{g}                   & \tikzmark{g1}\rV_{2, 2}                    & (1, 0), (1, 2), (3, 2)            & (3, 4), \mathbf{(4, 4)}, (5, 4)  & (5, 6), (7, 6), \mathbf{(7, 8)}\\\hline\hline
  \multirow{2}{*}{$\rV_{1, 1}$}       & \multirow{2}{*}{\tikzmark{h}}  & \tikzmark{h1}\rV_{1, 2}                & (0, 0), (2, 2), (4, 3)            & \mathbf{(4, 4)}                  & \mathbf{(4, 5)}, (6, 6), \mathbf{(8, 8)}\\\cline{4-6}
                                            &                                & \tikzmark{h2}\rV_{2, 2}                    & (0, 0), (2, 2), (4, 2)            & \mathbf{(4, 4)}                  & \mathbf{(4, 6)}, (6, 6), \mathbf{(8, 8)}\\\hline\hline
  \rV_{1, 2}                      & \tikzmark{i}                   & \tikzmark{i1}\rV_{2, 2}                    & (0, 0), (2, 2), (3, 2)            & (3, 4), \mathbf{(4, 4)}, (5, 4)  & \mathbf{(5, 6)}, (6, 6), \mathbf{(8, 8)}\\\hline
\end{tabular}
\begin{tikzpicture}[overlay, remember picture, yshift=.25\baselineskip, shorten >=.5pt, shorten <=.5pt]
  \draw [->] ([xshift=-1em, yshift=.5ex]{pic cs:j}) -- ([yshift=.5ex]{pic cs:j1});
  \draw [->] ([xshift=-1em, yshift=.5ex]{pic cs:j}) -- ([yshift=.5ex]{pic cs:j2});
  \draw [->] ([xshift=-1em, yshift=.5ex]{pic cs:j}) -- ([yshift=.5ex]{pic cs:j3}) node[pos=0.9, above] {\tiny $a$};
  \draw [->] ([xshift=-1em, yshift=.5ex]{pic cs:j}) -- ([yshift=.5ex]{pic cs:j11}) node[pos=0.9, above] {\tiny $b$};
  \draw [->] ([xshift=-1em, yshift=.5ex]{pic cs:j}) -- ([yshift=.5ex]{pic cs:j4});
  \draw [->] ([xshift=-1em, yshift=.5ex]{pic cs:j}) -- ([yshift=.5ex]{pic cs:j5});
  \draw [->] ([xshift=-1em, yshift=.5ex]{pic cs:j}) -- ([yshift=.5ex]{pic cs:j6});
  \draw [->] ([xshift=-1em, yshift=.5ex]{pic cs:j}) -- ([yshift=.5ex]{pic cs:j7});
  \draw [->] ([xshift=-1em, yshift=.5ex]{pic cs:j}) -- ([yshift=.5ex]{pic cs:j8});
  \draw [->] ([xshift=-1em, yshift=.5ex]{pic cs:j}) -- ([yshift=.5ex]{pic cs:j9});
  \draw [->] ([xshift=-1em, yshift=.5ex]{pic cs:k}) -- ([yshift=.5ex]{pic cs:k1});
  \draw [->] ([xshift=-1em, yshift=.5ex]{pic cs:k}) -- ([yshift=.5ex]{pic cs:k2});
  \draw [->] ([xshift=-1em, yshift=.5ex]{pic cs:l}) -- ([yshift=.5ex]{pic cs:l1});
  \draw [->] ([xshift=-1em, yshift=.5ex]{pic cs:l}) -- ([yshift=.5ex]{pic cs:l2});
  \draw [->] ([xshift=-1em, yshift=.5ex]{pic cs:l}) -- ([yshift=.5ex]{pic cs:l3});
  \draw [->] ([xshift=-1em, yshift=.5ex]{pic cs:l}) -- ([yshift=.5ex]{pic cs:l4});
  \draw [->] ([xshift=-1em, yshift=.5ex]{pic cs:l}) -- ([yshift=.5ex]{pic cs:l5});
  \draw [->] ([xshift=-1em, yshift=.5ex]{pic cs:m}) -- ([yshift=.5ex]{pic cs:m1});
  \draw [->] ([xshift=-1em, yshift=.5ex]{pic cs:a}) -- ([yshift=.5ex]{pic cs:a1});
  \draw [->] ([xshift=-1em, yshift=.5ex]{pic cs:a}) -- ([yshift=.5ex]{pic cs:a2});
  \draw [->] ([xshift=-1em, yshift=.5ex]{pic cs:b}) -- ([yshift=.5ex]{pic cs:b1});
  \draw [->] ([xshift=-1em, yshift=.5ex]{pic cs:b}) -- ([yshift=.5ex]{pic cs:b2});
  \draw [->] ([xshift=-1em, yshift=.5ex]{pic cs:b}) -- ([yshift=.5ex]{pic cs:b3});
  \draw [->] ([xshift=-1em, yshift=.5ex]{pic cs:d}) -- ([yshift=.5ex]{pic cs:d1});
  \draw [->] ([xshift=-1em, yshift=.5ex]{pic cs:d}) -- ([yshift=.5ex]{pic cs:d2});
  \draw [->] ([xshift=-1em, yshift=.5ex]{pic cs:e}) -- ([yshift=.5ex]{pic cs:e1});
  \draw [->] ([xshift=-1em, yshift=.5ex]{pic cs:f}) -- ([yshift=.5ex]{pic cs:f1});
  \draw [->] ([xshift=-1em, yshift=.5ex]{pic cs:g}) -- ([yshift=.5ex]{pic cs:g1});
  \draw [->] ([xshift=-1em, yshift=.5ex]{pic cs:h}) -- ([yshift=.5ex]{pic cs:h1});
  \draw [->] ([xshift=-1em, yshift=.5ex]{pic cs:h}) -- ([yshift=.5ex]{pic cs:h2});
  \draw [->] ([xshift=-1em, yshift=.5ex]{pic cs:i}) -- ([yshift=.5ex]{pic cs:i1});
  \draw [->] ([xshift=-1em, yshift=.5ex]{pic cs:j}) -- ([yshift=.5ex]{pic cs:j10});
\end{tikzpicture}
\caption{Asymptotic splittings of all enhancements shown in figure \ref{fig:enhancement2moduli}.
We assume $\hat{m} \ge 4$, otherwise not all enhancements can occur. A boldface $\mathbf{(m, n)}$ indicates that $V_{mn}$ contains some highest weight form $a^{mn}_{j_{mn}}$ defined around equation \eqref{eqn:highestWeightForm}. Note that we did not include the 16 cases $\rI_{0, 0}\, \to\,  \mathsf{Type\ B}$, since these are simply the one-modulus enhancements with all elements in $\cE$ of the form $\mathbf{(4, m)}$. The enhancement $\rI_{0, 1} \to \rI_{1, 2}$ has two different $\cE$ set configurations, and we distinguish them by adding small labels $a$ and $b$ on top of the arrows.} \label{tab:asymptoticSplittings}
\end{table}

Given the data summarized in table \ref{tab:asymptoticSplittings} it is not hard to derive the corresponding 
scalar potentials $V_{\rm M}$ using \eqref{VM-growth}. For completeness, we list the results in table \ref{potentials_list1}.
It is interesting to point out that all potentials 
obtained in this way actually come in pairs. There are two ways we find agreeing potentials, which we listed in table \ref{potentials_list1}. Firstly, note 
that some of the sets $\cE$ in table \ref{tab:asymptoticSplittings} are simply identical, as, for example, 
for the enhancements $\rI_{0, 1} \rightarrow \rIII_{0, 1}$ and $\rII_{0, 0} \rightarrow \rII_{1, 1}$.
Secondly, two potentials might agree if we exchange the names $s \leftrightarrow u$. 
This happens, for example, for the enhancements $\rII_{0, 1} \rightarrow \rV_{2, 2}$ and $\rIV_{0, 1} \rightarrow \rV_{2, 2}$.
Recall that all enhancements in table \ref{tab:asymptoticSplittings} are determined for fixed 
growth sector $\cR_{12}$ defined in \eqref{eqn:2moduliGrowthSector}, which allows for the limit of 
sending first $s \rightarrow \infty$ and then $u \rightarrow \infty$. However, 
we can also look at the other sector $\cR_{21}$, in which the roles of $s$ and $u$ are exchanged. 
This implies that a certain form of a potential can arise from two different enhancements depending 
on the considered growth sector, the chosen names $s,u$, and thus the order of limits. The physical significance of such phenomenon is not completely clear, but it appears to be partially related to the possibility of realising the combinations of enhancements that yield identical scalar potentials in geometry. This topic will be studied more systematically in future works.

\begin{table}[H]
\centering
\begin{tabular}{|RcL|C|}\hline
  \multicolumn{3}{|c|}{Enhancements}                                                                             & \textrm{Potential } V_{\mathrm{M}}\\\hline\hline
  \rI_{0, 1}\tikzmark{A1}                      & & \multirow{2}{*}{\tikzmark{A}$\rV_{1, 2}$} & \multirow{2}{*}{$\frac{c_1}{s} + \frac{c_2}{u^4} + \frac{c_3}{u^2} + c_4 u^2 + c_5 u^4  + c_6 s - c_0$}\\
  \rV_{1, 1}\tikzmark{A2}                          & &                                                     &\\\hline\hline

  \rI_{0, 1}\tikzmark{B1}                      & & \multirow{2}{*}{\tikzmark{B}$\rV_{2, 2}$}     & \multirow{2}{*}{$\frac{c_1}{us} + \frac{c_2}{u^4} + \frac{c_3}{u^2} + \frac{c_4 u}{s} + \frac{c_5 s}{u} + c_6 u^2 + c_7 u^4 + c_8 u s - c_0$}\\
  \rV_{1, 2}\tikzmark{B2}                      & &                                                     &\\\hline\hline

  \rI_{1, 1}\tikzmark{C1}                          & & \multirow{2}{*}{\tikzmark{C}$\rV_{2, 2}$}     & \multirow{2}{*}{$\frac{c_1}{s^2} + \frac{c_2}{u^4} + \frac{c_3}{u^2} + c_4 u^2 + c_5 u^4 + c_6 s^2 - c_0$}\\
  \rV_{1, 1}\tikzmark{C2}                          & &                                                     &\\\hline\hline

  \rII_{0, 1}\tikzmark{D1}                     & & \multirow{2}{*}{\tikzmark{D}$\rV_{2, 2}$}     & \multirow{2}{*}{$\frac{c_1}{u^3 s} + \frac{c_2}{u s} + \frac{c_3 u}{s} + \frac{c_4 u^3}{s} + \frac{c_5 s}{u^3} + \frac{c_6 s}{u} + c_7 u s + c_8 u^3 s - c_0$}\\
  \rIV_{0, 1}\tikzmark{D2}                     & &                                                     &\\\hline\hline

  \rI_{0, 1}\tikzmark{E1}                      & & \multirow{2}{*}{\tikzmark{E}$\rI_{1, 2}$} & \multirow{4}{*}{$\frac{c_1}{u s} + \frac{c_2}{u} + \frac{c_3 u}{s} + \frac{c_4 s}{u} + c_5 u + c_6 u s - c_0$}\\
  \rI_{0, 2}\tikzmark{E2}                      & &                                                     &\\\cline{1-3}
  \rI_{0, 1}\tikzmark{F1}                      & & \multirow{2}{*}{\tikzmark{F}$\rII_{1, 1}$}    &\\
  \rII_{0, 1}\tikzmark{F2}                     & &                                                     &\\\hline\hline

  \rI_{0, 1}\tikzmark{G1}                      & & \multirow{2}{*}{\tikzmark{G}$\rI_{1, 2}$} & \multirow{6}{*}{$\frac{c_1}{s} + \frac{c_2}{u^2} + c_3 u^2 + c_4 s - c_0$}\\
  \rI_{1, 1}\tikzmark{G2}                          & &                                                     &\\\cline{1-3}
  \rI_{0, 1}\tikzmark{H1}                      & & \multirow{2}{*}{\tikzmark{H}$\rIII_{0, 1}$}&\\
  \rIII_{0, 0}\tikzmark{H2}                    & &                                                     &\\\cline{1-3}
  \rII_{0, 0}\tikzmark{I1}                     & & \multirow{2}{*}{\tikzmark{I}$\rII_{1, 1}$}    &\\
  \rI_{1, 1}\tikzmark{I2}                      & &                                                     &\\\hline\hline

  \rI_{0, 1}\tikzmark{J1}                      & & \multirow{2}{*}{\tikzmark{J}$\rI_{2, 2}$}     & \multirow{4}{*}{$\frac{c_1}{u s} + \frac{c_2}{u^2} + \frac{c_3 u}{s} + \frac{c_4 s}{u} + c_5 u^2 + c_6 u s - c_0$}\\
  \rI_{1, 2}\tikzmark{J2}                      & &                                                     &\\\cline{1-3}
  \rI_{0, 1}\tikzmark{K1}                      & & \multirow{2}{*}{\tikzmark{K}$\rIII_{1, 1}$}&\\
  \rIII_{0, 1}\tikzmark{K2}                    & &                                                     &\\\hline\hline

  \rI_{0, 1}\tikzmark{L1}                      & & \tikzmark{L2}\rI_{0, 2}                   & \multirow{5}{*}{$\frac{c_1}{s} + \frac{c_2}{u} + c_3 u + c_4 s - c_0$}\\\cline{1-3}
  \rI_{0, 1}\tikzmark{L3}                      & & \tikzmark{L4}\rII_{0, 1}                  &\\\cline{1-3}
  \rII_{0, 0}\tikzmark{L5}                         & & \tikzmark{L6}\rII_{0, 1}                  &\\\cline{1-3}
  \rI_{0, 1}\tikzmark{L7}                      & & \multirow{2}{*}{\tikzmark{L9}$\rII_{0, 1}$}&\\
  \rII_{0, 0}\tikzmark{L8}                         & &                                                     &\\\hline\hline

  \rI_{0, 1}\tikzmark{M1}                      & & \tikzmark{M2}\rI_{1, 1}                       & \multirow{3}{*}{$\frac{c_1}{u s} + \frac{c_2 u}{s} + \frac{c_3 s}{u} + c_4 u s - c_0$}\\\cline{1-3}
  \rI_{0, 2}\tikzmark{M3}                      & & \tikzmark{M4}\rI_{2, 2}                       &\\\cline{1-3}
  \rII_{0, 1}\tikzmark{M5}                     & & \tikzmark{M6}\rIII_{0, 0}                 &\\\hline\hline

  \rI_{1, 1}\tikzmark{N1}                          & & \tikzmark{N2}\rI_{2, 2}                       & \multirow{5}{*}{$\frac{c_1}{s^2} + \frac{c_2}{u^2} + c_3 u^2 + c_4 s^2 - c_0$}\\\cline{1-3}
  \rI_{1, 1}\tikzmark{N3}                          & & \tikzmark{N4}\rIII_{1, 1}                 &\\\cline{1-3}
  \rIII_{0, 0}\tikzmark{N5}                    & & \tikzmark{N6}\rIII_{1, 1}                 &\\\cline{1-3}
  \rI_{1, 1}\tikzmark{N7}                          & & \multirow{2}{*}{\tikzmark{N9}$\rIII_{1, 1}$}&\\
  \rIII_{0, 0}\tikzmark{N8}                    & &                                                     &\\\hline\hline

  \rIII_{1, 1}\tikzmark{O1}                    & & \tikzmark{O2}\rV_{2, 2}                       &\rule[-.4cm]{0cm}{1cm}   \frac{c_1}{u^2 s^2} + \frac{c_2}{s^2} + \frac{c_3}{u^2} + \frac{c_4 u^2}{s^2} + \frac{c_5 s^2}{u^2} + c_6 u^2 + c_7 s^2 + c_8 u^2 s^2 - c_0\\\hline
\end{tabular}
\begin{tikzpicture}[overlay, remember picture, yshift=.25\baselineskip, shorten >=.5pt, shorten <=.5pt]
  \draw [->] ([yshift=.5ex]{pic cs:A1}) -- ([yshift=.9ex]{pic cs:A});
  \draw [->] ([yshift=.5ex]{pic cs:A2}) -- ([yshift=.1ex]{pic cs:A});
  \draw [->] ([yshift=.5ex]{pic cs:B1}) -- ([yshift=.9ex]{pic cs:B});
  \draw [->] ([yshift=.5ex]{pic cs:B2}) -- ([yshift=.1ex]{pic cs:B});
  \draw [->] ([yshift=.5ex]{pic cs:C1}) -- ([yshift=.9ex]{pic cs:C});
  \draw [->] ([yshift=.5ex]{pic cs:C2}) -- ([yshift=.1ex]{pic cs:C});
  \draw [->] ([yshift=.5ex]{pic cs:D1}) -- ([yshift=.9ex]{pic cs:D});
  \draw [->] ([yshift=.5ex]{pic cs:D2}) -- ([yshift=.1ex]{pic cs:D});
  \draw [->] ([yshift=.5ex]{pic cs:E1}) -- ([yshift=.9ex]{pic cs:E}) node[midway, above] {\tiny $a$};
  \draw [->] ([yshift=.5ex]{pic cs:E2}) -- ([yshift=.1ex]{pic cs:E});
  \draw [->] ([yshift=.5ex]{pic cs:F1}) -- ([yshift=.9ex]{pic cs:F});
  \draw [->] ([yshift=.5ex]{pic cs:F2}) -- ([yshift=.1ex]{pic cs:F});
  \draw [->] ([yshift=.5ex]{pic cs:G1}) -- ([yshift=.9ex]{pic cs:G}) node[midway, above] {\tiny $b$};
  \draw [->] ([yshift=.5ex]{pic cs:G2}) -- ([yshift=.1ex]{pic cs:G});
  \draw [->] ([yshift=.5ex]{pic cs:H1}) -- ([yshift=.9ex]{pic cs:H});
  \draw [->] ([yshift=.5ex]{pic cs:H2}) -- ([yshift=.1ex]{pic cs:H});
  \draw [->] ([yshift=.5ex]{pic cs:I1}) -- ([yshift=.9ex]{pic cs:I});
  \draw [->] ([yshift=.5ex]{pic cs:I2}) -- ([yshift=.1ex]{pic cs:I});
  \draw [->] ([yshift=.5ex]{pic cs:J1}) -- ([yshift=.9ex]{pic cs:J});
  \draw [->] ([yshift=.5ex]{pic cs:J2}) -- ([yshift=.1ex]{pic cs:J});
  \draw [->] ([yshift=.5ex]{pic cs:K1}) -- ([yshift=.9ex]{pic cs:K});
  \draw [->] ([yshift=.5ex]{pic cs:K2}) -- ([yshift=.1ex]{pic cs:K});
  \draw [->] ([yshift=.9ex]{pic cs:L1}) -- ([yshift=.9ex]{pic cs:L2});
  \draw [->] ([yshift=.1ex]{pic cs:L1}) -- ([yshift=.1ex]{pic cs:L2});
  \draw [->] ([yshift=.9ex]{pic cs:L3}) -- ([yshift=.9ex]{pic cs:L4});
  \draw [->] ([yshift=.1ex]{pic cs:L3}) -- ([yshift=.1ex]{pic cs:L4});
  \draw [->] ([yshift=.9ex]{pic cs:L5}) -- ([yshift=.9ex]{pic cs:L6});
  \draw [->] ([yshift=.1ex]{pic cs:L5}) -- ([yshift=.1ex]{pic cs:L6});
  \draw [->] ([yshift=.5ex]{pic cs:L7}) -- ([yshift=.9ex]{pic cs:L9});
  \draw [->] ([yshift=.5ex]{pic cs:L8}) -- ([yshift=.1ex]{pic cs:L9});
  \draw [->] ([yshift=.9ex]{pic cs:M1}) -- ([yshift=.9ex]{pic cs:M2});
  \draw [->] ([yshift=.1ex]{pic cs:M1}) -- ([yshift=.1ex]{pic cs:M2});
  \draw [->] ([yshift=.9ex]{pic cs:M3}) -- ([yshift=.9ex]{pic cs:M4});
  \draw [->] ([yshift=.1ex]{pic cs:M3}) -- ([yshift=.1ex]{pic cs:M4});
  \draw [->] ([yshift=.9ex]{pic cs:M5}) -- ([yshift=.9ex]{pic cs:M6});
  \draw [->] ([yshift=.1ex]{pic cs:M5}) -- ([yshift=.1ex]{pic cs:M6});
  \draw [->] ([yshift=.9ex]{pic cs:N1}) -- ([yshift=.9ex]{pic cs:N2});
  \draw [->] ([yshift=.1ex]{pic cs:N1}) -- ([yshift=.1ex]{pic cs:N2});
  \draw [->] ([yshift=.9ex]{pic cs:N3}) -- ([yshift=.9ex]{pic cs:N4});
  \draw [->] ([yshift=.1ex]{pic cs:N3}) -- ([yshift=.1ex]{pic cs:N4});
  \draw [->] ([yshift=.9ex]{pic cs:N5}) -- ([yshift=.9ex]{pic cs:N6});
  \draw [->] ([yshift=.1ex]{pic cs:N5}) -- ([yshift=.1ex]{pic cs:N6});
  \draw [->] ([yshift=.5ex]{pic cs:N7}) -- ([yshift=.9ex]{pic cs:N9});
  \draw [->] ([yshift=.5ex]{pic cs:N8}) -- ([yshift=.1ex]{pic cs:N9});
  \draw [->] ([yshift=.9ex]{pic cs:O1}) -- ([yshift=.9ex]{pic cs:O2});
  \draw [->] ([yshift=.1ex]{pic cs:O1}) -- ([yshift=.1ex]{pic cs:O2});
\end{tikzpicture}
\caption{Enhancements and their associated asymptotic scalar potential $V_{\rm M}$. In this table, we group together the enhancements that are simply identical or identical as we exchange the growth sector $\cR_{12}$ and $\cR_{21}$, i.e., exchange $s$ with $u$. In each box there are two arrows with the upper one valid for the growth sector $\cR_{12}$ and 
the lower one valid for the growth sector $R_{21}$. 
The double-arrow cases, e.g. $\rIII_{1, 1} \to \rV_{2, 2}$ in the last row, have the scalar potential $V_{\rm M}$ symmetric in $s$ and $u$. The coefficients $c_i$ with $i > 0$ are positive, while the sign of $c_0$ is undetermined.}
\label{potentials_list1}
\end{table}

\subsection{Main example: enhancement from type $\rII$ singularity} \label{mainexample}

In this subsection we focus on an enhancement from the type $\rII$ singularity, i.e. $\rII_{0, 1} \to \rV_{2, 2}$.
This is one case appearing in table \ref{tab:asymptoticSplittings} and we already noted around \eqref{special_cases} that it plays a special
role, since it involves Sen's weak coupling limit. In fact, we will see later that it precisely reproduces the potential and
de Sitter no-go result  of \cite{Hertzberg:2007wc}.

\begin{table}[H]
\centering
\begin{tabular}{R|C|C|C|C|C|C|C|C|C|}\cline{2-10}
                                             & \multicolumn{2}{c|}{$V_{\rm light}$} & \multicolumn{5}{c|}{$V_{\rm rest}$}             & \multicolumn{2}{c|}{$V_{\rm heavy}$}\\\hline
                                               \multicolumn{1}{|r|}{$\cE$}             & (3,0) & (3,2)                      & (3,4) & (3,6) & (4,4)      & (5,2) & (5,4) & (5,6) & (5,8)\\\hline
  \multicolumn{1}{|r|}{$V_{mn}$}             & V_{30} & V_{32}                      & V_{34} & V_{36} & V_{44}      & V_{52} & V_{54} & V_{56} & V_{58}\\\hline
  \multicolumn{1}{|r|}{$\dim{V_{mn}}$}       & 1      & 1                           & 1      & 1      & \hat{m} - 2 & 1      & 1      & 1      & 1\\\hline
  \multicolumn{1}{|r|}{$\textrm{Basis}$}     & v_{30} & v_{32}                      & v_{34} & v_{36} & v^{\kappa}  & v_{52} & v_{54} & v_{56} & v_{58}\\\hline
  \multicolumn{1}{|r|}{Flux number}          & f_6    & f_4                         & f_2    & f_0    & g_\kappa    & h_0    & h_1    & h_2    & h_3\\\hline
\end{tabular}
\caption{The data of the asymptotic splitting of the primitive middle cohomology $H^4_{\rm p}(Y_4, \bbR)$ associated with the enhancement $\rII_{0, 1} \to \rV_{2, 2}$. The basis and flux numbers of the subspace $V_{44}$ are denoted by $g_\kappa$ and $v^\kappa$ with $\kappa=1,\ldots,\hat m- 2$.
Note that we assume $\hat{m} \ge 4$, so all the subspaces are present in the asymptotic splitting.} \label{tab:2moduliDataAsymptoticSplitting}
\end{table}

Let us first record the asymptotic splitting associated to this enhancement. According to table \ref{tab:asymptoticSplittings}, we have $\cE_{\rm light} = \{(3, 0), (3, 2)\}$, $\cE_{\rm rest} = \{(3, 4), (3, 6), (4, 4), (5, 2), (5, 4)\}$, and $\cE_{\rm heavy} = \{(5, 6), (5, 8)\}$. The asymptotic splitting is then explicitly given by
\begin{equation}
  H^4_{\rm p}(Y_4,\mathbb{R}) = V_{30} \oplus V_{32} \oplus V_{34} \oplus V_{36} \oplus V_{44} \oplus V_{52} \oplus V_{54} \oplus V_{56} \oplus V_{58}\ ,
\end{equation}
where the dimension and basis of each subspace is summarized in table \ref{tab:2moduliDataAsymptoticSplitting} and we have also recorded our choice of notation for the flux numbers in the enhancement $\rII_{0, 1} \to \rV_{2, 2}$. The flux numbers are defined to be the coefficient of a flux $G_4$ in the basis shown in table \ref{tab:2moduliDataAsymptoticSplitting} to the asymptotic splitting, i.e.
\begin{equation} \label{eqn:2moduliG4splitting}
  G_4 = f_6 v_{30} + f_4 v_{32} + f_2 v_{34} + f_0 v_{36} + g_\kappa v^\kappa + h_0 v_{52} + h_1 v_{54} + h_2 v_{56} + h_3 v_{58}\ .
\end{equation}
The particular names of flux numbers are chosen for convenience of our discussion in section \ref{sec:example}
when we show that our formalism reproduces well-known existing no-go results.
Furthermore, taking into account the orthogonality relation \eqref{V-orth}, we normalize the basis such that
\begin{equation} \label{eqn:2ModuliBasisNormalisation}
   \pair{v_{30}, v_{58}} = \pair{v_{32}, v_{56}} = \pair{v_{34}, v_{54}} = \pair{v_{36}, v_{52}} = 1.
\end{equation}
The pairing in the basis $v^\kappa$ of $V_{44}$ will be denoted by $\eta^{\kappa \lambda}$. It is positive, i.e.~one has $\eta^{\kappa \lambda} g_\kappa g_\lambda > 0$ for non-zero $g_\kappa$.

Applying the asymptotic splitting of flux \eqref{eqn:2moduliG4splitting} and table \ref{tab:2moduliDataAsymptoticSplitting} to the asymptotic behavior of the scalar potential \eqref{VM-growth}, we have
\begin{equation} \label{VM_mainexample}
  V_{\rm M} \sim \frac{1}{\cV_4^3}\left(\frac{A_{f_6}}{u^3 s} + \frac{A_{f_4}}{u s} + \frac{A_{f_2} u}{s} + \frac{A_{f_0} u^3}{s} + \frac{A_{h_0} s}{u^3} + \frac{A_{h_1} s}{u} + A_{h_2} u s + A_{h_3} u^3 s + A_{44} - A_{\rm loc} \right),
\end{equation}
where $A_{\rm loc} = \pair{G_4, G_4}$ and the coefficients in the growth terms are defined according to our notation of flux numbers in table \ref{tab:2moduliDataAsymptoticSplitting} as follows
\begin{IEEEeqnarray*}{rClrClrCl}
  A_{f_6} & = &\Vert \rho_{30} \Vert^2_\infty\, ,\quad & A_{f_4} & = & \Vert \rho_{32} \Vert^2_\infty\, ,\quad & A_{f_2} & = &  \Vert \rho_{34} \Vert^2_\infty\, ,\\
  A_{f_0} & = &\Vert \rho_{36} \Vert^2_\infty\,, \quad & A_{h_0} & = & \Vert \rho_{52} \Vert^2_\infty\, ,\quad & A_{h_1} & = & \Vert \rho_{54} \Vert^2_\infty\, ,\\
  A_{h_2} & = &\Vert \rho_{56} \Vert^2_\infty\, ,\quad & A_{h_3} & = & \Vert \rho_{58} \Vert^2_\infty\,,\quad & A_{44}  & = & \Vert \rho_{44} \Vert^2_\infty\, .
\end{IEEEeqnarray*}
Note that all $A$'s are positive and still functions of the axions $\phi^1,\phi^2$ via the exponential in \eqref{xNG_2}. Setting
$\phi^i=0$ one finds that $ A_{f_6} \propto (f_6)^2$, $ A_{f_4} \propto (f_4)^2$ etc.
With the asymptotic splitting \eqref{eqn:2moduliG4splitting} and the normalization \eqref{eqn:2ModuliBasisNormalisation}, the tadpole condition \eqref{tadpole} can be expressed as
\begin{equation} \label{eqn:2moduliTadpole}
  \frac{\chi(Y_4)}{24} = \frac{1}{2}\pair{G_4, G_4} = f_6 h_3 + f_4 h_2 + f_2 h_1 + f_0 h_0 + \frac{1}{2} \eta^{\kappa \lambda} g_\kappa g_\lambda.
\end{equation}

Now we discuss the separation $G_4 = \hat{G}_4 + G_4^0$ of a flux $G_4$ into an unbounded part $\hat{G}_4$ and a remaining part $G_4^0$. First we deal with the unbounded component $\hat{G}_4$ which belongs to $V_{\rm light}$. By checking table \ref{tab:2moduliDataAsymptoticSplitting}, we see that the requirement $\hat{G}_4 \in V_{\rm light}$ implies that an unbounded flux $\hat{G}_4$ can contain components $f_6$ or $f_4$. Also the first orthogonality in \eqref{no-tadpole} and the massless condition \eqref{massless} on $\hat{G}_4$ are automatically satisfied because we ask for $\hat{G}_4 \in V_{\rm light}$.

Once an unbounded part $G_4$ is identified, the second condition in \eqref{no-tadpole} can be used to restrict the remaining part $G_4^0$. The general results are displayed in table \ref{tab:2ModuliHatGvsGr}. We explain its derivation in an example where we take $f_6$ as the unbounded flux, i.e. we set $\hat{G}_4 = f_6 v_{30}$.  Then, subtracting $\hat{G}_4$ from the splitting \eqref{eqn:2moduliG4splitting} we have the following form of $\widetilde{G}_4^0$ which needs further restriction
\begin{equation*}
  \widetilde{G}_4^0 = f_4 v_{32} + f_2 v_{34} + f_0 v_{36} + g_\kappa v^\kappa + h_0 v_{52} + h_1 v_{54} + h_2 v_{56} + h_3 v_{58}.
\end{equation*}
According to our normalization \eqref{eqn:2ModuliBasisNormalisation}, it is readily computed that
\begin{equation}
  \pair{\hat{G}_4, \widetilde{G}_4^0} = 2 f_6 h_3.
\end{equation}
Hence the second condition in \eqref{no-tadpole} implies $h_3 = 0$, i.e.
\begin{equation}
  G_4^0 = f_4 v_{32} + f_2 v_{34} + f_0 v_{36} + g_\kappa v^\kappa + h_0 v_{52} + h_1 v_{54} + h_2 v_{56}.
\end{equation}
In this way, we have separated the flux components in $G_4$ into an unbounded flux component $f_6$ and the remaining flux components $f_4, \ldots, h_2$, with the condition that the flux component $h_3 = 0$ is absent. Inserting the condition $h_3 = 0$ into the tadpole condition \eqref{eqn:2moduliTadpole}, we see that the tadpole condition is then satisfied by the remaining components
\begin{equation}
  f_4 h_2 + f_2 h_1 + f_0 h_0 + \frac{1}{2} \eta^{\kappa \lambda} g_\kappa g_\lambda = \frac{\chi(Y_4)}{24}.
\end{equation}
We can now repeat this analysis for all combinations of possible unbounded fluxes $f_6$ and $f_4$, we obtain table \ref{tab:2ModuliHatGvsGr}. This data will be used in section \ref{Fluxvacua} to determine the vacua of \eqref{VM_mainexample}. 

\begin{table}[H]
\centering
\begin{tabular}{|C|C|C|C|}\hline
  \hat{G}_4 & G_4^0                                  & \textrm{Condition on } G_4^0 & \textrm{Self-dual Pairs in } G_4^0\\\hline\hline
  f_6       & f_4, f_2, f_0, g_\kappa, h_0, h_1, h_2 & h_3 = 0                      & (f_4, h_2), (f_2, h_1), (f_0, h_0)\\\hline
  f_4       & f_6, f_2, f_0, g_\kappa, h_0, h_1, h_3 & h_2 = 0                      & (f_6, h_3), (f_2, h_1), (f_0, h_0)\\\hline
  f_6, f_4  & f_2, f_0, g_\kappa, h_0, h_1, h_2, h_3 & f_6 h_3 + f_4 h_2 = 0        & (f_2, h_1), (f_0, h_0)\\\hline
\end{tabular}
\caption{All possible ways of separating $G_4$ into an unbounded flux $\hat{G}_4$ and a remaining part $G_4^0$. The third column, condition on the remaining part, is coming from the second orthogonality in \eqref{no-tadpole}. The tadpole condition can be found by applying the third column to \eqref{eqn:2moduliTadpole} and it is satisfied by the remaining flux components. The forth column lists possible self-dual components inside $G_4^0$ which is introduced in section \ref{self-dual_asy} and will be used in section \ref{sec:parametric}.} \label{tab:2ModuliHatGvsGr}
\end{table}

\section{Asymptotic structure of flux vacua} \label{Fluxvacua}

In this section we will analyze the vacua structure of the flux-induced scalar potential in the strict asymptotic regimes of the field space. We will focus on asymptotic two-moduli limits of the form \eqref{limit} in the complex structure moduli space of a Calabi-Yau fourfold. These limits are characterized by two scalar fields, denoted as $s,u$, becoming large and the choice of a growth sector in \eqref{eqn:2moduliGrowthSector}, i.e. an order in the growth of the fields. We will select $\cR_{12}$ describing paths in which $s$ grows faster than $u$,  but the results for the other growth sector can be trivially found after exchanging the roles of $s$ and $u$ and renaming the coordinates. 

A complete classification of these two-moduli limits in the complex structure moduli space of a Calabi-Yau fourfold was performed in section \ref{general-two-moduli} together with the scalar potential arising in each case (see table \ref{potentials_list1}). 
Our starting point will, therefore, be the general asymptotic form of the flux potential derived in \eqref{VM-growth} and given by
\beq
V=\frac{1}{s^\alpha}\Big(\sum_{(m,n)\in \cE}  A_{mn} s^{m-4} u^{n-m}- A_{\rm loc}\Big)\\
   \equiv \frac{1}{s^\alpha}\Big(\sum_{i=1}^\cN A_{m_i n_i} s^{m_i-4} u^{n_i-m_i}- A_{\rm loc}\Big)\ .
\label{V}
\eeq
where the possible values for $(m,n)$ are given in table \ref{tab:asymptoticSplittings} and depend on the type of limit under consideration. Recall that one just has to plug the values $(m,n)$ of table \ref{tab:asymptoticSplittings} into eq. \eqref{V} to recover all possible potentials shown in table \ref{potentials_list1}. For later convenience, we have re-labelled the elements of $\cE$ as $(m_i,n_i)$, $i=1,\ldots,\cN$, where $\mathcal{N}$ is the number of different pairs $(m,n) \in \cE$ that can occur in each limit. 

Notice that the coefficients $A_{mn}$ are not arbitrary but depend on the integer fluxes and axions as in \eqref{Al}. However, we will leave them as free parameters in this section except for their sign, since they are restricted to be positive definite in the strict asymptotic regime (see eq.\eqref{Al}). This way, we can keep our analysis more general and our results will also apply to higher dimensional moduli spaces with $h^{2,1}>2$ in which there are more \emph{spectator fields} in addition to the two moduli becoming large. In those situations, the coefficients $A_{mn}$ will also be functions of these spectator fields, but the moduli scaling of $s$ and $u$ is expected to be the same. Interestingly, we will be able to formulate a no-go theorem for de Sitter only based on the scaling of $s$ and $u$ and independent of the concrete value of $A_{mn}$ as long as they remain positive. Only in section \ref{sec:axionDependence} we will specify again the concrete values for $A_{mn}$ to study axion stabilization and derive some universal results about backreaction effects in axion monodromy models. 

The reader might have also noticed that we have included an additional overall factor $1/s^\alpha$ in \eqref{V} in comparison to  \eqref{VM-growth}. This will allow us to map our results to Type IIA flux compactifications, in which an additional factor of the dilaton appears upon performing mirror symmetry and going to the Einstein frame of Type IIA, as reviewed in section \ref{orientifold_vacua}. This factor is known to be $1/s^3$ in the weak coupling limit, but we will leave the power also as a free parameter since  its value is undetermined for any of the other limits of our list beyond weak coupling.

Therefore, the general potential \eqref{V} includes all the asymptotic potentials arising in M-theory flux compactifications in a Calabi-Yau fourfold (and their corresponding F-theory/Type IIB duals) if we set $\alpha=0$, but can also describe other asymptotic string compactifications. The goal in this section is to take this general form of the asymptotic potential and analyze its vacuum structure. We will be particularly interested in whether this potential can admit any kind of vacuum at parametric control. Interestingly, since we have left the coefficients $A_{mn}$ as arbitrary parameters, the above potential can also potentially yield AdS vacua. This is impossible in F-theory/Type IIB flux compactifications as the coefficients $A_{mn}$ are correlated such that the potential is positive definite. However, it can occur in Type IIA flux compactifications where $A_{\rm loc}$ can receive contributions from other sources like metric fluxes or other components of NS flux which do not map to $H_3$ or $F_3$ fluxes in Type IIB. Hence, our general form of the potential will also allow us to study the conditions to get candidates for AdS vacua at parametric control. It is important to keep in mind, though, that they are only candidates in the sense that one should further check that the resulting values for $A_{mn}$ are compatible with some top-down string construction.

Since this section contains many different interesting results about the structure of asymptotic flux vacua, let us add here a short outline of what comes next. In sections \ref{sec:parametric} and \ref{sec:minimization} we will describe our strategy to determine the (non-)existence of vacua at parametric control.  We will then apply this strategy to a particular example corresponding to the familiar Sen's weak coupling limit and discuss the existence of dS and AdS vacua in section \ref{sec:example}. Afterwards, we will apply the same methodology to all possible limits classified in section \ref{general-two-moduli} and present the results for de Sitter in section \ref{sec:deSitter} and for AdS vacua in section \ref{sec:AdS}. 

\subsection{Flux ansatz and parametric control\label{sec:parametric}}

In this paper, we are interested in the presence of critical points at parametric control, i.e. for parametrically large field values of the scalars $s,u$. Let us recall that this is an additional constraint we need to impose as the asymptotic flux potential of \eqref{V} can in general yield vacua at finite values of $s,u$ that are not necessarily large. Furthermore, we need to require to stay in a growth sector in order to be consistent with the strict asymptotic approximation, so the ratio $s/u$ also needs to be large.

As it will become more clear through the following sections, we find that it is impossible to get any critical point at parametrically large field values of $s,u$ if all fluxes are bounded by tadpole cancellation. Therefore, it becomes necessary to add some unbounded flux that can be adjusted to be large in the asymptotic limit. For this reason, we will assume the following general Ansatz for the fluxes,
\beq
\label{ansatz1}
G_4=\hat G_4+G_4^0
\eeq
where $\hat G_4$ is an asymptotically massless unbounded flux with respect to the remaining background fluxes in $G_4^0$. This special class of fluxes were introduced in section \ref{unboundedflux}. We require them to be massless, in addition to unbounded, so that they only violate the self-duality condition mildly and it is restored in the limit. The fluxes in $G_4^0$ cannot be scaled to be large, but still must be consistent with generating a critical point at parametric control. We will consider two options, self-dual fluxes or more general fluxes with the same asymptotic scaling, as described in the following.\\

\noindent\textbf{Self-dual fluxes $G_4^0$ :}\newline
Let us first consider in $G_4^0$ only fluxes that are self-dual in the strict asymptotic regime. 
The self-duality condition on the $G_4$-flux in the strict asymptotic regime was given in  \eqref{self-dual_sl2}. In order to simplify the discussion and highlight the main properties we  assume that every subspace $V_{mn}$ is one-dimensional, except for $V_{44}$.  Because of the  property \eqref{eqn:CinftySwapV} of the operator $C_\infty$ and its relation to the $C_{\rm sl(2)}$ operator \eqref{general-norm-growth} (see also \eqref{eqn:defnCSL2}), given an $(m, n) \in \cE$, the minimal form of a non-vanishing self-dual flux should be
\begin{equation} \label{Gself-d}
  G_4 = g_{m n} v^{m n} + g_{8 - m\, 8 - n} v^{8 - m\, 8 - n}\ ,
\end{equation}
where $v^{mn}$ is the basis vector of $V_{mn}$ and no sum over $m,n$ is taken in \eqref{Gself-d}.
The self-dual condition \eqref{eqn:SDcondition} on such $G_4$ further specializes  for the case of two moduli into the following form
\begin{equation} \label{eqn:simpleSD2moduli}
  s^{m - 4} u^{n - m} g_{m n} \cK_{m n} = g_{8 - m\, 8 - n}\ ,
\end{equation}
where $\cK_{m n} = \Vert v^{m n} \Vert_\infty^2$. One realizes immediately that in most cases, if we impose two such conditions then both moduli $s$ and $u$ are fixed and one can find (finitely) many vacua by just imposing the self-duality conditions. All these vacua are Minkowski as the vacuum energy vanishes. It remains to check whether solving such self-dual conditions stabilizes the moduli inside the strict asymptotic regime, where $s/u$ and $u$ are required to be large. Since the product of flux components $g_{m n} g_{8 - m\, 8 - n}$ contributes to the tadpole condition, we see that it is not possible to make both $s/u$ and $u$ parametrically large. So there are actually no vacua at parametric control using only self-dual fluxes.

We now turn on only one pair of self-dual components and allow for an unbounded massless flux as in \eqref{ansatz1}. In this case, we can rewrite the self-dual condition \eqref{eqn:simpleSD2moduli} into
\begin{equation} \label{us-relation}
  u = \left(\frac{g_{8 - m\, 8 - n}}{g_{m n} \cK_{m n}}\right)^{\frac{1}{n - m}} s^\beta,
\end{equation}
where $\beta = \frac{4 - m}{n - m}$. This actually imposes a correlation between two moduli in terms of fluxes that are bounded by tadpole condition.  The possibilities for the exponent $\beta$ are:
\begin{enumerate}
\item  $\beta < 0\quad$ The modulus $u$ grows inversely with $s$, so it is not possible to get vacua at parametrically large field values of both $s$ and $u$.
\item $\beta = 0\quad$ The modulus $u$ is completely fixed into a ratio of flux numbers bounded by the tadpole condition, so it is not possible to make it parametrically large.
\item $0 < \beta < 1\quad$ In this case there is no obstruction to make $s/u$ and $u$ large, but only one combination is fixed. \label{enum:goodCase}
\item $\beta \ge 1\quad$ This case also includes $\beta = \infty$, where $s$ is fixed into a ratio of flux numbers bounded by the tadpole condition. It is not possible to make $s/u$ arbitrarily large to be consistent with the strict asymptotic approximation.
\end{enumerate}
We can then see that only case \ref{enum:goodCase} could yield vacua at parametric control. Since only one combination of $s, u$ can be stabilized with the self-dual pair of fluxes, the other combination needs to be fixed  by turning on some massless unbounded flux components so that we can dial both $s/u$ and $u$ into large values.
To see this we substitute $u$ given in terms of $s$ by \eqref{us-relation} back into the scalar potential and minimize the remaining one-variable potential with respect to $s$. The potential reads 
\beq
V\propto \frac{1}{s^\alpha} \sum_{(\hat m,\hat n)\in \hat\cE} \left( \hat A_{\hat m\hat n}\ s^{\hat m-4+\beta(\hat n-\hat m)} - \hat{A}_{\rm loc} \right)\, ,
\eeq
where the sum only involves now unbounded massless fluxes. 
Recall that $\hat m \le 4$ and $\hat n < 4$ are required for the flux to be massless. Interestingly, this potential can never yield de Sitter critical point for case \ref{enum:goodCase} in which $0 < \beta < 1$, since all the terms involve negative powers of $s$.
In F-theory/Type IIB flux compactifications, the contribution from $A_{\rm loc}$ cancels with the contribution from the pair of self-dual fluxes such that $\hat{A}_{\rm loc} = 0$. However, if we insist of keeping $\hat{A}_{\rm loc}$ as a free parameter so that it survives some negative contribution to the potential (as could occur in Type IIA flux compactifications), the potential \eqref{V} might also have AdS vacua. We check that, for all potentials in table \ref{potentials_list1}, only the enhanced limit $\rII_{0, 1} \to \rV_{2, 2}$ could yield an AdS vacuum at parametric control with $0 < \beta < 1$. This case indeed corresponds to the famous large volume and weak coupling limit in IIA. We will explain in more detail this vacuum in section \ref{sec:example}.\\

\noindent\textbf{General flux $G_4^0$ :}\newline
Next, we will consider a more general situation in which any flux can appear in $G^0_4$, but still keeping the condition that the vacuum is at parametric control. Non self-dual fluxes can arise, for instance, from backreaction effects of localized sources in the string compactification.
A way to implement the condition of parametric control is to require that all terms in the potential that are necessary to stabilize the moduli  should scale in the same way as $s,\, u\rightarrow \infty$. In other words, we will look for solutions of the form
\beq
s\sim \lambda^p\ ,\quad u\sim \lambda^q \ ,
\label{control}
\eeq
with $p,q$ positive such that $\lambda$ can be taken to be parametrically large. Each potential term will then scale as
\beq
V_i\sim A_i\lambda^{r_i}\ , \quad r_i=(m_i - 4 - \alpha)p+(n_i - m_i)q\ ,
\label{Vscaling}
\eeq
if $A_i$ corresponds to a contribution from $G_4^0$.
For massless unbounded fluxes in $\hat G_4$, there is no such a constraint as the flux can always be scaled up to yield the desired asymptotic scaling  with $\lambda$.

We then require that a solution of the type \eqref{control} must be found only using unbounded massless fluxes and terms in $G_4^0$ yielding the same value for $r$ and therefore scaling the same way with $\lambda$. This guarantees that the solution will still exist in the limit $\lambda\rightarrow \infty$, i.e.~at parametrically large values of $s, u$.
 As a final check if a solution is found, we need to require that $p>q$ in order to have the ratio $s/u$ large and be consistent with the strict asymptotic approximation. Similar scaling arguments to look for parametrically controlled vacua have also been recently used in \cite{Roupec:2018mbn,Junghans:2018gdb} for the weak string coupling limit in Type IIA compactifications. Let us stress that the non-trivial part of our analysis does not lie in applying such scaling arguments, but rather in identifying the asymptotic potentials that can arise in a valid flux compactification.

We will show the results for our main example in section \ref{sec:example}, and then for all possible two-moduli limits in sections \ref{sec:deSitter} and \ref{sec:AdS} in  the case of dS and AdS vacua respectively. But first, let us discuss our method to solve the minimization equations  in a systematic and convenient way.

\subsection{Minimization conditions\label{sec:minimization}}

In order to study the existence of extrema of the potential \eqref{V} we will translate the existence problem 
into a more convenient formulation using methods from linear optimization. 
To begin with, we note that the extrema of $V$ are determined by the conditions
\beqa
\label{minimize}
u \partial_u V & = & \sum_{i = 1}^{\mathcal{N}} (n_i - m_i) V_{i}=0 \ ,\\
s \partial_s V & = & \sum_{i = 1}^{\cN}(m_i - 4 - \alpha)V_i + \alpha V_{\cN+1}=0\ ,\label{minimize2}
\eeqa
where we have defined 
\beq
V_i\equiv \frac{A_{m_i n_i}}{s^\alpha} s^{m_i-4} u^{n_i-m_i}\ ,\quad V_{\cN+1}= \frac{A_{\rm loc}}{s^\alpha}\ ,
\eeq
 such that $V=\sum_{i=1}^{\cN} V_i- V_{\cN+1}$. 
 We have introduced $V_{\cN+1}$ in order to treat $A_{\rm loc}$ in analogy with other terms by associating it a scaling $m=n=0$.
 We further define
 \beq
   V_0 \equiv |V|_{\partial V =0}\ ,
 \eeq 
 i.e.~introduce the absolute value of the potential at this extremum. The definition of $V_0$ implies that at the extremum one has
\beq
\label{V0}
\sum_{i=1}^{\cN} V_i- V_{\cN+1}\pm V_0=0\ ,
\eeq
where the positive sign implies an Anti-de Sitter extremum while the negative sign a de Sitter extremum. The equations \eqref{minimize}, \eqref{minimize2}, and \eqref{V0} can be packed into the following homogeneous system $Av=0$ with
\beq
\mathcal{A} = \left(
\begin{array}{ccccc}
  m_1 - 4 - \alpha & m_2 - 4 - \alpha & \cdots & \alpha & 0\\
  n_1 - m_1        & n_2 - m_2        & \cdots & 0      & 0\\
  1                & 1                & \cdots & -1     & \pm 1
\end{array}\right)
\eeq
and $v^T=(V_0 ,\, V_1 , \, V_2, \dots,\, V_{\cN+1})$. 
Notice that we will have as many columns as contributions $V_i$ to the potential with different $m_i,n_i$ plus one.

We can now use Stiemke's theorem which states that either a linear homogeneous system $Av =0$ possesses a solution with all variables positive or there exists a linear combination of the equations that has all non-negative coefficients, one or more of which are positive. Applied to our problem, one thus finds that either 
there exists a $v^T=(V_0,\ldots,V_{\cN+1})$ such that
\beq
\mathcal{A} v=0\ ,\quad V_\kappa >0\, , \ \kappa = 0,\ldots,\cN+1\ ,
\eeq
or there exists a $\pi^T = (a,b,c) \neq 0$ such that
\beq
( \mathcal{A}^T \pi)_\kappa \geq 0 \, , \ \kappa = 0,\ldots,\cN+1\ .
\label{Stiemke}
\eeq
This second condition will be much easier to prove, and will allows us generalize some no-go theorems about de Sitter.
Stiemke's theorem thus implies that \eqref{V} has no extremum with cosmological constant $\mp V_0$ if the following system of inequalities is feasible, 
i.e.~has a non-trivial solution,
\beqa
\label{inequalities}
a(m_i - 4 - \alpha) + b(n_i - m_i) + c & \geq & 0\ ,\\
\alpha a - c                           & \geq & 0\ ,\\
\pm c                                  & \geq & 0\ ,
\eeqa
where let us recall that $+$ stands for AdS and $-$ for dS.  In other words, there will \textit{not} be a dS (AdS) critical point if one can find $a,b\in \mathbb{R}$ such that \eqref{inequalities} is satisfied 
for every pair $(m_i,n_i)$ and $c\leq 0$ ($c\geq 0$). Notice that it only makes sense to impose the second inequality it if $V_{\cN}\neq 0$.
Analogously, if no non-trivial solution is found to \eqref{inequalities}, then the system has a critical point which is a solution of the minimization conditions \eqref{minimize}. In order to determine whether it corresponds to a minimum or a maximum one would need to further study the Hessian matrix $\partial_i\partial_j V$. However, in this paper, we will restrict ourselves to analyze the presence of critical points in general.

\subsection{Parametrically controlled vacua for the main example\label{sec:example}}

In this section we will analyze the presence of asymptotic flux vacua at parametric control for a particular example: the enhancement $\rII_{0, 1} \to \rV_{2, 2}$ in a two-dimensional moduli space. 
This enhancement is one of the possible limits appearing in table~\ref{tab:asymptoticSplittings} and 
served as our main example in section~\ref{mainexample}. 
The importance of this example arises from the fact that it corresponds to the well known Sen's weak coupling limit and large complex structure limit in Type IIB. It can also be mapped to Type IIA at weak coupling and large volume, which will allow us to recover some no-go theorems for de Sitter vacua 
found in Type IIA compactifications \cite{Hertzberg:2007wc,Flauger:2008ad}. We will not find new results in this section, but it will serve us to exemplify the methodology that we will later apply to the other asymptotic limits of the moduli space of a Calabi-Yau fourfold.

The Sen weak coupling limit ($s\rightarrow \infty$) corresponds to a Type $\rII_{0, 1}$ singular divisor \cite{Donagi:2012ts,Clingher:2012rg}. When intersecting with a Type $\rIV_{0, 1}$ corresponding the large complex structure point  ($u\rightarrow \infty$), it enhances to a Type $\rV_{2, 2}$  singularity of codimension-two at the intersection. 
The values of $m,n$ consistent with this type of singularity are given in table \ref{tab:asymptoticSplittings} and imply a scalar potential of the form
\beq \label{mainex_potential}
V \sim \frac{1}{s^\alpha}\left(\frac{A_{f_6}}{u^3 s} + \frac{A_{f_4}}{u s} + \frac{A_{f_2} u}{s} + \frac{A_{f_0} u^3}{s} + \frac{A_{h_0} s}{u^3} + \frac{A_{h_1} s}{u} + A_{h_2} u s + A_{h_3} u^3 s \pm  A_{\rm loc} \right),
\eeq
as already stated in \eqref{VM_mainexample}. From the perspective of Type IIB perturbative string theory ($\alpha=0$), the fluxes denoted as $f_p$ ($h_p$) correspond to different components of R-R flux $F_3$ (NS-NS flux $H_3$).
as discussed in section \ref{orientifold_vacua}. When mapping the potential to Type IIA flux compactifications at weak coupling and large volume (so $\alpha=3$), it is important that we stay in the growth sector with $s/u$ large, so that the 10d string coupling $g_s$ remains small (see \eqref{def-taurho}). The R-R flux $F_3$ maps to R-R fluxes $F_p$ in Type IIA (that is why we have chosen the notation), while only $h_0$ maps to a NS flux in Type IIA. The other components have a more exotic interpretation in terms of geometric $h_1$ and non-geometric $h_2,h_3$ fluxes in Type IIA . In fact, the moduli dependence of the term proportional to $h_1$ can also arise as a contribution from the six-dimensional Ricci scalar in case the manifold has positive curvature, or from KK monopoles. The term $A_{\rm loc}$ has the right moduli dependence of a contribution from O6-planes. Since only the moduli dependence matters and not the specific value of the coefficients $A_{mn}$, our results will apply to compactifications involving any of these ingredients or any other object exhibiting the same moduli dependence as the above terms. See  e.g.\cite{Haque:2008jz,deCarlos:2009fq,Wrase:2010ew,Danielsson:2012by,Blaback:2013ht,Andriot:2016xvq,Andriot:2017jhf,Andriot:2018ept,Blaback:2018hdo,Shukla:2019dqd} for works attempting to construct classical de Sitter vacua using  these ingredients. Notice that other types O$q$-planes or NS5-branes are not captured in this setup, as their moduli dependence does not have a geometric interpretation in terms of $G_4$-fluxes in M/F-theory.

First of all, let us prove that in case that all fluxes are bounded, i.e.~cannot take arbitrarily large values, it is impossible to have any AdS or dS extrema 
at parametric control. 
In order to get a solution at parametric control, we will apply Stiemke's theorem only to those terms that can scale with the same power of $\lambda$ in \eqref{Vscaling}. The groups of terms that give rise to the same asymptotic scaling are:
\beqa
\label{p1}
p=q\ : &\qquad ({f_4},{h_0})\,, \quad ({f_2},{h_1},A_{\rm loc})\, ,\quad ({f_0},{h_2})\ ,\\
p=2q\ : &\qquad ({f_2},{h_0})\, , \quad ({f_0},{h_1})\ , \\
p=3q\ : &\qquad ({f_0},{h_0},A_{\rm loc})\ .
\label{p6}
\eeqa
In particular, the pairs $(f_0,h_0)$ and $(f_2,h_1)$ correspond to self-dual pair of fluxes that exhibit the same asymptotic scaling that the negative term $A_{\rm loc}$.
We can now check whether \eqref{inequalities} can be satisfied for any of the above groups. The answer is that we can always find a solution to  Stiemke's problem, meaning that there is no way to solve the minimization conditions at parametric control. Hence, there is nor AdS or dS minimum at parametric control if all fluxes are bounded.

It is not hard to see, however, that the preceding analysis is too restrictive to establish a general no-go statement, 
since it neglects the possibility to also adjust the fluxes to become large in the asymptotic limit. 
As explain in section \ref{unboundedflux}, fluxes are expected to be bounded if they contribute to the tadpole cancellation condition. 
However,  there is a special class of fluxes, namely the unbounded massless fluxes introduced in section \ref{unboundedflux},
that do not contribute to the tadpole condition and violate the self-duality mildly. 
For this reason, in section \ref{mainexample}, we identified all possible unbounded massless fluxes compatible with the singular limit taken for our main example (see table \ref{tab:2ModuliHatGvsGr}). In the following, we will consider a total flux of the form \eqref{ansatz1} where we allow for unbounded massless fluxes $\hat G_4$ in addition to the group of flux terms \eqref{p1}-\eqref{p6} denoted as $G_4^0$.
This means that we can always scale up $\hat G_4$ to achieve the desired asymptotic scaling on $\lambda$ fixed by the scaling of the fluxes in $G_4^0$. Schematically, the procedure to find an AdS or dS minimum at parametric control goes as follows:
\begin{enumerate}
\item We select the fluxes in $G_4^0$ such that all of them exhibit the same asymptotic scaling. This imposes an extra constraint in table \ref{tab:2ModuliHatGvsGr}. In this case, there are six possibilities given by equations \eqref{p1}-\eqref{p6}.
\item We add any massless flux $\hat G_4$ which is unbounded with respect to the choice of $G_4^0$, following table \ref{tab:2ModuliHatGvsGr}.
\item We check whether the Stiemke's problem \eqref{inequalities} has a non-trivial solution.
\end{enumerate}

\begin{table}[H]
\centering
\begin{tabular}{|C|C|C|C|C|}\hline
\rule[-.27cm]{0cm}{.7cm}  {\rule[-.1cm]{0cm}{.34cm} G_4^0} & {\hat G_4}    & s^\alpha V_0  & \text{AdS vacuum}        & \text{dS vacuum}\\\hline\hline
\rule[-.25cm]{0cm}{.67cm} ({f_4},{h_0})                        & {f_6},{f_4} & \lambda^{-2q} & \textrm{No}              & \textrm{No}\\\hline
\rule[-.25cm]{0cm}{.67cm} ({f_2},{h_1},A_{\rm loc})            & {f_6},{f_4} & \lambda^{0}   & \textrm{Yes if }\alpha>0 & \textrm{No}\\\hline
\rule[-.25cm]{0cm}{.67cm} ({f_0},{h_2})                        & {f_6}         & \lambda^{2q}  & \textrm{No}              & \textrm{No}\\\hline
\rule[-.25cm]{0cm}{.67cm} ({f_2},{h_0})                        & {f_6},{f_4} & \lambda^{-q}  & \textrm{No}              & \textrm{No}\\\hline
\rule[-.25cm]{0cm}{.67cm} ({f_0},{h_1})                        & {f_6},{f_4} & \lambda^{q}   & \textrm{No}              & \textrm{No}\\\hline
\rule[-.25cm]{0cm}{.67cm} ({f_0},{h_0},A_{\rm loc})            & {f_6},{f_4} & \lambda^{0}   & \textrm{Yes if }\alpha>0 & \textrm{No}\\\hline
\end{tabular}
\caption{All possible fluxes that have the potential to provide a minimum at parametric control.} \label{resultsIIA}
\end{table}

The results are summarized in table \ref{resultsIIA}.  Interestingly, the addition of the unbounded massless fluxes allow us to find now AdS but not dS vacua. 
This is expected from previous results in the literature \cite{Hertzberg:2007wc}, since the scalar potential \eqref{mainex_potential} agrees with the 
one in \cite{Grimm:2004ua,Hertzberg:2007wc} when taking $\alpha=3$.  Hence, we recover the no-go theorems for de Sitter  \cite{Hertzberg:2007wc,Flauger:2008ad,Wrase:2010ew} in Type IIA flux compactifications at weak coupling and large volume based on the moduli scaling of the potential, including RR, NS and metric fluxes, O6-planes and even positive curvature. We also slightly generalize it by including geometric and non-geometric fluxes yielding the moduli dependence associated to $h_1$, $h_2$ and $h_3$. The requirement of keeping parametric control of the vacuum is what usually fails in previous classical de Sitter construction attempts, as also recently noticed in \cite{Junghans:2018gdb,Banlaki:2018ayh}.

Regarding the AdS vacua, let us recall that they only appear thanks to leaving $A_{\rm loc}$ free instead of completing a perfect square. Therefore, they are not exactly the mirror duals of the Type IIB potentials with $G_3$-flux, but there should be some additional contribution to $A_{\rm loc}$.
In such a case, we find that there are only two possible candidates for AdS vacua at parametric control as long as $\alpha>0$.  
These two possibilities indeed correspond to the pair of what would-be self-dual fluxes in IIB in the case that $A_{\rm loc}$ was not a free parameter, but they loose such an interpretation in IIA. 

Let us first consider the group of flux terms $({f_0},{h_0},A_{\rm loc})$ with unbounded ${f_4}$. This is precisely the combination of fluxes used in \cite{DeWolfe:2005uu} to get supersymmetric AdS vacua at parametrically large volume and small coupling in massive Type IIA. Taking $A_{f_4}\sim \lambda^2$ and $\alpha=3$, the moduli and the potential energy will scale asymptotically as
\beq \label{scaling_IIA}
  \quad s\sim \lambda^{3/2}\ , u\sim \lambda^{1/2}\ :\qquad  \frac{s}{u} \sim \lambda  \ , \quad V\sim  \frac{1}{\lambda^{9/2}}\ .
\eeq
This implies that if one makes $A_{f_4}$ large enough, $\lambda^{1/2} > \gamma \gtrsim 1$, the vacua indeed lie in the growth sector \eqref{eqn:2moduliGrowthSector}
and the nilpotent orbit approximation \eqref{Pinil} is valid.  
Actually, one finds that for this setting one can make $\gamma$ stepwise larger when increasing the flux $A_{f_4}$.
In this limit the 
strict asymptotic approximation using the sl(2)-norm \eqref{general-norm-growth} becomes more accurate, such that 
the existence of the considered vacua can indeed be trusted.
However, let us stress that there could be other reasons for which this vacuum cannot be lifted to a true top-down string theory construction. Here, we are only checking if the scaling of the moduli is adequate to generate a vacuum at parametric control.

The Type IIA setting has several interesting features that follow from the scaling behavior \eqref{scaling_IIA}.
Firstly, one sees that the Hubble scale $H=\sqrt{V}/M_p^2$ becomes parametrically small when 
sending $\lambda\rightarrow \infty$. Secondly, as stressed in~\cite{DeWolfe:2005uu} these Type IIA solutions also enjoy a separation 
of scales between the Hubble scale and the Kaluza-Klein scale. The KK scale in Type IIA Calabi-Yau compactifications is given by
\beq
M_{KK}=\frac{g_sM_p}{(\cV^{\rm A}_{s})^{2/3}}\sim \frac{M_p}{st^{1/2}}\sim \lambda^{-7/4}\ ,
\eeq
implying $H/M_{KK}\sim \lambda^{-1/2}\rightarrow 0$. This would go against the strong versions of the AdS conjectures put forward in \cite{Alday:2019qrf,Lust:2019zwm}. In principle, it also seems possible to get a similar result using the unbounded massless flux $f_6$ instead of $f_4$ but this possibility should, however, be discarded when we further impose axion stabilization as discussed in section \ref{sec:axionDependence}.

The other possible candidate for AdS vacuum arises from considering the group of terms $({f_2},{h_1},A_{\rm loc})$ with unbounded ${f_4}$ or ${f_6}$ (as in \cite{Camara:2005dc}). However, in this case $s$ and $u$ scale the same way, implying that $s/u$ cannot be made large and the strict asymptotic approximation fails. Hence, this vacuum cannot be trusted in our setup. Let us mention, though, for completeness, that the vacuum energy and the KK scale also scale the same way at the asymptotic limit in this example, $H\sim M_{KK}\sim \lambda^{-3/2}$ , implying that there would not be scale separation unlike in the previous example.

\subsection{No-go results for de Sitter at parametric control\label{sec:deSitter}}

The power of using the theory of limiting MHS, is that it allows us to go beyond the singularities corresponding to large volume and weak coupling and study the asymptotic vacua structure for any other type of limit in a systematic way. As explained, the type of limit will determine the moduli scaling of the flux potential by providing the values of $m,n$ in \eqref{V} that are allowed in each case. In this section, we will generalize our previous results to other types of singularities in the Calabi-Yau four-fold as long as they can be understood as the singular limit of only two moduli becoming large. These two moduli can correspond to any two complex structure moduli of the fourfold, so either bulk complex structure, dilaton or 7-brane moduli in Type IIB.
All possible singular limits of this type have been classified in table \ref{tab:asymptoticSplittings} and the potentials have been explicitly written in table \ref{potentials_list1}. 
We will take the same ansatz for the fluxes as in \eqref{ansatz1}, including some unbounded massless fluxes $\hat G_4$ in addition to fluxes with the same asymptotic scaling in $G^0_4$. This guarantees that the minima of the potential (if any) will occur at parametrically large field values of the moduli.
We find the following no-go theorem:

\noindent\mybox{lightgray}{\textbf{No-go statement:} There is no dS critical point at parametric control near any two large field limit of a Calabi-Yau fourfold in the strict asymptotic approximation if the scalar potential $V$ vanishes at the limit $s,u\rightarrow \infty$. }

Let us recall that this no-go is valid for any possible two large field limit of a Calabi-Yau fourfold. Hence, our results go beyond previous no-go theorems found at the large volume and weak string coupling limits of Type II CY compactifications \cite{Hertzberg:2007wc,Flauger:2008ad,Wrase:2010ew,Banlaki:2018ayh,Roupec:2018mbn,Junghans:2018gdb,Andriot:2019wrs}. Generically our settings, if they have a Type II interpretation at all, will yield situations in which one is not at weak string coupling. The systematics to argue for the validity of our statement, though, is very similar to the one taken for previous no-go theorems in which the moduli scaling of individual terms in the potential is exploited. 

Our no-go also goes beyond the famous Maldacena-Nu\~nez no-go theorem \cite{Maldacena:2000mw} in the context of four-dimensional $\cN = 1$ compactifications, which is based on solving the equations of motion of the internal geometry when there are only $p$-form gauge fluxes. Our starting scalar potential in M-theory includes higher derivative terms, as we also include the term depending on the Euler characteristic of the Calabi-Yau fourfold \eqref{tadpole}.\footnote{See \cite{Grimm:2014xva,Grimm:2014efa,Grimm:2015mua,Grimm:2017pid} for a 
complete treatment of M-theory higher-derivative terms relevant at this order.} Under M/F-duality and the application of mirror symmetry, 
terms are well-known to map to effects arising, for example, from O6-planes. Furthermore, a certain $G_4$-flux component maps under this 
duality chain to the Romans mass. In addition we have further generalized our discussion by allowing for independent potential terms. This 
can prevent cancellations and correlations between the different terms assumed in the analysis of \cite{Maldacena:2000mw}. 

To avoid confusion, let us clearly list the assumptions that enter in the derivation of the above de Sitter no-go theorem. We require:
\begin{itemize}
\item Only two fields, denoted as $s$ and $u$, become large although the moduli space can be higher dimensional.
\item Parametric control: the vacuum should survive in the asymptotic limit as explained around \eqref{control}, i.e. for parametrically large field values of $s$ and $u$.
\item Strict asymptotic approximation: we only keep the leading asymptotic growth of each term of the potential, as explained below \eqref{general-norm-growth}.
\item The potential should vanish asymptotically in the limit $s,u\rightarrow \infty$. 
\end{itemize}

The first three assumptions will be relaxed in future work. As for the last one, it should be understood more as a consistency constraint to keep control of the compactification. Only self-dual fluxes satisfy the equations of motion of the Calabi-Yau, but have vanishing potential. To keep the analysis as general as possible, we have allowed for any type of flux that could ever be present,  which implies that we are also allowing for some breaking of the self-duality condition. However, we impose that the potential should still vanish asymptotically so this breaking is mild and can be understood as a perturbation over the warped Calabi-Yau geometry. Otherwise, it seems to us that the potential should not be trusted if it diverges at the large field limit. Let us recall that this is a very mild assumption and most likely not enough to guarantee consistency of the scalar potentials we study. However, since we already get a no-go theorem for de Sitter, there is no need of reducing even further the list of examples by imposing further constraints like satisfying the equations of motion of the internal geometry, which is obviously a much harder task. 

We can conclude that, for the moment, our findings are compatible with a generalized Dine-Seiberg problem \cite{Dine:1985he}, conjectured in \cite{Obied:2018sgi,Ooguri:2018wrx}, valid for any asymptotic limit of a string compactification, forbidding the presence of de Sitter vacua at parametric control.

Finally, we can also check the (asymptotic) de Sitter conjecture \cite{Obied:2018sgi} in our setting. This conjecture, not only implies the absence of dS vacua, but goes beyond it by providing a bound on the slope of the potential that also disfavors slow roll inflation. More precisely, it was conjectured in \cite{Obied:2018sgi,Ooguri:2018wrx} that there is an 
order one constant $\gamma$ such that  
\beq \label{deS-conj}
  |\nabla V| \equiv | (\partial_{z^K} V) G^{K \bar L} (\partial_{\bar z^L} V) |^{1/2} \geq \gamma V\ .
\eeq

In the remaining of this subsection we ignore the axion dependence, and write all partial derivatives with respective to saxions $\partial_i = \frac{\partial}{\partial s^i}$.
Let us assume that we can establish the following bound
\beq \label{bound1}
   f^{-2} \geq (\kappa^i_k s^k) G_{ij} (\kappa^j_l s^l)\ ,
\eeq 
where $\kappa^i_k$ is some constant matrix which we will determine below. Then we can use Cauchy-Schwarz to show the following 
estimate 
\beq
   (\partial_i V G^{ij} \partial_j V)^{1/2}   \geq f  (\partial_i V G^{ij} \partial_j V )^{1/2}   ((\kappa^i_k s^k) G_{ij} (\kappa^j_k s^k))^{1/2} 
    \geq f  \partial_j V   (\kappa^j_k s^k) 
\eeq
Therefore, if the following inequality holds
\beq
\label{bound2}
f  \partial_j V   (\kappa^j_k s^k) \geq  \gamma V
\eeq
then also the conjectured de Sitter bound is satisfied.

Let us next recall the Stiemke's problem we are using to prove for the absence of dS vacua,
\beqa
 \label{inequality}
a(m_i - 4 - \alpha) + b(n_i - m_i) + c & \geq & 0\ ,\\
\alpha a - c                           & \geq & 0\ .
\eeqa
If this system of inequalities has a solution with $c\leq 0$ ($c\geq 0$), then the potential does not have a dS (AdS) extremum. Notice that the first inequality also implies that
\beq
\sum_{i=1}^\cN (a(m_i - 4 - \alpha) + b(n_i - m_i) + c)V_i - (c - \alpha a)V_{\cN+1}\geq 0
\eeq
as $V_i,V_{\cN+1}>0$. This further implies
\beq
as\partial_s V+bu\partial_u V\geq -cV
\label{dV}
\eeq
which corresponds to \eqref{bound2} with
\beq
\label{abc}
    \kappa^s_s  = a \ , \qquad \kappa_u^u = b\ , \qquad c=-\gamma/f
\eeq
and all others vanishing. Hence, as long as  \eqref{bound1} is satisfied, we can show  that the conjecture \eqref{deS-conj} holds by using the Stiemke's inequality  \eqref{inequality} again.

Let us then check \eqref{bound1}. The leading behavior of the metric can be computed from the asymptotic form of the K\"ahler potential in \eqref{Ksl2}, obtaining
\beq
g_{t_1\bar t_1}=\frac{d_1}{s^2}\ ,\quad g_{t_2\bar t_2}=\frac{d_2-d_1}{u^2}\ 
\eeq
where $d_1,d_2$ are integers characterizing the singularity type as discussed after \eqref{Ksl2}. Therefore, by only using this leading term of the metric, it is trivial to check that the bound \eqref{bound1} gets saturated for
\beq
f^{-2}= a d_1 + b (d_2-d_1)
\eeq
The next to leading order terms for the metric will be further suppressed in the 
asymptotic regime.
Combining this with \eqref{abc} we get that the parameter in the de Sitter conjecture is given by
\beq
\gamma^2=|c|^2/(a d_1 + b (d_2-d_1))
\eeq
where $a,b,c$ are constrained to satisfy \eqref{dV}. We have already checked that it is always possible to find some values of $a,b,c$ such that the Stiemke's inequalities \eqref{inequality}, and thus \eqref{dV}, are satisfied for all two-moduli limits of the Calabi-Yau fourfold. The remaining question is whether this solution implies $\gamma\sim \mathcal{O}(1)$. For this reason we  check if the system \eqref{inequality} has a solution with  $\gamma>1$, which is a stronger condition.
Interestingly, we find that there is always such a solution, implying that the bound \eqref{bound1} is always satisfied for any limit as long as $d_1\neq0$ and/or $d_2-d_1\neq 0$ and $\alpha=0$ in \eqref{V}. If $d_1=d_2=0$, then the enhanced singularity is of Type I, meaning that it is at finite distance, while $\alpha=0$ selects the potentials coming from Type IIB/F-theory flux compactifications. If $\alpha\neq 0$ the bound \eqref{bound2} is only satisfied if both $d_1\neq 0$ and $d_2-d_1\neq 0$. However, this bound is a stronger condition than \eqref{deS-conj} so it does not imply that de Sitter conjecture is not satisfied but only that we cannot determine its fate by considering only the leading term of the field metric.

To sum up, we find that the de Sitter bound \eqref{deS-conj} is satisfied for any asymptotic limit in F-theory flux compactifications which is at infinite distance in the complex structure moduli space of a Calabi-Yau fourfold. This nicely matches with the argument in  \cite{Ooguri:2018wrx} that relates the de Sitter Conjecture and the Distance Conjecture, as the latter only concerns infinite distance regimes. For finite distance singularities, the bound  \eqref{bound1} is not necessarily satisfied to leading order so the analysis becomes more difficult and we leave it for future work. 

\subsection{Candidates for AdS minima at parametric control\label{sec:AdS}}

Let us analyze the conditions to get AdS vacua at parametric control. First of all, let us stress again that an AdS vacuum is not possible in Type IIB/F-theory Calabi-Yau compactifications as the potential is definite positive. Even if there is a negative contribution coming from localized sources, it always completes a perfect square when imposing tadpole cancellation. However, an AdS vacuum might appear when dualizing the setup to Type IIA and assuming additional contributions to the tadpole cancellation conditions that do not necessarily impose anymore the completion of the perfect square. These additional sources can correspond for example to other fluxes that do not simply map to $G_3$ fluxes in Type IIB. They will modify the value of the coefficients $A_{mn}$ and usually make $A_{\rm loc}$ to also depend on the complex structure moduli, but the moduli dependence of each term on the IIA dilaton $s$ and the IIA K\"ahler modulus $u$ is expected to be the same. Since we are not specifying the value of the coefficients $A_{mn}$ here, this possibility is automatically incorporated in our analysis. Furthermore, there is an additional overall dilaton factor appearing in the dualization process that makes the negative contribution $V_{\rm loc}$ to become moduli dependent. This moduli dependence of the negative contribution is essential to get AdS vacua, as we will see. 

The main observation of this section is that, in order to get a vacuum at parametric control, it is necessary to have an unbounded massless flux $\hat G_4$ satisfying the properties in \eqref{no-tadpole} and \eqref{massless}. Otherwise, in the absence of this flux, it is easy to check that the inequalities \eqref{inequalities} always admit a solution with $c\leq 0$ implying the absence of AdS vacua at parametric control. In this section, we identify the unbounded massless fluxes $\hat G_4$ that arise at the different limits of table \ref{tab:asymptoticSplittings} for a given choice of $G_4^0$. Recall that the background fluxes in $G_4^0$ are chosen to have the same asymptotic scaling in order to yield minima at parametric control. From all possible combinations of fluxes, there are only seventeen yielding a candidate for AdS vacua at parametric control as long as $\alpha>0$, listed in table \ref{tab:AdSresults}. Notice that we are only checking for extrema of the potential, so they could correspond to either minima or maxima. However, even if $s$ and $u$ can be made parametrically large, we need to also check that $s/u$ is large so that we remain in a growth sector \eqref{eqn:2moduliGrowthSector} and the strict asymptotic approximation is valid.

For this purpose, we need to provide the asymptotic scaling of the moduli at the large field limit. This scaling of the moduli, as well as the scaling of the vacuum energy, can be determined even without providing the explicit solution for the scalars at the minimum, as we explain in the following. Let us denote  $\hat A_{\hat m\hat n}$ as the flux coefficient associated to $\hat G_4$ and $A^0_{m_in_i}$ the ones corresponding to $G^0_4$. The potential reads
\beq
V= \frac{1}{s^\alpha}\left(\frac{\hat A_{\hat m\hat n}}{s^{4 - \hat m}u^{\hat m - \hat n}}-A_{\rm loc}\right)+ \frac{1}{s^\alpha}\frac{A^0_{m_in_i}}{s^{4 - m_i}u^{ m_i -  n_i}}
\eeq
where all terms must scale the same way asymptotically in order to survive at the large field limit and yield a minimum at parametric control. 
Taking into account that we can scale up the flux  $A_{\hat f}\sim \lambda^2$ and denoting the scaling of the moduli as
\beq
\label{scaling}
s\sim \lambda^p\ ,\quad u\sim \lambda^q
\eeq
with $p,q>0$, we get that the following equalities should hold true,
\beqa
2 + p(\hat m - 4) + q(\hat n - \hat m) = 0\label{2pq}\\
(m_i - 4)p + (n_i - m_i)q = 0 \ .
\label{mqnp}
\eeqa
This guarantees that all terms scale the same way with $\lambda$  in the limit $\lambda\rightarrow \infty$. Furthermore, if an AdS solution exists, the vacuum energy will necessarily scale as
\beq
|V_0|\sim \lambda^{-\alpha p}\ .
\eeq

Notice that $p,q$ are uniquely determined due to \eqref{2pq} and \eqref{mqnp}, so they can be determined case by case. 
In table \ref{tab:AdSresults} we have included two columns with the asymptotic scaling of $s$ and $u$ in each case. Remarkably, only one case allows for $s/u$ large, meaning that the other solutions cannot actually be trusted as they go away from the strict asymptotic regime. Interestingly, this single solution with $s/u$ large corresponds to the familiar case with $f_0,h_0$ fluxes and unbounded $f_4$ or $f_6$ in the enhancement $\rII_{0, 1} \to \rV_{2, 2}$. This is the example already found in Type IIA flux compactifications in \cite{DeWolfe:2005uu,Camara:2005dc}, and was discussed in great detail in section \ref{sec:example}. It is quite remarkable that there are not other AdS vacua at parametric control appearing at any of the other limits of the Calabi-Yau fourfold.

It has been recently conjectured that AdS vacua with scale separation are in the swampland \cite{Gautason:2015tig,Lust:2019zwm}. This means that there should be an infinite tower of states with mass of the same order than the vacuum energy. If this tower corresponds to a KK tower, it further implies that there is no scale separation between the external and internal dimensions. In Type IIA Calabi-Yau compactifications at weak coupling, the KK scale is given by
\beq
\label{MKK}
M_{KK}=\frac{g_sM_p}{\nu^{2/3}}\sim \frac{M_p}{su^{1/2}}\sim \lambda^{-p-q/2}
\eeq
where we have replaced the asymptotic scaling of the moduli \eqref{scaling} in the last step.
This would imply the following ratio with respect to the vacuum energy,
\beq
\label{sep}
\frac{H}{M_{KK}}\sim \lambda^{-\alpha p/2+p+q/2}
\eeq
where we have defined $H\equiv \sqrt{|V_0|}/M_p$.
A scale separation would then be possible if $p>q/(\alpha-2)$.  Unfortunately, we cannot determine $\alpha$ in general. We only know that $\alpha=3$ at the large volume and weak coupling point, which corresponds to the enhanced singularity of our main example in section \ref{sec:example}. In that case, scale separation occurs since $p>q$, i.e. the dilaton $s$ grows faster than the volume $u$.  Hence, for $\alpha=3$ the condition of being in the strict asymptotic regime is correlated to exhibit some scale separation.

In table \ref{tab:AdSresults} we have included a column specifying the value of \eqref{sep} at each of the limits yielding AdS vacua. Interestingly, none  of them would exhibit scale separation except for the typical example of weak coupling and large volume of Type IIA mentioned above and discussed more carefully around eq.\eqref{scaling_IIA}. However, it is important to remark that the use of the KK scale \eqref{MKK} beyond the weak coupling limit is questionable and the results of this last column should not be taken very seriously. 
An alternative way to define a cut-off scale valid at any infinite distance singularity, regardless whether it occurs at weak coupling or large volume,  could be by means of the Swampland Distance Conjecture. At each infinite distance singularity, there will be an infinite tower of states becoming exponentially light, and the cut-off of the effective theory is given at most by the species scale of this tower. This tower has been identified in a systematic way for every infinite distance singular limit of Calabi-Yau threefolds in \cite{GPV,Grimm:2018cpv,Corvilain:2018lgw} and we leave the analogous analysis for fourfolds for future work. It would be interesting to check if any of the examples in table \ref{tab:AdSresults} could enjoy a scale separation between the vacuum energy and this SDC cut-off.

In fact, there seems to be an even deeper relation between these AdS vacua and the Distance Conjecture. We have seen that an unbounded massless flux is required to get candidates for AdS vacua at parametric control. The presence of this type of fluxes has the same mathematical origin than the presence of an infinite massless tower of stable charged states at the large field limit. The `masslessness' condition for which the Hodge norm $||\hat G_4||^2$ should asymptotically vanish is equivalent to the condition required in \cite{GPV} for a charged BPS state to become massless at the singular limit in a Calabi-Yau threefold. Furthermore, the condition to be `unbounded' resembles the condition of stability for the BPS state \cite{GPV}. The infiniteness of the tower would correspond, though, to whether the flux coefficient $\hat A_{\hat m\hat n}=||\rho_{\hat m\hat n}(\hat G_4,\phi)||_\infty$ depends on the axionic fields.

\begin{table}[H]
\centering
\begin{tabular}{|C|C|C|C|C|C|}\hline
  G_4^0                                  & \hat{G}_4 & s             & u             & H/M_{KK}\\\hline
  (3,4),(4,4),(5,4)                      & (3,2)     & \lambda^1     & \lambda^1     & \lambda^0\\
  (3,4),(4,4),(5,4)                      & (4,3)     & \lambda^2     & \lambda^2     & \lambda^{0}\\
  (3,4),(4,4),(5,4)                      & (4,2)     & \lambda^1     & \lambda^1     & \lambda^0\\
  (3,4),(4,4),(5,4)                      & (4,0)     & \lambda^{1/2} & \lambda^{1/2} & \lambda^0\\
  (3,4),(4,4),(5,4)                      & (3,3)     & \lambda^2     & \lambda^2     & \lambda^0\\
  (3,4),(4,4),(5,4)                      & (2,2)     & \lambda^1     & \lambda^1     & \lambda^0\\ 
  (3,4),(4,4),(5,4)                      & (3,0)     & \lambda^{1/2} & \lambda^{1/2} & \lambda^0\\
  \rowcolor{lightgray} (3,6),(4,4),(5,2) & (3,0)     & \lambda^{1}   & \lambda^{1/3} & \lambda^{-1/3}\\
  \rowcolor{lightgray} (3,6),(4,4),(5,2) & (3,2)     & \lambda^{3/2} & \lambda^{1/2} & \lambda^{-1/2}\\
  (2,4),(4,4),(6,4)                      & (2,0)     & \lambda^{1/2} & \lambda^{1/2} & \lambda^0\\
  (2,4),(4,4),(6,4)                      & (4,2)     & \lambda^1     & \lambda^1     & \lambda^0\\
  (2,4),(4,4),(6,4)                      & (2,2)     & \lambda^1     & \lambda^1     & \lambda^0\\
  (1,2),(4,4),(7,6)                      & (1,0)     & \lambda^{1/3} & \lambda^{1}   & \lambda^{1/3}\\
  (1,2),(4,4),(7,6)                      & (3,2)     & \lambda^{1/2} & \lambda^{3/2} & \lambda^{1/2}\\
  (3,4),(4,4),(5,4)                      & (1,0)     & \lambda^{1/2} & \lambda^{1/2} & \lambda^0\\
  (3,4),(4,4),(5,4)                      & (1,2)     & \lambda^1     & \lambda^1     & \lambda^0\\
  (3,4),(4,4),(5,4)                      & (0,0)     & \lambda^{1/2} & \lambda^{1/2} & \lambda^0\\\hline
\end{tabular}
\caption{All possible combinations of flux terms yielding an AdS extremum (assuming $\alpha>0$). In the last column we have replaced $\alpha=3$ to relate to Type IIA perturbative string theory. The notation has been chosen according to table \ref{tab:asymptoticSplittings} in which we provide the integers $\ell=(m,n)\in \cE$ associated to each flux term. Only the shaded examples present $s/u$ large, consistent with the strict asymptotic approximation.}\label{tab:AdSresults} 
\end{table}

\section{Asymptotic structure of flux vacua: axion dependence} \label{sec:axionDependence}

In the previous section we have discussed the stabilization of the fields $s^i$, corresponding to the imaginary part of 
$t^i = \phi^i + i s^i$, by studying the potential \eqref{V}.  The goal of this section is to also include the dependence on 
the axions $\phi^i$. Firstly, we will discuss the constraints that arise upon imposing stabilization via fluxes for the candidate 
AdS minima discussed in section \ref{sec:AdS}. Secondly, we will derive some universal backreaction effects that appear when displacing the axions at large field values and discuss their implications for axion monodromy inflationary  models.

\subsection{Axion stabilization} \label{sec:axion}

So far we have studied the minimization of the potential with respect to the saxions $s^i$ and 
ensured that the vacua are at large values of $s^i$. The axions do not need to be stabilized at large field values 
to have a minimum at parametric control. Hence, even if we have an axionic flat direction, this could be stabilized 
by higher order or non-perturbative corrections to the scalar potential. This implies that, in order to derive no-go 
theorems for de Sitter vacua at parametric control, it is sufficient to study stabilization of the saxions. 
Clearly, if we aim to find a fully-fledged minimum, it is crucial to study axion stabilization as well. For this reason 
it is interesting to study the fate of the AdS extrema found in section \ref{sec:AdS} upon studying axion stabilization. It turns out that 
minimization of the potential with respect to the axions imposes additional constraints on  
the values of the limiting flux norm $A_\Bell=\|\rho_\Bell(G_4,\phi)\|^2_\infty$ that can 
invalidate some of the AdS extrema found previously.

Let us repeat for convenience the asymptotic form of the scalar potential in the strict asymptotic approximation. In the limit of two (saxionic) fields becoming large with $\frac{s}{u} > \gamma ,u > \gamma$, the potential reads 
\beq \label{V-recall}
V=\frac{1}{s^\alpha} \Big( \sum_{(m,n)\in \cE}   \underbrace{ \|\rho_{mn}(G_4,\phi,\psi)\|_\infty^2}_{A_{mn}} \, s^{m-4} u^{n-m}- V_{\rm loc} \Big)
\eeq
where  $\phi \equiv \phi^1$ and $\psi \equiv \phi^2$ are the axionic partners of $s$ and $u$ respectively, as in \eqref{2fieldlimit}. The $\rho_{mn}$ arise as in \eqref{xNG_2} from the split into the vector spaces $V_{mn}$. In this section we will make the replacement $N_i \rightarrow N_i^-$ in \eqref{xNG_2}, as this will simplify our 
discussion significantly. The operator $N_i^-$ was introduced in \eqref{triples} as part of the commuting $\slt$-triples  in section \ref{flux-split}. We note that using $N_i$ would induce new mixed terms that are, however, suppressed
in the strict asymptotic regime. Moreover, we expect that the conclusions of section \ref{sec_backreaction} are not 
altered under the exchange $N_i \leftrightarrow N_i^-$. Therefore, we will now consider 
\beq
\label{rho3_N-}
\rho^-(G_4,\phi)= e^{-\phi^i N^-_i}G_4=\sum_{(m,n)\in \cE} \rho^-_{mn}\, .
\eeq
In order to proceed it will be convenient to use an explicit basis of $V_{mn}$ denoted by $v^{mn}_{j_{mn}}$
as in section \ref{self-dual_asy}. 
We will show in the following how such a basis can be constructed by starting with some highest weight states, 
and applying the successive action of the lowering operators $N^-_i$.

Firstly, we recall that given an $\slt$-algebra with generators $\{N^-, Y, N^+\}$ as  in \eqref{triples}, every (finite dimensional) irreducible representation is isomorphic to a vector space generated by a highest weight vector $\hat a^{p+4}$, defined by demanding that $(N^-)^p \hat a^{p+4} \ne 0$ while $(N^-)^{p + 1} \hat a^{p+4} = 0$, and 
its images under $N^j$. In other words the irreducible representation can be written as
\beq
 \spanC{\hat a^{l+4}, N^- \hat a^{l+4}, \ldots, (N^-)^l \hat a^{l+4}}\, .
\eeq
A general representation of this $\slt$-algebra is then given by a direct sum of irreducible representations. 
Therefore it suffices to specify a set of highest weight vectors to fix a representation of the $\slt$-algebra.

In the case of two-moduli case introduced in section \ref{flux-split}, we have two copies of commuting $\slt$-algebras acting on 
$H^4_{\rm p}(Y_4, \bbR)$, turning it into a representation of two $\slt$-algebras. In order to specify a basis for $H^4_{\rm p}(Y_4, \bbR)$, we introduce the highest weight vectors $\hat a^{p+4,q+p+4}_{\kappa} \in V_{p+4,q+p+4}$ with $p,q\geq 0$. These states are characterized by the highest powers $p,q$ of $N_1^-$ and $N_2^-$ that are not annihilating $\hat a_\kappa \equiv \hat a^{p+4,q+p+4}_{\kappa}$ as
\begin{align} \label{eqn:highestWeightForm}
  (N_1^-)^{p} \hat a_{\kappa} \ne 0\, ,& \qquad  (N_1^-)^{p+1} \hat a_{\kappa}= 0\, ,\\
  (N_2^-)^{q} \hat a_{\kappa} \ne 0\, ,& \qquad (N_2^-)^{q+1} \hat a_{\kappa}= 0\, .\nonumber
\end{align}
The index $\kappa$ labels how many such highest weight states exist for the considered splitting. For example, there 
could be multiple $\hat a_\kappa$ in one $V_{p+4,q+p+4}$. In mathematical terms 
these highest weight states capture the information about the primitive part of $V_{p+4,q+p+4}$. 
Let us next discuss how the highest weight vectors span the spaces $V_{m,n}$.  Each vector spaces $V_{mn}$ is 
defined to be the simultaneous eigenspace of $Y_1$ and $Y_1 + Y_2$. 
Using the $\slt$-algebra we can generate a special basis $v^{mn}_{j_{mn}}$ of $V_{m,n}$ by acting with $N_1^-$ and $N^-_2$ on 
 all highest weight vectors as 
\begin{equation} \label{eqn:basisNa}
  \left\{ v^{m,n}_{j_{mn}} \right\}_{j_{mn} = 1}^{\dim V_{m,n}} = \left\{ (N_1^-)^a (N_2^-)^b \ \hat a^{m+2a,n+2a+2b}_{\kappa}
   \right\}\, ,
\end{equation}
where we have to use all highest weight states and therefore also collect the possible choices for the index $\kappa$. 
Before using this basis in studying the axions, it is worthwhile to add two observations. Firstly, in a Calabi-Yau fourfold there is 
always a highest weight vector $a_0$ which belongs to $V_{4 + d_1, 4 + d_2}$, where $d_1,d_2$ are integers characterizing the singularity type as discussed after \eqref{Ksl2}.
Secondly, we note that the basis given by \eqref{eqn:basisNa} is \emph{not} yet compatible with our normalization \eqref{v-orth<>}, 
and we would reverse some signs for some of the basis vectors to ensure compatibility. It turns out the that 
normalization will not be relevant in this section and it suffices to use the basis \eqref{eqn:basisNa}.

Let us now return to our discussions of the axion-couplings appearing in \eqref{V-recall}.
We first expand the $\rho_{mn}$ into the basis \eqref{eqn:basisNa} writing 
\beq
    \rho_{mn}^- =   \sum_{j_{mn}} \varrho^{j_{mn}}_{m,n}(\phi,\psi)\, v^{m,n}_{j_{mn}}\ ,  \qquad {\small \text{no sum over $(m,n)$}}\ ,
\eeq
where $\varrho_{mn}(\phi,\psi)$ are the axion-dependent coefficient functions.
We now show by using 
\eqref{eqn:basisNa} that 
\beq
  \partial_{\phi} \varrho_{m,n}^{j_{mn}}=-\varrho_{m+2,n+2}^{j_{mn}}\ ,\qquad \partial_\psi \varrho_{m,n}^{j_{mn}} =-\varrho_{m,n+2}^{j_{mn}}\ ,
\eeq
which holds due to the fact that the axions $\phi, \psi$ appear through an 
exponential factor $e^{-\phi^i N_i}$ in \eqref{rho3_N-}.
It is now straightforward to minimize the scalar potential \eqref{V-recall} with respect to the 
axions $(\phi,\psi)$. We first rewrite it in terms of the $\varrho_{m,n}^{j_{mn}}$ 
as in \eqref{V4forms}. The minimization conditions then read
\beqa
\label{min1}
\partial_\phi V=-\frac{2}{s^{\alpha} }\sum_{(m,n)\in \cE}\sum_{\substack{i_{mn}\\j_{mn}}}  Z^{mn}_{i_{mn}, j_{mn}}
 \, \varrho_{m,n}^{i_{mn}}\, \varrho_{m+2,n+2}^{j_{mn}}=0\ ,\\
\partial_\psi V=-\frac{2}{s^{\alpha} } \sum_{(m,n)\in \cE}\sum_{\substack{i_{mn}\\j_{mn}}} 
Z^{mn}_{i_{mn}, j_{mn}}\, \varrho_{m,n}^{i_{mn}}\, \varrho_{m,n+2}^{j_{mn}}=0\ ,
\label{min2}
\eeqa
with the asymptotic from of $Z^{mn}_{i_{mn}, j_{mn}}$ given in \eqref{Zsl2}.
From these conditions \eqref{min1} and \eqref{min2}, it is eminent 
that the stabilization of axions by fluxes imposes additional relations between the different $\varrho_{m,n}^{j_{mn}}$-functions and, therefore, in the coefficients $A_{mn}$. For instance, if an axion appears only through one function $\varrho_{m',n'}^{j_{m'n'}}$, the above minimization conditions imply that this 
$\varrho_{m',n'}^{j_{m'n'}}$ has to vanish at the minimum. This determines the vacuum expectation value of the axion in terms of the internal fluxes, but also implies that all terms proportional $\varrho_{m',n'}^{j_{m'n'}}$ are absent when studying the stabilization with respect to the saxions. 
Therefore, extrema of the potential that arise from self-dual fluxes found in section \ref{sec:AdS} might disappear when imposing these further constraints as some flux terms might not be present anymore.

For concreteness, let us illustrate the implications of these constraints in our main example of section \ref{sec:example}. Using \eqref{rho3_N-} and \eqref{eqn:2moduliG4splitting} we get
\beqa
\varrho_{30} & = & f_6 - f_4 \psi + \frac12 f_2 \psi^2 - \frac16 f_0 \psi^3 - h_0 \phi + h_1 \phi \psi - \frac12 h_2 \phi \psi^2 + \frac16 h_3 \phi \psi^3 \, ,\quad \varrho_{58} =h_3\, ,\\
\varrho_{32} & = & f_4 - f_2 \psi + \frac12 f_0 \psi^2 - h_1 \phi + h_2 \phi \psi - \frac12 h_3 \phi \psi^2 \, ,\quad \varrho_{56} = h_2 - h_3 \psi\, ,\\
\varrho_{34} & = & f_2 - f_0 \psi - h_2 \phi + h_3 \phi \psi \, ,\quad \varrho_{54} = h_1 - h_2 \psi + \frac12 h_3 \psi^2\, ,\\
\varrho_{36} & = & f_0 - h_3\phi \, ,\quad \varrho_{52} = h_0 - h_1\psi + \frac12 h_2 \psi^2 - \frac16 h_3 \psi^3\, ,
\eeqa
which matches with the $\varrho$-functions coupled to the three-form gauge fields obtained from dimensionally reducing Type II compactification in \cite{4forms}. 
In section \ref{sec:example} we found only two possible candidates for AdS vacua at parametric control, shown in table \ref{resultsIIA}. It can be checked that if we want to keep the unbounded flux term $V_{f_6}$ in the last row of the table, then we also need to turn on some $h_1,h_2$ or $h_3$ flux. Otherwise, \eqref{min1} implies that $\varrho_{30}=0$ at the minimum. For $V_{f_4}$, there are no new restrictions appearing. 
The analysis for the other types of asymptotic limits should be performed analogously. However, as explained in section \ref{sec:AdS}, this example was the only one leading to an AdS vacua at parametric control consistent with the growth sector, so we conclude the analysis here.

\subsection{Backreaction in axion monodromy inflation} \label{sec_backreaction}

It is also interesting to study the implications of the form \eqref{V-recall} of the scalar potential for axion monodromy inflation \cite{Silverstein:2008sg,McAllister:2008hb}.
In such models one axion is displaced far from its minimum and then rolls down to its true vacuum. 
In order that such a model can be implemented, one would like to slowly roll down the scalar potential along an almost purely axionic direction to keep control of the potential over large field excursions. However, backreaction effects can be very important and must be properly taken into account \cite{McAllister:2014mpa}. In the context of F-term axion monodromy models \cite{Marchesano:2014mla,Hebecker:2014eua,Blumenhagen:2014gta,Ibanez:2014kia} in Calabi-Yau compactifications, this constitutes a real challenge \cite{Hebecker:2014kva,Blumenhagen:2014nba} as the problem is linked to the difficulties of achieving significant mass hierarchies. In particular, as pointed out in  \cite{Baume:2016psm} and further analysed in \cite{Valenzuela:2016yny,Blumenhagen:2017cxt}, a large displacement of an axion $\phi$ can severely modify the saxion vevs which backreact on the kinetic axionic term and
substantially reduce the field range. In those papers, it was found by analyzing various examples that, in typical F-term axion monodromy models in Calabi-Yau compactifications, the saxion vev behaves at large field as
\beq
\label{backreaction}
\langle s\rangle \sim \lambda \phi
\eeq
implying the following backreacted kinetic term for the axion
\beq
\mathcal{L}\supset \frac{1}{s^2}(\partial \phi)^2 \sim \frac{1}{\lambda^2 \phi^2}(\partial \phi)^2
\eeq
and only a logarithmic growth of the proper field distance $\Delta\phi\sim \frac{1}{\lambda}\log \phi$. Furthermore, as predicted by the Swampland Distance Conjecture, large field distances are accompanied by an exponential drop-off of the quantum gravity cut-off due to an infinite tower of states becoming massless as $s\rightarrow \infty$. Due to \eqref{backreaction}, a large displacement of $\phi$ implies necessarily a large displacement on the saxion $s$, so the quantum gravity cut-off behaves as
\beq
\label{cutoff}
\Lambda_{QG}\sim \exp{(-\lambda \Delta \phi)}
\eeq
spoiling inflation at distances $\Delta\phi>\lambda$. It was argued \cite{Baume:2016psm} that $\lambda$ is an order one parameter in Planck units if the axion corresponds to the closed string sector of Type II compactifications. More generally, $\lambda$ might be related to the mass hierarchy between the axion and the saxion \cite{Valenzuela:2016yny}, allowing for some room to get large field ranges, although this mass hierarchy seems very difficult to get in fully-fledged global string compactifications and is usually incompatible with keeping the moduli masses below the Kaluza-Klein scale \cite{Blumenhagen:2017cxt}. It is still an open question whether this mass hierarchy can truly be obtained in a well controlled string compactification.

Although promising, this analysis of the backreaction in axion monodromy is very model dependent and is missing some general understanding of  the underlying reason for which the minimization of the potential should always imply \eqref{backreaction} at large field. Interestingly, we can now revisit this issue by taking advantage of the universal tools that the mathematical machinery of asymptotic Hodge theory provides. This will allow us to prove \eqref{backreaction} for most of the two-parameter large field limits arising in the Calabi-Yau compactification studied in the previous sections and, more importantly, provide the underlying geometric reason for such a linear backreaction at large field. 

Let us first state the observation that aim to show in the following. 
We consider two-parameter field limits with saxion-axion pairs $(s, \phi)$ and $(u, \psi)$. Our main focus will be on the $(u, \psi)$-pair, since the arguments are 
essentially identical for the $(s, \phi)$-pair. We first extract the leading potential $V^{(\psi)}$, obtained by keeping 
the term in each $A_{mn}$ in \eqref{V-recall} that is dominant for large $\psi$.
Below we will identify the two-parameter limits in which $V^{(\psi)}$ enjoys the following homogeneity property
\begin{equation} \label{eqn:homogeneousV}
  V^{(\psi)}\big(s, \zeta u; \phi, \zeta \psi\big) = \zeta^h \ V^{(\psi)}\big(s,u; \phi, \psi \big)\, ,
\end{equation}
for some homogeneous degree $h$. Let us now assume that $V^{(\psi)}$ has a extremum $\langle u \rangle >0$, i.e.~one demands that 
\begin{equation}
  0 = \partial_{u} V^{(\psi)} \big|_{u =  \langle u \rangle}\, .
  \label{dV_1}
\end{equation}
Then, assuming that the scalar potential $V^{(\psi)}$ is a polynomial in $u, 1/u$, and $\psi$, we find that  $\langle u \rangle $ satisfies the linear-backreaction relation
\begin{equation}
\label{linear}
  \langle u \rangle \sim \lambda \psi\, ,
\end{equation}
as in \eqref{backreaction}. Therefore, our target is to check the homogeneity  property \eqref{eqn:homogeneousV} at leading order in $\psi$ for all possible two-parameter enhancements.

We note that the intuition for the property \eqref{eqn:homogeneousV} to hold is rather simple. Notice first 
that the axion $\psi$ is always accompanied with a power of $N_2^-$, since it only appears via $\rho^-(G_4, \phi) = e^{-\phi N_1^- - \psi N_2^-} G_4$, see \eqref{rho3_N-}. 
Now one can use the fact that $N_2^-(V_{m,n}) \subset V_{m, n - 2}$, which is a simple consequence of the $\slt$-algebra, that the image of any basis vector $v^{n,m}_{j_{mn}}$ under $N_2^-$ will be proportional to $v^{m,n - 2}_{j_{m\, n-2}}$. 
We then deduce that while a flux along the basis vector $v^{mn}_{j_{mn}}$ yields a term proportional to $s^{m - 4} u^{n - m}$ in the scalar potential, the vector $N_2 v^{m,n}_{j_{mn}}$ will 
induce a term proportional to $s^{m - 4} u^{n - m - 2}$. In other words, the action of $N_2^-$ reduces the power of $u$ by $2$ in the scalar potential.  Since one $N_2^-$ is accompanied by a $\psi$, in the scalar potential term will be proportional to $\psi^2$, which precisely compensates the reduced $u$-power. This suggests that the scalar potential indeed can admit the homogeneity  behavior \eqref{eqn:homogeneousV}, at least if the potential is not generated by a too degenerate set of highest weight states $a^{m,n}_{j_{mn}}$ as we see in the remainder of the subsection.

Following the discussion on the $\slt$-representations in subsection \ref{sec:axion}, we now expand $G_4$ in the special basis generated 
from highest weight vectors as in \eqref{eqn:basisNa},
\begin{equation}
  G_4 = \sum_{(m, n) \in \cE} g_{m,n} v^{m,n}\, .
\end{equation}
In this expansion we have suppressed the sum over $j_{mn}$ to simplify the notation. This simplification will not alter our discussion about axion backreaction.
The crucial point is that the special basis \eqref{eqn:basisNa} allows us to split $G_4$ into 
terms as 
\beq \label{G4-dec}
   G_4 = \sum_\kappa G_4 (\hat a_\kappa)\ ,
\eeq
where $\hat a_\kappa$ are the highest weight vectors introduced in \eqref{eqn:highestWeightForm} and $G_4 (\hat a_\kappa)$ 
is the part of $G_4$ whose basis elements are only generated by $\hat a_\kappa$. Crucially, the decomposition \eqref{G4-dec} is
orthogonal with respect to the norms $|| \cdot ||_\infty$ and $|| \cdot ||_{\rm sl(2)}$ discussed in section \ref{aym_HodgeNorm}. It will therefore suffice 
to discuss the potential induced by the individual components $G_4 (\hat a_\kappa)$ and then add the various terms together. 

The next step is to carry out the expansion \eqref{rho3_N-} of the flux $\rho^-(G_4, \phi)$ into the special basis \eqref{eqn:basisNa}. It is straightforward to compute
\begin{align}
  \rho^-(G_4, \phi)   & = \sum_{a, b} \sum_{(m, n) \in \cE} \frac{(-1)^{a + b}}{a!b!} \phi^a \psi^b g_{mn} (N_1^-)^a (N_2^-)^b\, v^{m,n}\nonumber\\
                    & = \sum_{a, b} \sum_{(m', n') \in \cE} \frac{(-1)^{a + b}}{a!b!} \phi^a \psi^{b} g_{m' + 2a, n' + 2a + 2b} \, v^{m',n'}\, ,
\end{align}
where in the last equality we have shifted the sum over $(m, n)$, so we obtain the flux component
\begin{equation} \label{eqn:rhoExpandAxion}
  \varrho_{mn} = \sum_{a, b} \frac{(-1)^{a + b}}{a!b!} \phi^a \psi^{b} g_{m + 2a, n + 2a + 2b}\, ,
\end{equation}
for each $(m, n) \in \cE$.
For  each $ \varrho_{mn}$ we now extract the terms that have the leading growth in $\psi$ and then determine their contributions in the scalar potential using \eqref{general-norm-growth} in the strict asymptotic regime. 
Let us denote by $b_{mn}$ the highest power of $\psi$ appearing in \eqref{eqn:rhoExpandAxion} for which $g_{m + 2a, n + 2a + 2b_{mn}} \ne 0$. 
This implies the leading $\psi$ contribution in $\varrho_{mn}$ is given by 
\begin{equation} \label{eqn:rhoExpandLeading}
  \varrho_{mn} \sim \sum_{a} \frac{(-1)^{a + b_{mn}}}{a! b_{mn}!} \phi^a \psi^{b_{mn}} g_{m + 2a, n + 2a + 2b_{mn}}\, .
\end{equation}
In the strict asymptotic regime the leading scalar potential $V^{(\psi)}$ will then take the schematic form
\begin{equation} \label{eqn:leadingRhoV}
  V^{(\psi)}(u,\psi) \cong \sum_{(m, n) \in \cE} \psi^{2b_{mn}} u^{n - m}\ ,
\end{equation}
where the factor $2$ in the $\psi$-power arises due to the norm-squared appearing in the asymptotic growth expression \eqref{general-norm-growth}. Note that 
we have omitted all factors that are not related to $\psi$ and $u$. It is eminent that this $V^{(\psi)}(u,\psi)$ is not necessarily homogeneous
and rather one finds 
\begin{equation} \label{eqn:leadingRhoV_scale}
  V^{(\psi)}(\zeta u, \zeta \psi) \cong \sum_{(m, n) \in \cE}\zeta^{n - m + 2b_{mn}}\ \psi^{2b_{mn}} u^{n - m}\ ,
\end{equation}
whether or not one can factor out $\zeta$ as an overall scaling depends on the $b_{mn}$.

In order to identify the situations in which $V^{(\psi)}(u,\psi)$ given in \eqref{eqn:leadingRhoV} is actually homogeneous, we need to further 
characterize the exponents $n - m + 2b_{mn}$. Here the split \eqref{G4-dec} becomes important. Since 
the potential splits into a sum in this decomposition it will suffice to discuss one of the terms depending on one of the 
highest weight vectors $\hat a \equiv \hat a_{\kappa'}$. In other words, we study the potential generated by 
the flux $G_4(\hat a)$ and later piece all potentials together. It will also be important to  
introduce the highest power $\mu \equiv \mu(G(\hat a))$ of $N_2^-$ that does not annihilate $G_4(\hat a)$ as 
\begin{equation} \label{eqn:defMu}
  (N_2^-)^\mu G_4(\hat a) \ne 0,\quad (N_2^-)^{\mu + 1} G_4(\hat a) = 0\, .
\end{equation}
The axion $\psi$ appears in the scalar potential generated by this flux if  $\mu > 0$.
Let us now note that $g_{m + 2a, n + 2a + 2b_{mn}}$ entering the leading term in \eqref{eqn:rhoExpandLeading} is associated to the basis vector $v^{m + 2a, n + 2a + 2b_{mn}}$. Since we are concerned with the $G_4(\hat a)$ part of the potential, we know that this 
basis element can be obtained by acting on the highest weight vector $\hat a$ by acting with $(N_1^-)^c$, $(N_2^-)^d$ for some 
$c,d \geq 0$ as in \eqref{eqn:basisNa}. This implies that $\hat a \in  V_{m + 2a + 2c, n + 2a + 2b_{mn} + 2c + 2d}$ such that
\begin{equation} \label{va_rel}
  v^{m + 2a, n + 2a + 2b_{mn}} = (N_1^-)^{c } (N_2^-)^{d} \hat a\ .
\end{equation}
By using the definition \eqref{eqn:highestWeightForm} we know that the highest power $q \equiv q(\hat a)$ of $N_2^-$ that does not annihilate the highest weight 
vector $\hat a$ is given by 
\beq \label{qmn_exp}
    q = n - m + 2b_{mn} + 2d\ .
\eeq
Since $\mu$ is defined to be the highest power of $N_2^-$ that does not annihilate $G_4(\hat a)$, one also 
has 
\beq \label{N2onv}
 (N_2^-)^\mu v^{m + 2a, n + 2a + 2b_{mn}} \neq 0\ , \qquad  (N_2^-)^{\mu+1} v^{m + 2a, n + 2a + 2b_{mn}} = 0\ ,
\eeq 
since otherwise its flux component will not survive in the leading order of $\psi$ in $\varrho_{mn}$. 
Expressing the basis vector using the highest weight vector $\hat a$ by inserting \eqref{va_rel}
and using that \eqref{va_rel} contains $d$ additional powers of $N_2^-$ we infer that \eqref{N2onv} implies
\begin{equation} \label{eqn:homogeneousPower}
  \mu + d = q \ .
\end{equation}
Inserting \eqref{qmn_exp} into this expression we find the relation
\begin{equation} \label{mn_to_q}
  n - m + 2b_{mn} = 2\mu(G_4(\hat a)) - q(\hat a)\ .
\end{equation}
This equation completely determines the highest power of $\psi$ appearing in \eqref{eqn:leadingRhoV}. 
Note also that the left-hand side of this expression is the scaling of the individual terms in \eqref{eqn:leadingRhoV_scale},
while the right-hand side depends on the highest power $q(\hat a)$ of $N^-_2$ that does not annihilate $\hat a$
and the highest power $\mu(G_4(\hat a))$ of $N^-_2$ that does not annihilate $G_4(\hat a)$. 
In other words, we have translated the question of homogeneity into a condition on the highest weight state $\hat a$ and the 
flux $G_4(\hat a)$. By plugging this into \eqref{eqn:leadingRhoV_scale}, the flux scalar potential satisfies
\beq \label{Vpsi-split}
   V^{(\psi)}(\zeta u; \zeta \psi) = \sum_{\kappa} \zeta^{2\mu(G_4(\hat a)) - q(\hat a)}\  V^{(\psi)}(G_4(\hat a_\kappa))\ ,
\eeq
where we have used that we can split the leading flux scalar potential as 
$ V^{(\psi)}(u;\psi) = \sum_{\kappa}   V^{(\psi)}(G_4(\hat a_\kappa))\ $
since the involved norms split orthogonally with respect to the split \eqref{G4-dec}.  We can now determine under what circumstances the potential becomes homogeneous at large field as in \eqref{eqn:homogeneousV}.

The simplest case in which the homogeneity property \eqref{eqn:homogeneousV} of $V^{(\psi)}$ is realized arises when we assume that $G_4$ 
contains only flux directions generated from \textit{a single} $\hat a$. This implies that the sum \eqref{G4-dec} only contains a single term. 
In this case the homogeneity is immediate from \eqref{mn_to_q}, 
since there is just a single $q=q(\hat a)$ and each term in \eqref{eqn:leadingRhoV_scale} has the same power $\zeta^{2\mu - q}$.
Our main example discussed in section \ref{mainexample} displays such behavior, as all spaces $V_{mn}$ (except for $V_{44}$, which is not relevant here\footnote{The constant terms $V_{44}$ and $V_{\rm loc}$ always break the homogeneity property. However, it can be shown that, as long as there is some flux with a growth proportional to a positive power of $u$, then these constant terms only involve a subleading correction to \eqref{eqn:leadingRhoV} which becomes negligible for $\psi\gg 1$.}) can be spanned by basis vectors built from $\hat a = a_0$ and one has $q(\hat a) = d_2-d_1=3$.\footnote{Strictly speaking one has to transform the 
$a_0$ into its sl(2)-analog denote by $\tilde a_0$ in \cite{Grimm:2018cpv}. With the notation defined in \eqref{eqn:highestWeightForm}
one can also write this element as $a^{4 + d_1, 4 + d_2}$ with $(d_1, d_2) = (1, 4)$.} Clearly, our main example is very 
special in this respect. However, the simple homogeneity argument extends to many other fluxes also in other enhancements. Interestingly, cases in which there is a single highest weight state $\hat a_\kappa$ can be understood as arising from a superpotential in a two-dimensional moduli space. In these cases, all the flux terms arise from $\hat a = a_0$ and the linear backreaction is automatically satisfied.

On the other hand, the homogeneity is not automatic if \eqref{G4-dec} contains parts from different heights weight vectors $\hat a_\kappa$. This can occur, for instance, when there are more moduli than those sent to a limit,  and whose dependence is typically hidden in the value of $a_0$.
Assuming that the basis elements in $G_4$ are built from two 
highest weight vectors $\hat a_1, \hat a_2$, we need to check whether or not 
\beq  \label{check_cond}
    2\mu(G_4(\hat a_1))- q(\hat a_1) = 2\mu(G_4(\hat a_2)) - q(\hat a_2)\ . 
\eeq
In order to check if this condition can be violated we inspect Table \ref{tab:asymptoticSplittings} and read of the possible 
$q=y-x$ (and  $p= x-4$) of the highest weight vectors $\hat a_\kappa = \hat a_{\kappa}^{x,y}$ in each enhancement. 
The easiest way to violate \eqref{check_cond} is to consider the cases with $\mu=0$, in which some terms of the 
scalar potential are independent of $\psi$, and pick two appropriate highest weight vectors from 
Table \ref{tab:asymptoticSplittings}. 
More involved are situations in which $\mu>0$. In these cases, one identifies that only special 
fluxes in the enhancements  $\rI_{0, 1} \to \rI_{2, 2} $, 
$\rI_{0, 1} \to \rIII_{1, 1}$, $ \rI_{0, 1} \to \rV_{2, 2}$ can violate \eqref{check_cond}.\footnote{The 
enhancements violating  the to \eqref{check_cond} analog condition in the $(s,\phi)$ coordinates for $\mu>0$ are 
$ \rI_{1, 2} \to \rI_{2, 2},\ \rIII_{0, 1} \to \rIII_{1, 1}$, and $  \rV_{1, 2} \to \rV_{2, 2} $.} 

Let us note that violating \eqref{check_cond} does not imply that the linear relation \eqref{backreaction} is necessarily violated. 
In fact, we can proceed to order the terms $V^{(\psi)}(G_4(\hat a_\kappa))$ in \eqref{Vpsi-split} by their scaling with $\zeta$ and denote 
the highest weight component with maximal $2\mu - q$ by $\hat a_1$. Clearly, if the condition $\partial_u V^{(\psi)}(G_4(\hat a_1))=0$
allows to fix $u$ to a vacuum $\langle u \rangle_{1}$ then one has a linear backreaction $\langle u \rangle_{1} \sim \lambda_1 \psi$ as in \eqref{backreaction}. The additive terms appearing in the full $V^{(\psi)}$ are then only yielding sub-leading corrections that are proportional 
to $1/\psi^n$, $n\geq 0$. In other words, also in these more involved situations, one cannot avoid a leading term in $\langle u \rangle$ 
proportional to the axion at large field. 

It is also important to emphasize that the leading term in the axions for each flux term has the same coefficient given by the same internal flux, so it can be factorized out and plays no role in the minimization process. This implies that $\lambda$ in \eqref{backreaction} becomes a parameter $\lambda\sim \cO(1)$ independent of the fluxes for the case of $h^{3,1}=2$. Hence, in this case, one cannot use the fluxes to tune the parameter to be small, and the backreaction issues found in Calabi-Yau threefolds seem to be also present in the complex structure moduli space of Calabi-Yau fourfolds. If there are more moduli than those taken to the limit, $\lambda$ could also depend on these spectator moduli, but its precise numerical value and how much it can be tuned goes beyond the scope of this work. This nicely links to the results obtained in \cite{Hebecker:2014kva}. Interestingly, these properties 
remain to be true when replacing $N^-_i$ by $N_i$, i.e.~when returning to the original expression for the scalar potential, since the leading terms will keep their characteristic behavior. This further strengthens the deep link of these homogeneity properties to the underlying geometric structure
and deserves more study in the future.  

If the homogeneity result persists in general, it clearly has important implications for axion monodromy inflation. The fact that  $\lambda\sim \cO(1)$ implies that the backreaction cannot be delayed and that the exponential drop-off of the cut-off \eqref{cutoff} will occur as soon as the axionic field takes transplanckian field values. In this sense, inflating along an axionic direction does not allow one to travel further than inflating along the saxion, and both types of trajectories are sensitive to the exponential drop-off of the cut-off predicted by the Swampland Distance Conjecture \cite{Ooguri:2006in}. This is consistent with the refined Distance Conjecture \cite{Klaewer:2016kiy} and the transplanckian censorship \cite{Bedroya:2019snp}.
Let us recall, though, that we are only studying gradient-flow trajectories satisfying \eqref{dV_1}, while there could be other type of trajectories yielding successful inflation for a few times $M_p$. Hence, although highly constraining the structure of the asymptotic potentials, the phenomenological impact of our result is unclear. In any case, we find remarkable that the linear backreaction found in \cite{Baume:2016psm,Valenzuela:2016yny,Blumenhagen:2017cxt} is indeed tied to a deep underlying  mathematical structure arising at the asymptotic limits, which allow us to check the large field behavior of gradient flow trajectories in a model independent way and test in very general terms  the swampland conjectures \cite{Ooguri:2006in,Klaewer:2016kiy,Bedroya:2019snp,Draper:2019utz} that disfavor transplanckian field ranges.

\section{Conclusions}
\label{sec:con}

Motivated by the recent swampland conjectures on de Sitter and Anti-de Sitter vacua in string theory 
and progress on the Swampland Distance conjecture, we initiated in this work the systematic study of 
flux compactification at asymptotic regions in field spaces. Such asymptotic flux compactifications 
turn out to be remarkably constrained by the arising universal structure at the boundaries 
of geometric moduli spaces. This structure is described by asymptotic Hodge theory and 
corresponds to the appearance of so-called limiting mixed Hodge structures at each limit. While 
generally these constructions are mathematically involved, we have exploited two of their features in this work 
that are directly useful in flux compactifications: (1) the asymptotic expression for the Hodge norm and 
the asymptotic flux potential can be determined and systematically approximated, (2) the appearing regimes 
and asymptotic behaviors can be classified using $\slt^{\hat n}$-representation theory. 
Importantly, these statements are true for any Calabi-Yau fourfold and hence allow us to infer general 
conclusions about the validity of the swampland conjectures and common features of all effective theories arising 
in these asymptotic regimes.   

In order to systematically study the asymptotic flux scalar potentials we have focused in this work on 
F-theory compactifications on Calabi-Yau fourfolds with $G_4$ and then generalized the configurations to allow for non-positive 
definite potentials as they occur in Type IIA flux compactifications. We classified all possible 
two-field limits in such settings and determined all flux induced scalar potentials that can occur 
in the strict asymptotic regime. It is important to stress that these potentials are rather constrained 
and it seems hard to infer simple rules for their construction without referring to the underlying 
asymptotic Hodge theory. With this set of scalar potentials at hand, we were able to show 
that none of them possesses de Sitter vacua, at least, when demanding parametric control and looking at 
scaling limits of the coordinates. This allows us to establish a new no-go theorem for de Sitter in section \ref{sec:deSitter} extending the existing literature. This no-go is in accord 
with the recent asymptotic de Sitter conjecture \cite{Ooguri:2018wrx} and we showed that the latter is indeed satisfied if 
one focuses on infinite distance limits in F-theory flux compactifications. We did, however, not show that the bound on the potential suggested in 
\cite{Obied:2018sgi,Garg:2018reu,Ooguri:2018wrx} is satisfied for finite distance singular limits, but rather leave this as an interesting task for future research.  

It is interesting to highlight that our asymptotic approach sheds new light on flux vacua that have been 
investigated in the past \cite{Grana:2005jc,Douglas:2006es}. We have seen that imposing self-duality 
on the fluxes imposes simple conditions on the large moduli in the strict asymptotic regime, since the
associated Hodge operator identifies pairwise eigenspaces of the underlying $\slt$-structure. 
Self-duality in the vacuum ensures consistency with the equations of motion of F-theory and M-theory and leads to 
Minkowski vacua in these settings. To violate this condition only minimally, we introduced the notion of 
asymptotically massless flux by imposing that its Hodge norm vanishes when taking the asymptotic limit. Such 
fluxes can be unbounded by the tadpole constraint and are crucial when trying to engineer chains of vacua 
with parametrically controlled stabilized moduli. In fact, in our generalized settings these unbounded asymptotically massless 
fluxes allow to identify infinite chains of candidate AdS vacua at parametric control, in analogy to the $F_4$ flux component in the Type IIA vacua of \cite{DeWolfe:2005uu}. 
 It turns out 
that the demand for parametric control actually yields the conditions for being in the strict asymptotic 
regime and hence ensures self-consistency of our approximation. 
However, even if these candidate vacua seemingly become increasingly well controlled as one approaches large field values, more work would be required to ensure consistency of the global compactification. While we believe that our findings illuminate the 
underlying structure, which is also key in the Type IIA vacua of  \cite{DeWolfe:2005uu}, our analysis does not show the 
existence of these vacua. In fact, we have pointed out that the geometric requirements to have 
infinite chains of candidate AdS vacua appear to be similar to the ones relevant for the Swampland Distance Conjecture \cite{GPV,Grimm:2018cpv}, 
which puts conditions on the validity of effective theories. We hope that this refined understanding will eventually help to 
elucidate the status of such chains of AdS vacua. 

Last but not least, we also analyze the axion dependence on the scalar potentials and obtain universal features about the large field behavior of gradient flow trajectories. Deeply linked to the underlying mathematical structure, we get that the potential becomes to leading order a homogeneous function at large field for any asymptotic limit, implying a linear backreaction on the moduli when displacing the axions. This provides the geometric origin of the backreaction pointed out in \cite{Baume:2016psm} and extends it to Calabi-Yau fourfolds, so it can also potentially apply to D7-brane moduli. We also find that the parameter controlling the backreaction is flux-independent so it cannot be tuned small for $h^{3,1}=2$, constraining the length of these trajectories to transplanckian values before the effective theory breaks down. This sheds new light to the open debate \cite{Hebecker:2014kva,Blumenhagen:2014nba,Baume:2016psm,Valenzuela:2016yny,Bielleman:2016olv,Landete:2017amp,Blumenhagen:2017cxt,Kim:2018vgz}  about backreaction issues in F-term axion monodromy  \cite{Marchesano:2014mla,Hebecker:2014eua,Blumenhagen:2014gta,Ibanez:2014kia,Ibanez:2014swa} and provides evidence for the refined Distance Conjecture   \cite{Ooguri:2006in,Klaewer:2016kiy}. It would be interesting to study how much of this story can be extrapolated to other compactifications \cite{McAllister:2014mpa}. Let us remark, though, that our current asymptotic analysis does not actually establish strong constraints on inflation at the moment. Such large field inflationary models often only require axion displacements of a few orders $M_p$ which are not necessarily excluded even if $\lambda\sim \mathcal{O}(1)$  \cite{Scalisi:2018eaz}, in agreement with current experimental bounds.

Our findings immediately suggest several interesting and tractable problems to address in the future. 
To begin with, it would be desirable to extend the analysis of two-field limits to sending more fields 
to be large. The classification of all possible appearing structures would then allow one to analyze all 
possible flux scalar potentials, at least, in the strict asymptotic regime. It is curious to see if the no-goes 
on de Sitter vacua and the constructions of Anti-de Sitter vacua can be generalized, possibly by employing 
inductive arguments. A challenging but exciting task is to then leave the strict asymptotic approximation
and show that the findings persist.  
This would require to include corrections containing fractions of 
coordinates and induce numerous mixed terms into the flux scalar potentials.  
Furthermore, we are not including the effect of the warping on the geometry, so it would be interesting to study how this warping factor could modify the results. Eventually it is also desirable to generalize the classification of scalar potentials by going beyond Calabi-Yau manifolds. 
Let us stress that the used machinery, based on asymptotic Hodge theory, is not restricted to Calabi-Yau manifolds. In fact, it is actually algebraic in nature and not even requires the existence of 
an underlying geometric setting. 

In conclusion, the presented paper might be viewed as only a first step towards a much bigger goal of classifying the scalar potentials that can arise in string compactifications. The universal mathematical structure that emerges in the asymptotic regimes might not only allow us to test the Swampland Conjectures, but also yield new universal patterns and constraints that any low energy effective theory should satisfy to be consistent with a UV string theory embedding. Even if we are restricted to the asymptotic limits of the moduli space, let us recall that these regions correspond to regimes in which approximate global symmetries, weakly coupled gauge theories, and an Einstein gravity description typically arise. If one could show that the realization of these properties is necessarily tied to these asymptotic regimes, the systematic analysis of these limits could have important implications for phenomenology. For the moment, this universal structure indeed hints that there should be an underlying physical reason as of why all the effective field theories arising at these limits share some common features. This could be related to the restoration of global symmetries, the notion of emergence, the ubiquitous presence of string dualities, or something still to be discovered.

\subsubsection*{Acknowledgments}

We would like to thank David Andriot, Frederik Denef, Markus Dierigl, Damian van de Heisteeg, Luis Ib\'a\~nez, Fernando Marchesano, Miguel Montero, Eran Palti, and Yannik Zimmermann for very useful discussions and correspondence.
IV is supported by the Simons Foundation Origins
of the Universe program. 

\appendix
\section{Brief summary of the underlying mathematical machinery} \label{app:A}
In this section, we briefly introduce the mathematical machinery, theory of degenerating variation of Hodge structure, behind the asymptotic splitting \eqref{split-Vell} and the growth estimation \eqref{general-norm-growth}. More information on this theory for physicists can be found in \cite{GPV, Grimm:2018cpv, Corvilain:2018lgw, Grimm:2019wtx}. The original mathematical papers are \cite{Schmid, CKS, MR817170} and the discussion of enhancements among singularity types can be found in \cite{Kerr2017}.

While the whole machinery is very general, for the sake of concreteness, let us focus on the primitive cohomology $H^4_{\rm p}(Y_4, \bbC)$ of a Calabi-Yau fourfold $Y_4$. When the fourfold $Y_4$ is smooth, the cohomology $H^4_{\rm p}(Y_4, \bbC)$ enjoys a Hodge decomposition
\begin{equation} \label{eqn:HodgeDecomposition}
  H^4_{\rm p}(Y_4, \bbC) = \bigoplus_{p + q = 4} H^{p, q},
\end{equation}
where $H^{q, p} = \conj{H^{p, q}}$ and we denote by $H^{2, 2}$ the primitive part of the space of harmonic $(2, 2)$-forms. The above decomposition depends on the complex structure on the fourfold $Y_4$. As one deforms the complex structure while keeping $Y_4$ smooth, one varies the Hodge decomposition \eqref{eqn:HodgeDecomposition}. This is described by the theory of variation of Hodge structure on the vector space $H^4_{\rm p}(Y_4, \bbC)$.

As discussed in the main text of this paper, for interesting physics to occur, one often needs to push the complex structure moduli to certain limit in the complex structure moduli space, to the extent that one is left with a singular Calabi-Yau fourfold $Y_4$. When this happens, mathematicians showed that the cohomology $H^4_{\rm p}(Y_4, \bbC)$ of a singular Calabi-Yau usually cannot support a Hodge decomposition like \eqref{eqn:HodgeDecomposition}. Instead, another structure, called the limiting mixed Hodge structure replaces the role of the Hodge decomposition \eqref{eqn:HodgeDecomposition}. This structure is commonly defined in terms of filtrations, but here we refer to a characterization of such a structure showing its similarity with the Hodge decomposition \eqref{eqn:HodgeDecomposition}. A more precise description in terms of filtrations is provided in appendix \ref{app:B}.

To define a limiting mixed Hodge structure, one first fix the dimension of various subspaces $h^{4 - q, q} = \dim H^{4 - q, q}$ in \eqref{eqn:HodgeDecomposition}.
Then a limiting mixed Hodge structure on the primitive cohomology $H^4_{\rm p}(Y_4, \bbC)$ is given by a decomposition (called \emph{Deligne splitting})
\begin{equation} \label{eqn:DeligneSplitting}
  H^4_{\rm p}(Y_4, \bbC) = \bigoplus_{0 \le p, q \le 4} I^{p, q},
\end{equation}
where a generalized conjugation property given by \eqref{eqn:DeligneSplittingConjugation} on $I^{p, q}$ and $I^{q, p}$ holds. Moreover, certain conditions on the dimensions of the subspaces $I^{p, q}$ in \eqref{eqn:DeligneSplitting} have to be satisfied. They are
\begin{align}
  \dim I^{p, q} = \dim I^{q, p}, \quad \dim I^{p, q} = \dim I^{4 - q, 4 - p}, & \textrm{ for all } p, q,\nonumber\\
  \dim I^{p, q} \le \dim I^{p + 1, q + 1},                                    & \textrm{ for } p + q \le 2, \label{eqn:DimensionRules}\\
  \sum_{p = 0}^4 \dim I^{p, q} = h^{4 - q, q},                                & \textrm{ for all } q.\nonumber
\end{align}

As discussed in \cite{Kerr2017} and exemplified for Calabi-Yau threefolds in \cite{Grimm:2018cpv}, the conditions on the dimensions of the subspaces \eqref{eqn:DimensionRules} are enough to classify all\footnote{In fact, these conditions can fully classify the $\bbR$-split limiting mixed Hodge structures up to some change of basis on the middle cohomology of a Calabi-Yau threefold. On Calabi-Yau fourfolds, there might be complications \cite{Kerr2017}. We expect that these complications do not change much of our physical conclusions, and we will address these complications in future work.} possible ($\bbR$-split) limiting mixed Hodge structures \eqref{eqn:DeligneSplitting} on the middle cohomology $H^4_{\rm p}(Y_4, \bbC)$ up to some change of basis. Since the conditions are about numerical dimensions, it is handy to record these numbers on a $5 \times 5$ lattice, with the left-bottom corner representing $p = 0, q = 0$, and $p$ grows to the right horizontally while $q$ grows upwards vertically. These grids recording the dimensions of $I^{p, q}$ are called Hodge-Deligne diamonds. As an example, we show the list of Hodge-Deligne diamonds in the two-moduli example discussed in section \ref{sec:singularitiesAndEnhancements}. The results are given in table \ref{tab:HDenumeration}. Note how the diamonds reflect the conditions in \eqref{eqn:DimensionRules}.

\begin{table}[p]
  \centering
    \begin{tabularx}{0.8\textwidth}{|c|DDD|}\hline
      \multirow{6}{*}{\raisebox{-20ex}{$\rI$}}
      & $\rI_{0, 0}$                    & $\rI_{0, 1}$                    & $\rI_{0, 2}$\\
      & $\scriptstyle{(\hat{m} \ge 0)}$ & $\scriptstyle{(\hat{m} \ge 2)}$ & $\scriptstyle{(\hat{m} \ge 4)}$\\
      & \begin{tikzpicture}[scale=0.6] 
          \draw[step = 1, gray, ultra thin] (0, 0) grid (4, 4);

          \draw[fill] (0, 4) circle[radius=0.03];
          \draw[fill] (0.95, 3.05) circle[radius=0.03];
          \draw[fill] (1.05, 2.95) circle[radius=0.03];
          \draw (2, 2) node {$\scriptscriptstyle \hat{m}$};
          \draw[fill] (2.95, 1.05) circle[radius=0.03];
          \draw[fill] (3.05, 0.95) circle[radius=0.03];
          \draw[fill] (4, 0) circle[radius=0.03];
        \end{tikzpicture}&
        \begin{tikzpicture}[scale=0.6] 
          \draw[step = 1, gray, ultra thin] (0, 0) grid (4, 4);

          \draw[fill] (0, 4) circle[radius=0.03];
          \draw[fill] (1, 3) circle[radius=0.03];
          \draw[fill] (2, 3) circle[radius=0.03];
          \draw[fill] (1, 2) circle[radius=0.03];
          \draw (2, 2) node {$\scriptscriptstyle \hat{m} - 2$};
          \draw[fill] (3, 2) circle[radius=0.03];
          \draw[fill] (2, 1) circle[radius=0.03];
          \draw[fill] (3, 1) circle[radius=0.03];
          \draw[fill] (4, 0) circle[radius=0.03];
        \end{tikzpicture}&
        \begin{tikzpicture}[scale=0.6] 
          \draw[step = 1, gray, ultra thin] (0, 0) grid (4, 4);

          \draw[fill] (0, 4) circle[radius=0.03];
          \draw[fill] (1.95, 3.05) circle[radius=0.03];
          \draw[fill] (0.95, 2.05) circle[radius=0.03];
          \draw[fill] (2.05, 2.95) circle[radius=0.03];
          \draw[fill] (1.05, 1.95) circle[radius=0.03];
          \draw (2, 2) node {$\scriptscriptstyle \hat{m} - 4$};
          \draw[fill] (2.95, 2.05) circle[radius=0.03];
          \draw[fill] (1.95, 1.05) circle[radius=0.03];
          \draw[fill] (3.05, 1.95) circle[radius=0.03];
          \draw[fill] (2.05, 0.95) circle[radius=0.03];
          \draw[fill] (4, 0) circle[radius=0.03];
        \end{tikzpicture}\\
      & $\rI_{1, 1}$                    & $\rI_{1, 2}$                    & $\rI_{2, 2}$\\
      & $\scriptstyle{(\hat{m} \ge 1)}$ & $\scriptstyle{(\hat{m} \ge 3)}$ & $\scriptstyle{(\hat{m} \ge 2)}$\\
      & \begin{tikzpicture}[scale=0.6] 
          \draw[step = 1, gray, ultra thin] (0, 0) grid (4, 4);

          \draw[fill] (0, 4) circle[radius=0.03];
          \draw[fill] (1, 3) circle[radius=0.03];
          \draw[fill] (3, 3) circle[radius=0.03];
          \draw (2, 2) node {$\scriptscriptstyle \hat{m}$};
          \draw[fill] (1, 1) circle[radius=0.03];
          \draw[fill] (3, 1) circle[radius=0.03];
          \draw[fill] (4, 0) circle[radius=0.03];
        \end{tikzpicture}&
        \begin{tikzpicture}[scale=0.6] 
          \draw[step = 1, gray, ultra thin] (0, 0) grid (4, 4);

          \draw[fill] (0, 4) circle[radius=0.03];
          \draw[fill] (2, 3) circle[radius=0.03];
          \draw[fill] (1, 2) circle[radius=0.03];
          \draw[fill] (3, 3) circle[radius=0.03];
          \draw (2, 2) node {$\scriptscriptstyle \hat{m} - 2$};
          \draw[fill] (1, 1) circle[radius=0.03];
          \draw[fill] (3, 2) circle[radius=0.03];
          \draw[fill] (2, 1) circle[radius=0.03];
          \draw[fill] (4, 0) circle[radius=0.03];
        \end{tikzpicture}&
        \begin{tikzpicture}[scale=0.6] 
          \draw[step = 1, gray, ultra thin] (0, 0) grid (4, 4);

          \draw[fill] (0, 4) circle[radius=0.03];
          \draw[fill] (2.95, 3.05) circle[radius=0.03];
          \draw[fill] (3.05, 2.95) circle[radius=0.03];
          \draw (2, 2) node {$\scriptscriptstyle \hat{m}$};
          \draw[fill] (0.95, 1.05) circle[radius=0.03];
          \draw[fill] (1.05, 0.95) circle[radius=0.03];
          \draw[fill] (4, 0) circle[radius=0.03];
        \end{tikzpicture}\\
      \hline
      \multirow{3}{*}{\raisebox{-9.5ex}{$\rII$}}
      & $\rII_{0, 0}$                   & $\rII_{0, 1}$                   & $\rII_{1, 1}$\\
      & $\scriptstyle{(\hat{m} \ge 0)}$ & $\scriptstyle{(\hat{m} \ge 2)}$ & $\scriptstyle{(\hat{m} \ge 1)}$\\
      & \begin{tikzpicture}[scale=0.6] 
          \draw[step = 1, gray, ultra thin] (0, 0) grid (4, 4);

          \draw[fill] (1, 4) circle[radius=0.03];
          \draw[fill] (0, 3) circle[radius=0.03];
          \draw[fill] (1, 3) circle[radius=0.03];
          \draw (2, 2) node {$\scriptscriptstyle \hat{m}$};
          \draw[fill] (3, 1) circle[radius=0.03];
          \draw[fill] (4, 1) circle[radius=0.03];
          \draw[fill] (3, 0) circle[radius=0.03];
        \end{tikzpicture}&
        \begin{tikzpicture}[scale=0.6] 
          \draw[step = 1, gray, ultra thin] (0, 0) grid (4, 4);

          \draw[fill] (1, 4) circle[radius=0.03];
          \draw[fill] (0, 3) circle[radius=0.03];
          \draw[fill] (2, 3) circle[radius=0.03];
          \draw[fill] (1, 2) circle[radius=0.03];
          \draw (2, 2) node {$\scriptscriptstyle \hat{m} - 2$};
          \draw[fill] (3, 2) circle[radius=0.03];
          \draw[fill] (2, 1) circle[radius=0.03];
          \draw[fill] (4, 1) circle[radius=0.03];
          \draw[fill] (3, 0) circle[radius=0.03];
        \end{tikzpicture}&
        \begin{tikzpicture}[scale=0.6] 
          \draw[step = 1, gray, ultra thin] (0, 0) grid (4, 4);

          \draw[fill] (1, 4) circle[radius=0.03];
          \draw[fill] (0, 3) circle[radius=0.03];
          \draw[fill] (3, 3) circle[radius=0.03];
          \draw (2, 2) node {$\scriptscriptstyle \hat{m}$};
          \draw[fill] (1, 1) circle[radius=0.03];
          \draw[fill] (4, 1) circle[radius=0.03];
          \draw[fill] (3, 0) circle[radius=0.03];
        \end{tikzpicture}\\
      \hline
      \multirow{3}{*}{\raisebox{-9.5ex}{$\rIII$}}
      & $\rIII_{0, 0}$                  & $\rIII_{0, 1}$                  & $\rIII_{1, 1}$\\
      & $\scriptstyle{(\hat{m} \ge 2)}$ & $\scriptstyle{(\hat{m} \ge 4)}$ & $\scriptstyle{(\hat{m} \ge 3)}$\\
      & \begin{tikzpicture}[scale=0.6] 
          \draw[step = 1, gray, ultra thin] (0, 0) grid (4, 4);

          \draw[fill] (2, 4) circle[radius=0.03];
          \draw[fill] (0.95, 3.05) circle[radius=0.03];
          \draw[fill] (1.05, 2.95) circle[radius=0.03];
          \draw[fill] (0, 2) circle[radius=0.03];
          \draw (2, 2) node {$\scriptscriptstyle \hat{m} - 2$};
          \draw[fill] (4, 2) circle[radius=0.03];
          \draw[fill] (2.95, 1.05) circle[radius=0.03];
          \draw[fill] (3.05, 0.95) circle[radius=0.03];
          \draw[fill] (2, 0) circle[radius=0.03];
        \end{tikzpicture}&
        \begin{tikzpicture}[scale=0.6] 
          \draw[step = 1, gray, ultra thin] (0, 0) grid (4, 4);

          \draw[fill] (2, 4) circle[radius=0.03];
          \draw[fill] (1, 3) circle[radius=0.03];
          \draw[fill] (0, 2) circle[radius=0.03];
          \draw[fill] (2, 3) circle[radius=0.03];
          \draw[fill] (1, 2) circle[radius=0.03];
          \draw (2, 2) node {$\scriptscriptstyle \hat{m} - 4$};
          \draw[fill] (3, 2) circle[radius=0.03];
          \draw[fill] (2, 1) circle[radius=0.03];
          \draw[fill] (4, 2) circle[radius=0.03];
          \draw[fill] (3, 1) circle[radius=0.03];
          \draw[fill] (2, 0) circle[radius=0.03];
        \end{tikzpicture}&
        \begin{tikzpicture}[scale=0.6] 
          \draw[step = 1, gray, ultra thin] (0, 0) grid (4, 4);

          \draw[fill] (2, 4) circle[radius=0.03];
          \draw[fill] (1, 3) circle[radius=0.03];
          \draw[fill] (0, 2) circle[radius=0.03];
          \draw[fill] (3, 3) circle[radius=0.03];
          \draw (2, 2) node {$\scriptscriptstyle \hat{m} - 2$};
          \draw[fill] (1, 1) circle[radius=0.03];
          \draw[fill] (4, 2) circle[radius=0.03];
          \draw[fill] (3, 1) circle[radius=0.03];
          \draw[fill] (2, 0) circle[radius=0.03];
        \end{tikzpicture}\\
      \hline
      \multirow{3}{*}{\raisebox{-9.5ex}{$\rIV$}}
      & $\rIV_{0, 1}$                   & &\\
      & $\scriptstyle{(\hat{m} \ge 2)}$ & &\\
      & \begin{tikzpicture}[scale=0.6] 
          \draw[step = 1, gray, ultra thin] (0, 0) grid (4, 4);

          \draw[fill] (3, 4) circle[radius=0.03];
          \draw[fill] (2, 3) circle[radius=0.03];
          \draw[fill] (1, 2) circle[radius=0.03];
          \draw[fill] (0, 1) circle[radius=0.03];
          \draw (2, 2) node {$\scriptscriptstyle \hat{m} - 2$};
          \draw[fill] (4, 3) circle[radius=0.03];
          \draw[fill] (3, 2) circle[radius=0.03];
          \draw[fill] (2, 1) circle[radius=0.03];
          \draw[fill] (1, 0) circle[radius=0.03];
        \end{tikzpicture}
      &
      &\\
      \hline
      \multirow{3}{*}{\raisebox{-9.5ex}{$\rV$}}
      & $\rV_{1, 1}$                    & $\rV_{1, 2}$                    & $\rV_{2, 2}$\\
      & $\scriptstyle{(\hat{m} \ge 1)}$ & $\scriptstyle{(\hat{m} \ge 3)}$ & $\scriptstyle{(\hat{m} \ge 2)}$\\
      & \begin{tikzpicture}[scale=0.6] 
          \draw[step = 1, gray, ultra thin] (0, 0) grid (4, 4);

          \draw[fill] (1, 3) circle[radius=0.03];
          \draw[fill] (4, 4) circle[radius=0.03];
          \draw[fill] (3, 3) circle[radius=0.03];
          \draw (2, 2) node {$\scriptscriptstyle \hat{m}$};
          \draw[fill] (1, 1) circle[radius=0.03];
          \draw[fill] (0, 0) circle[radius=0.03];
          \draw[fill] (3, 1) circle[radius=0.03];
        \end{tikzpicture}&
        \begin{tikzpicture}[scale=0.6] 
          \draw[step = 1, gray, ultra thin] (0, 0) grid (4, 4);

          \draw[fill] (2, 3) circle[radius=0.03];
          \draw[fill] (1, 2) circle[radius=0.03];
          \draw[fill] (4, 4) circle[radius=0.03];
          \draw[fill] (3, 3) circle[radius=0.03];
          \draw (2, 2) node {$\scriptscriptstyle \hat{m} - 2$};
          \draw[fill] (1, 1) circle[radius=0.03];
          \draw[fill] (0, 0) circle[radius=0.03];
          \draw[fill] (3, 2) circle[radius=0.03];
          \draw[fill] (2, 1) circle[radius=0.03];
        \end{tikzpicture}&
        \begin{tikzpicture}[scale=0.6] 
          \draw[step = 1, gray, ultra thin] (0, 0) grid (4, 4);

          \draw[fill] (4, 4) circle[radius=0.03];
          \draw[fill] (2.95, 3.05) circle[radius=0.03];
          \draw[fill] (3.05, 2.95) circle[radius=0.03];
          \draw (2, 2) node {$\scriptscriptstyle \hat{m}$};
          \draw[fill] (0.95, 1.05) circle[radius=0.03];
          \draw[fill] (1.05, 0.95) circle[radius=0.03];
          \draw[fill] (0, 0) circle[radius=0.03];
        \end{tikzpicture}\\\hline
    \end{tabularx}
  \caption{Sixteen possible Hodge-Deligne diamonds with $h^{3, 1} = 2$, corresponding to 16 singularity types given in table \ref{tab:singularityTypes}. We denote $h^{2, 2} = \hat{m}$ and $i^{p, q} = \dim I^{p, q}$. Then the number of dots around the lattice point at $(p, q)$ represents the value of $i^{p, q}$, and the label at $(2, 2)$ represents the value of $i^{2, 2}$. The subscripts under a type are recording $i^{3, 3}$ and $i^{3, 3} + i^{3, 2}$, respectively.} \label{tab:HDenumeration}
\end{table}

The enhancement relations of the form $\mathsf{Type\ A}  \to \mathsf{Type\ B} $ are then derived by decomposing the Hodge-Deligne diamond of $\mathsf{Type\ A} $ and recombining into the diamond of $\mathsf{Type\ B} $ in a way coherent with the $\slt$-triples \eqref{triples}. Precise statements of the recipe can be found in \cite{Kerr2017} and exemplified in \cite{Grimm:2018cpv}. For the two-moduli case, the enhancement network is displayed in figure \ref{fig:enhancement2moduli}. When we derive each enhancement relation, we obtain the asymptotic splitting \eqref{split-Vell} simultaneously. Explicit example of this asymptotic splitting in Calabi-Yau threefolds can be found in \cite{Grimm:2019wtx}. We can also characterize the asymptotic splitting as follows. Let $\bigoplus I^{p, q}_{\mathsf{A}}$ and $\bigoplus I^{r, s}_{\mathsf{B}}$ denote the Deligne splitting of $\mathsf{Type\ A} $ and $\mathsf{Type\ B} $, respectively. Then
\begin{equation}
  V_{mn} \cong \left(\bigoplus_{p + q = m} I^{p, q}_{\mathsf{A}}\right) \cap \left(\bigoplus_{r + s = n} I^{r, s}_{\mathsf{B}}\right),
\end{equation}
as complex vector spaces, where $V_{mn}$ is understood to be the complexification. Note that in the above expression it is valid to take the intersection on the RHS as $I^{p, q}_{\mathsf{A}}$ and $I^{r, s}_{\mathsf{B}}$ are subspaces of the same vector space, $H^4_{\rm p}(Y_4, \bbR)$. This finishes our brief discussion on the underlying mathematical machinery.

\section{Norms associated with some special Hodge structures} \label{app:B}
In this section, we introduce norms associated with various special Hodge structures, including the asymptotic norm $\Vert \cdot \Vert^2_\infty$, which appears in the coefficients in the asymptotic Hodge norm \eqref{general-norm-growth}, and Weil operators $C_\infty, C_{\rm sl(2)}$ inducing these norms.  We will adopt an approach different from appendix \ref{app:A}, by introducing everything in terms of filtrations so that the interested reader can compare the statements with mathematical literature \cite{Schmid, CKS}.

First we recall the definition of polarized Hodge structure on the primitive cohomology $H^4_{\rm p}(Y_4, \bbC)$. To ease the notation, we denote the underlying integral cohomology $H_\bbZ := H^4_{\rm p}(Y_4, \bbZ)$ and its complexification $H := H^4_{\rm p}(Y_4, \bbC)$. A polarized Hodge structure of weight $4$ on $H$ is given by a decreasing filtration
\begin{equation} \label{eqn:HodgeFiltration}
  0 \subs F^4 \subs F^3 \subs F^2 \subs F^1 \subs F^0 = H,
\end{equation}
such that
\begin{equation} \label{eqn:4-opposed}
  F^p \oplus \conj{F}^{5 - p} \cong H,\quad \textrm{ for all } p.
\end{equation}
We denote this Hodge structure on $H$ by $F$ when there is no danger of confusion. With such definition of a Hodge structure, one can rewrite it into the form of Hodge decomposition in \eqref{eqn:HodgeDecomposition} by setting the subspaces $H^{p, q} = F^{p} \cap \conj{F}^q$, where $p + q = 4$.

A Hodge structure $F$ given by \eqref{eqn:HodgeFiltration} also defines a Weil operator $C_F$, which is a linear automorphism of $H$ such that when restricted to the subspace $H^{p, q} = F^{p} \cap \conj{F}^q$, it acts as a scalar multiplication by $i^{p - q}$: For every $v \in H^{p, q}$, define $C_F(v) = i^{p - q} v$. A polarization form of the structure \eqref{eqn:HodgeFiltration} is given by an integer-valued bilinear form on $H_\bbZ$, $S: H_\bbZ \times H_\bbZ \to \bbZ$ and extended to the whole complexified space $H$ linearly, such that $S(F^p, F^{5 - p}) = 0$ for all $p$ and $S(C_F(v), \conj{v}) > 0$ for all non-zero $v \in H$.

To compare with the existing discussion in section \ref{aym_HodgeNorm}, we see that taking $F^p = \oplus_{r \ge p} H^{r, 4 - r}$ in the Hodge decomposition \eqref{eqn:HodgeDecomposition} gives us a Hodge filtration. Also the role of the Hodge star operator on the Calabi-Yau fourfold is played by the Weil operator, and the polarization form is given by the intersection bilinear form $\pair{v, w} = \int_{Y_4} v \wedge w$.

When we move into limits in the complex structure moduli space, in general the cohomology $H$ will not support the existence of a pure Hodge structure. But we can still study the behavior of the Hodge structure in the limit. The machinery allowing such study is given by the theory of limiting mixed Hodge structures, which is developed in \cite{Schmid, CKS}. The essential tool is a special generalization of pure Hodge structures, called limiting mixed Hodge structures. Let us briefly introduce such structures.

To define a limiting mixed Hodge structure on $H$, one still needs to specify a decreasing filtration
\begin{equation}
  0 \subs F^4 \subs F^3 \subs F^2 \subs F^1 \subs F^0 = H,
\end{equation}
but we do \emph{not} impose conjugation property \eqref{eqn:4-opposed} on these $F^p$ subspaces.

Furthermore, a new ingredient, weight filtration comes into the game. In the context of limiting mixed Hodge structures associated with limits of Hodge structure \eqref{eqn:HodgeFiltration} of weight $4$, the weight filtration is given by the \emph{monodromy weight filtration} $W(N)$ depending on a given real nilpotent operator $N$,
\begin{equation}
  0 \subs W_0(N) \subs W_1(N) \subs \cdots \subs W_8(N) = H_\bbR,
\end{equation}
where $H_\bbR := H^4_{\rm p}(Y_4, \bbR)$. The monodromy weight filtration is defined as the unique increasing filtration on $H_\bbR$ such that
\begin{IEEEeqnarray}{rCl}
  N W_k(N)                               & \subs                           & W_{k - 2}(N),\\
  N^k: \frac{W_{4 + k}(N)}{W_{3 + k}(N)} & \overset{\sim}{\longrightarrow} & \frac{W_{4 - k}(N)}{W_{3 - k}(N)},
\end{IEEEeqnarray}
for all $k$. There is a compatibility condition on the filtrations $F$ and $W$: On each graded quotient $\frac{W_{k}(N)}{W_{k - 1}(N)}$, the filtration $F$ induces a pure Hodge structure of weight $k$. Precise discussions on the definition of mixed Hodge structures can be found in \cite{CKS}. We often denote a limiting mixed Hodge structure by $(F, W(N))$ or simply $(F, N)$.

The filtrations $F$ and $W$ are related to the $I^{p, q}$ splittings discussed in appendix \ref{app:A}. In fact, the splitting $H = \bigoplus I^{p, q}$ is defined to be the unique splitting\cite{CKS} such that, for all $p, q, k$,
\begin{align}
  F^p      & = \bigoplus_{r \ge p} I^{r, s},\\
  W_k      & = \bigoplus_{r + s \le k} I^{r, s},\\
  I^{p, q} & = \conj{I^{q, p}} \mod \bigoplus_{\substack{r < p\\s < q}} I^{r, s}.\label{eqn:DeligneSplittingConjugation}
\end{align}

The `big mod' in the last condition \eqref{eqn:DeligneSplittingConjugation} looks annoying and those mixed Hodge structures without this `big mod' deserves a special name. A mixed Hodge structure such that its $I^{p, q}$-splitting satisfies $I^{p, q} = \conj{I^{q, p}}$ for all $p, q$ is said to be $\bbR$-split. To every mixed Hodge structure, Deligne\cite{CKS, MR1265541} constructed a real operator $\delta$ such that the mixed Hodge structure $(e^{-i \delta}F, W)$ is $\bbR$-split. This operator $\delta$ is unique with certain properties, and we refer the reader to \cite{CKS} for full discussion.

The limiting mixed Hodge structures and pure Hodge structures in limits are related by the nilpotent orbit theorem. Recall that locally the limit in the complex structure moduli space is given in local coordinates $t^1, \ldots, t^{\hat{n}}$ by sending $t^1, \ldots, t^{\hat{n}} \to i\infty$. To each singular locus $t^j \to i\infty$ there is an associated nilpotent operator $N_i$, the logarithm of the monodromy operator. We usually record the change of the $(4, 0)$-form $\Omega$ in a variation of Hodge structure on the primitive middle cohomology of a Calabi-Yau fourfold by period integrals. To describe the content of nilpotent orbit theorem, one needs to describe the dependence of the full Hodge filtration $F(t)$ on the complex structure moduli. The nilpotent orbit theorem tells us that the varying Hodge filtration $F(t)$ in the limit $t \to i\infty$ can be approximated by the so-called nilpotent orbit $F_{\rm nil}(t)$, which is a filtration given by
\begin{equation}
  F_{\rm nil}(t) = e^{\sum_{i} t^i N_i} F_{\rm nil},
\end{equation}
where $F_{\rm nil}$ is the decreasing filtration defining a mixed Hodge structure, i.e., it does not necessarily satisfy \eqref{eqn:4-opposed}. In section \ref{asympt-limits}, the vector generating the subspace $F_{\rm nil}^4$ is denoted by $a_0$. The filtration $F_{\rm nil}$ and the monodromy weight filtration $W^{\hat{n}} := W(N_1 + \cdots + N_{\hat{n}})$ together define the limiting mixed Hodge structure $(F_{\rm nil}, W^{\hat{n}})$ associated to the degeneration of Hodge structure $F(t)$. From the discussion in the last paragraph, there is an operator $\delta$ associated to this limiting Hodge structure such that $(e^{-i \delta}F_{\rm nil}, W^{\hat{n}})$ is $\bbR$-split.

In section \ref{flux-split}, we also discussed that when a singularity enhancement occurs, a set of commuting $\slt$-triples with nilnegative elements $N^-_1, \ldots, N^-_{\hat{n}}$ can be associated to such an enhancement. One conclusion of the $\mathrm{sl}(2)$-orbit theorem in \cite{CKS} states that, the filtration defined by
\begin{equation}
  F_{\infty} = e^{i \sum_{i = 1}^{\hat{n}} N_i^-} e^{-i \delta} F_{\rm nil},
\end{equation}
actually satisfies \eqref{eqn:4-opposed}: $F_{\infty}^p \oplus \conj{F}^{5 - p}_{\infty} \cong H$, for all $p$, and is polarized by the intersection bilinear form $\pair{\,\cdot\, , \,\cdot\,}$. In other words, $F_{\infty}$ is a pure Hodge filtration of weight $4$ polarized by $\pair{\,\cdot\, , \,\cdot\,}$. Let $C_{\infty}$ be its Weil operator, then the asymptotic norm is defined by
\begin{equation}
  \Vert v \Vert_\infty^2 = \pair{C_{\infty}v, \conj{v}}.
\end{equation}

What remains to be defined is the operator $C_{\rm sl(2)}$. This is the operator that brings the $s^i/s^{i + 1}$ scaling into the norm estimate \eqref{general-norm-growth}. It is defined via the help of another operator
\begin{equation}
  e(s) := \prod_{j = 1}^{\hat{n}} \exp\left\{\frac{1}{2} \log(s^j) Y_j\right\},
\end{equation}
where $Y_i$ are the neutral elements in the commuting $\slt$-triples \eqref{triples}, and $e(s)$ operates on each subspace $V_\Bell$ as a scalar multiplication by 
\begin{equation*}
  \big(\frac{s^1}{s^2}\big)^{\frac{l_1 - 4}{2}} \cdots \big(\frac{s^{\hat{n} - 1}}{s^{\hat{n}}}\big)^{\frac{l_{\hat{n} - 1} - 4}{2}} (s^{\hat{n}})^{\frac{l_{\hat{n}} - 4}{2}}.
\end{equation*}
Then the operator $C_{\rm sl(2)}$ is defined as
\begin{equation} \label{eqn:defnCSL2}
  C_{\rm sl(2)} := e^{-1}(s) C_\infty e(s).
\end{equation}
It is clear that the norm defined under $C_{\rm sl(2)}$ is given by expression \eqref{general-norm-growth}. Note also that for any real vector $v$ one has $\Vert v \Vert^2_{\rm sl(2)} = \pair{C_{\rm sl(2)} v, v} = \pair{C_\infty e.v, e.v}$, as $e$ is an isometry of the polarization pairing. So our condition \eqref{self-dual_sl2} on asymptotic self-dual fluxes $G_4$ with respect to $C_{\rm sl(2)}$ can also be written as a self-duality condition on $e.G_4$ with respect to $F_\infty$.

\bibliographystyle{jhep}
\bibliography{AntiDeSitter}

\end{document}